\documentclass[twocolumn,superscriptaddress,preprintnumbers,amsmath,amssymb,prd]{revtex4}

\usepackage{graphicx}% Include figure files
\usepackage{dcolumn}% Align table columns on decimal point
\usepackage{bm}% bold math
\usepackage{indentfirst} 
\usepackage{color}
\usepackage{CJK}
\usepackage{booktabs}
\usepackage{float}
\usepackage[colorlinks,linkcolor=blue,urlcolor=blue,citecolor=blue]{hyperref}

\usepackage{array} 
\usepackage{caption}

\usepackage{tikz,xcolor}

\definecolor{lime}{HTML}{A6CE39}
\DeclareRobustCommand{\orcidicon}{
	\begin{tikzpicture}
	\draw[lime, fill=lime] (0,0) 
	circle [radius=0.16] 
	node[white] {{\fontfamily{qag}\selectfont \tiny ID}};
	\draw[white, fill=white] (-0.0625,0.095) 
	circle [radius=0.007];
	\end{tikzpicture}
	\hspace{-2mm}
}
\foreach \x in {A, ..., Z}{%
	\expandafter\xdef\csname orcid\x\endcsname{\noexpand\href{https://orcid.org/\csname orcidauthor\x\endcsname}{\noexpand\orcidicon}}
}
\foreach \x in {A, ..., Z}{%
	\expandafter\xdef\csname orcid\x\endcsname{\noexpand\href{https://orcid.org/\csname orcidauthor\x\endcsname}{\noexpand\orcidicon}}
}

\begin{document}
\begin{CJK*}{UTF8}{gbsn}

\title{Scaling-Based Quantization of Spacetime Microstructure}
%\title{Spacetime Fluctuation and Quantization}
%\title{Anisotropic Scaling Fluctuations and Quantization of Planck-Scale Spacetime}
%\title{Scale Operators and Quantization of Spacetime Microstructure}
%\title{Hierarchical Scale-Space Geometry and Quantum Fluctuations}
%\title{Quantum Spacetime as a Hierarchy of Scale Manifolds}
%\title{Quantizing Spacetime Geometry via Anisotropic Scaling Fluctuations at the Planck Scale}

\author{Weihu Ma(马维虎)\orcidA{}}
    \email{maweihu@fudan.edu.cn}% Your name
%    \affiliation{Institute of Modern Physics, Fudan University, Shanghai 200433, People's Republic of China}
    \affiliation{Key Laboratory of Nuclear Physics and Ion-beam Application (MOE), Institute of Modern Physics, Fudan University, Shanghai 200433, China}
    \affiliation{Shanghai Research Center for Theoretical Nuclear Physics, NSFC and Fudan University, Shanghai 200438, China}
\author{Yu-Gang Ma(马余刚)\orcidB{}}
    \email{mayugang@fudan.edu.cn}% Your name
    \affiliation{Key Laboratory of Nuclear Physics and Ion-beam Application (MOE), Institute of Modern Physics, Fudan University, Shanghai 200433, China}
    \affiliation{Shanghai Research Center for Theoretical Nuclear Physics, NSFC and Fudan University, Shanghai 200438, China}
    \affiliation{School of Physics, East China Normal University, 200241, Shanghai, China}

\date{\today} % Leave empty to omit a date
\begin{abstract}
Planck-scale physics challenges the classical smooth-spacetime picture by introducing quantum fluctuations that imply a nontrivial spacetime microstructure. We present a framework that encodes these fluctuations by promoting local scale factors, rather than the metric tensor, to fundamental dynamical variables while preserving general covariance. The construction employs a two-tiered hierarchy of scale manifolds, comprising a first-order manifold of scale coordinates and a second-order manifold of fluctuation amplitude coordinates. On the first-order manifold, we formulate differential geometry, field equations, and a canonical quantization procedure. The theory yields a geometric renormalization-group flow for scale variables and implies spacetime discreteness at the microscopic level. By constructing a quadratic action and performing spectral decomposition with a stabilizing potential, we obtain discrete modal degrees of freedom quantized as harmonic oscillators. The framework proposes a microscopic description for zero-point energy of spacetime and explores implications for vacuum energy and ultraviolet regularization, suggesting a potential dynamical mechanism that could ameliorate the cosmological constant problem. Main results include a generalized uncertainty relation with scale-dependent coefficients, locally scaled Klein-Gordon and Dirac equations, geodesic equations for scale spacetime, and a microscopic area operator whose state counting is consistent with the Bekenstein-Hawking entropy. This work develops a scale-based quantization procedure, providing a foundation for further mathematical analysis and phenomenological tests of spacetime quantization.
\end{abstract}

\keywords{Spacetime fluctuation\sep Scale operator Quantization\sep Scale manifold\sep Micro area operator \sep Black-hole entropy\sep Scaling renormalization group}

\maketitle

\section{Introduction}
The reconciliation of quantum mechanics with general relativity remains one of the most profound challenges in fundamental physics. At the Planck scale, where the characteristic length is approximately \( l_p \sim 10^{-35} \) meters, classical concepts of spacetime geometry break down due to intrinsic quantum fluctuations. These fluctuations render the smooth manifold picture of spacetime inadequate, necessitating a reformulation of geometry and measurement consistent with quantum principles.

Various leading approaches to quantum gravity offer distinct perspectives on the microscopic structure of spacetime. Loop quantum gravity posits that space is fundamentally discrete, with areas and volumes arising from networks of quantized spin states \cite{Ashtekar1986,Jacobson1988,Rovelli1990,rovelli2004,Rovelli2008}. String theory instead describes all particles and interactions as different vibrational modes of one-dimensional strings propagating in a higher-dimensional background \cite{Green1987,Polchinski1998,Becker2007,Dine2007,green2012}. The asymptotic safety program conjectures that the gravitational coupling approaches a nontrivial ultraviolet fixed point under renormalization, ensuring predictivity at all scales \cite{Addazi2022,Niedermaier2006}. Noncommutative geometry replaces classical coordinates with noncommuting operators, encoding geometry algebraically and potentially smoothing out short‐distance singularities \cite{Connes1994}. Finally, causal set theory models spacetime as a discrete, partially ordered set of events, where the causal relations themselves constitute the fundamental structure \cite{sorkin2005,Loll2019}. Recent research in quantum gravity reveals various new approaches and theories \cite{Partanen2025,Klauder,Oriti,Berenstein,Shojai,Carlip,Harlow,Pawlowski,Braunstein,Giesel,King,Bajardi,Chakraborty,Saueressig,Bianchi}. Although these studies have greatly improved our understanding of physics at the Planck scale, an unified framework that connects geometric fluctuations to observable phenomena remains elusive.

Ho\v{r}ava-Lifshitz gravity \cite{Horava2009} introduces an anisotropic scaling between space and time, characterized by a dynamical critical exponent $z>1$, which renders the theory power‐counting renormalizable in the ultraviolet regime while recovering general relativity in the infrared limit.  By breaking full diffeomorphism invariance down to the subgroup preserving a preferred foliation of spacetime, the model allows higher-order spatial derivative terms (up to $2z$ derivatives) without introducing ghost instabilities, thereby improving the UV behavior of the graviton propagator.  Subsequent studies have explored its cosmological implications, demonstrating that the modified dispersion relations can lead to novel early-universe dynamics and potential observational signatures in the cosmic microwave background and gravitational-wave spectra \cite{Mukohyama2010}. Despite its appealing features, Ho\v{r}ava-Lifshitz gravity involves certain subtleties. The reduced diffeomorphism symmetry leads to an additional scalar graviton mode, whose role in low-energy dynamics has been widely explored \cite{Charmousis2009,Blas2009}. The restoration of Lorentz invariance in the infrared typically involves specific mechanisms, and various approaches have been proposed to address strong coupling and define a consistent Hamiltonian structure \cite{Papazoglou2009,Blas2011}. These aspects continue to guide the development and refinement of the theory.

Recent advances in observational cosmology and gravitational-wave astronomy have begun to place stringent bounds on possible deviations from classical general relativity at Planckian scales.  In particular, the analysis of the first direct detection of gravitational waves by LIGO and Virgo (GW150914) has been used to test the propagation speed, dispersion relations, and polarization content of gravitational radiation, finding agreement with Einstein's theory to within parts in $10^{-15}$ \cite{LIGO2016}.  Complementarily, measurements of the cosmic microwave background by the Planck satellite have constrained the spectrum of primordial perturbations and the tensor-to-scalar ratio, thereby limiting the parameter space of inflationary models that incorporate quantum-gravity-induced corrections to the primordial power spectrum \cite{Planck2018}.  Together, these empirical results impose critical benchmarks for any theory of quantum gravity that predicts modifications to spacetime dynamics or fundamental dispersion relations.  

The study of physics at the Planck scale has garnered significant attention due to its implications for understanding the fundamental nature of the universe. At the heart of this challenge lies the incompatibility between the smooth, deterministic spacetime geometry of general relativity and the inherently probabilistic, fluctuating nature of quantum systems at microscopic scales. At the Planck scale, quantum fluctuations challenge the classical concept of spacetime as a smooth continuum, revealing a complex microstructure that defies traditional models. The Heisenberg uncertainty principle \cite{Heisenberg1927} implies that spacetime coordinates themselves become uncertain at this scale, resulting in fluctuations in microscopic lengths. This fundamental uncertainty renders traditional geometric concepts inadequate for an accurate description. Our prior article \cite{weihu} laid the foundation by introducing a scale-based spacetime framework, in which local scaling functions were incorporated into key physical equations, including the geodesic, Einstein, Klein-Gordon, and Dirac equations, to reflect the effects of microscopic spacetime fluctuations. While that work focused primarily on reformulating classical geometric and field-theoretic structures in a scale-sensitive manner, the present study represents a significant extension by formulating a two-tiered scale geometry (via scale fluctuating factors $ a_\mu$ and $ b_\mu$), introducing canonical quantization of spacetime fluctuations, and constructing scale geometry dynamics on these quantum scale manifolds. This allows us to construct a framework for the quantization and renormalization of fluctuating spacetime itself.
The current research advances our understanding of Planck-scale physics by extending the scale-based theoretical framework proposed in prior work, while introducing novel quantization for characterizing quantum spacetime microstructure. 

Building on the general quantum-gravity landscape outlined above, this work introduces several key innovations. First, we define local anisotropic scaling measurements \(L^\alpha(X^\alpha)\) and their associated scale fluctuating factors \( a_\alpha\) and \( b_\alpha\). Second, these scale factors naturally give rise to two successive scale manifolds---a first-order scale manifold \((\mathcal{M}^{( a)},\hat g)\) with coordinates \(X^\alpha\) and a second-order amplitude manifold \((\mathcal{M}^{( b)},\tilde g)\) with coordinates \( a_\alpha\)---each endowed with its own Levi-Civita connection and geodesic equations governing the evolution of scaling spacetime. 
Third, by formulating Einstein-Hilbert actions and performing a canonical quantization of the fluctuation factors, this framework aims to understand how intrinsic quantum fluctuations reshape the local spacetime geometry.

The paper is organized as follows.
\textbf{Section II} introduces the operational principle of measuring infinitesimal intervals against fixed reference lengths and develops the dual-measurement picture that implies a discrete microscopic spacetime. 
In \textbf{Section III}, we define fluctuation-dependent differential operators, derive the deformed commutator algebra and its generalized uncertainty relations, and present locally scaled Klein-Gordon and Dirac equations. 
\textbf{Section IV} establishes Lorentz and general covariance in the scaling geometry and embeds the scale fluctuating factors $a_\mu$ into a vierbein formulation. 
\textbf{Section V} derives variational first- and second-order geodesic equations on the scale and amplitude manifolds. 
In \textbf{Section VI}, we construct a scaling Einstein-Hilbert Lagrangian and perform canonical quantization of the first-order fluctuating factors $a_\mu$, quantized to $\hat a_\alpha$. 
\textbf{Section VII} defines a micro-area operator and applies it to black-hole entropy counting. 
Finally, \textbf{Section VIII} discusses implications, limitations, and directions for further work, and \textbf{Section IX} concludes.

Because the defined scale fluctuating factors $a_\alpha$ transform differently from ordinary tensor components (see Section~IV), we adopt a careful, context-dependent convention for Einstein summation throughout the paper; the precise rules and illustrative examples are collected right after defining the scale factor $a_\alpha$ in Section~II.

\section{Micro-measurement}
\subsection{The principle of Micro-measurement}
As the micro measurement principle stated in \cite{weihu}, at the microscopic scale, measuring the distance between two infinitesimally close spacetime points, denoted by \( dx \), requires comparison with a fixed reference length. If \( dx \) is stable, its measurement depends solely on the reference. However, if \( dx \) fluctuates -- due to quantum effects as suggested by the Heisenberg uncertainty principle -- its measurement relative to the reference becomes a scale-dependent function, i.e., scaling function $L^\alpha(X^\alpha)$. In quantum gravity, such linear and nonlinear fluctuations reflect spacetime's dynamic nature at small scales. Describing spacetime microelements through scaling functions provides a framework for probing their microstructure and developing new mathematical tools to study it.

In quantum gravity regimes, the fluctuations undermine the applicability of classical smooth geometry and introduce intrinsic uncertainties in position measurements. To model this behavior, we consider differential microelements $dx^\alpha$ relative to reference lengths $dx_0^\alpha$ as micromeasurement, where $\alpha$ denotes orthonormal frame coordinates. At the Planck scale, quantum fluctuations of spacetime cause measurements of position and distance to be indescribable by classical geometry, and we introduce a dynamic scale function $L^\alpha(X^\alpha)$ to quantify this uncertainty.

The micro measurements of the  fluctuating spacetime are quantified through scaling functions:
\begin{equation}
r_{l^\alpha} = \frac{dx^\alpha}{dx_0^\alpha} = L^\alpha(X^\alpha(\tau))
\label{eq:microeq1}
\end{equation}
where $L^\alpha(X^\alpha(\tau))$ describe deviations from reference measurements $dx_0^\alpha$, denoted by spacetime scales \(X^{\alpha}\). The latter can be regarded as a scale of freedom for measuring the microscopic length of a local region. These functions capture quantum fluctuations in spacetime at the Planck scale. The micro variations of $dx^\alpha$, i.e., $d(dx^\alpha)\sim dL^\alpha$, are defined as the fluctuations, providing a way to model the geometry of spacetime as it fluctuates due to quantum effects. 
For proper–length measurements, we introduce the relation
\begin{equation}
r_l = \frac{d\lambda}{d\lambda_0} = \tau(X(s))
\end{equation}
where \( d\lambda \) denotes the infinitesimal physical proper-length element along a worldline in the fluctuating spacetime, and \( d\lambda_{0} \) is the corresponding reference proper length.
The function \( \tau(X) \) is a scalar scaling function that characterizes how microscopic spacetime fluctuations modify proper-length measurements. It plays a role analogous to the directional scaling functions \( L^{\alpha}(X^{\alpha}) \), but is applied to the Lorentz-invariant proper scale.
In this expression, \( X = X(s) \) indicates that the proper scale is evaluated along the worldline parameterized by the proper-time scale \( s \). The parameter \( s \) is the affine parameter associated with a timelike worldline.
The relative measurement becomes:
\begin{equation}
\frac{dx^\alpha}{d\lambda} = \frac{L^\alpha(X^\alpha(\tau))}{\tau(X(s))}
\end{equation}
with gauging $dx_0^\alpha = d\lambda_0$. A natural candidate for the reference scale $dx_0^\alpha$ is the Planck length in quantum gravity regimes, which we adopt in what follows. Notably, these reference differentials $dx_0^\alpha$ ($d\lambda_0$) remain constant in differentiation processes, representing dimensional constants rather than dynamic variables.

Based on the above definitions, the spacetime microstructure measurements manifest through three distinct types: 
(1) \textit{Static mode} with $L^\alpha = \text{constant}$ ($\tau = \text{constant}$), representing fluctuation-free measurements that reduce to $L^\alpha = 1$ (static finite interval) under single-axis normalization, and $L^\alpha = 0$ (trivial measurement) define a continuous spacetime, defining any small interval;
(2) \textit{Linear fluctuation mode} governed by $L^\alpha = K_1^\alpha X^\alpha + K_2^\alpha$ ($\tau = K_1 X + K_2$), where micro-lengths measuring linearly with scales, simplifiable to $L^\alpha = X^\alpha$ through reference transformations; 
(3) \textit{Nonlinear regime} characterized by general functional dependencies $L^\alpha = L^\alpha(X^\alpha)$ ($\tau = \tau(X)$), accommodating complex or discrete spacetime variations. 
Isotropy emerges when identical scaling functions $L^\alpha$ operate across all dimensions. The scaling function with non-trivial (except $L^\alpha=0$) measurement is applied to describe the spacetime fluctuation measurement. 

To analyze infinitesimal changes in these scale factors under rescaling, define 
\begin{equation}
d\bar{x}^{\alpha}=\bar{L}^{\alpha}dx^{\alpha}_{0}=L^{\alpha}(\zeta^{\alpha}X^{\alpha})dx^{\alpha}_{0}=L^{\alpha}(\bar X^{\alpha})dx^{\alpha}_{0}
\end{equation}
with $\bar{X}^\alpha=\zeta^\alpha X^\alpha$ and $\zeta^{\alpha}(X^\alpha)$ being the $\alpha-axis$ rescale factor of scale transformation. From $d(dx^{\alpha})=d\bar{x}^{\alpha}-dx^{\alpha}=(\frac{d\bar{x}^{\alpha}}{dx^{\alpha}}-1)dx^{\alpha}=(\frac{\bar{L}^{\alpha}}{L^{\alpha}}-1)dx^{\alpha}$ and $d(dx^{\alpha})=\frac{dL^{\alpha}}{dX^{\alpha}}dX^{\alpha}dx_{0}^{\alpha}$, we get 
\begin{equation}
\begin{aligned}
&\frac{dX^{\alpha}}{dx^{\alpha}}= a_{\alpha}\frac{1}{dx_{0}^{\alpha}}
\end{aligned}
\label{eq:defination}
\end{equation}
with 
\begin{equation}
\begin{aligned}
& a_{\alpha}=\frac{\bar{L}^{\alpha}/L^{\alpha}-1}{dL^{\alpha}/dX^{\alpha}}.
\label{eq:firstorder}
\end{aligned}
\end{equation}
Eq. (\ref{eq:defination}) can also be expressed as 
\begin{equation}
\begin{aligned}
&dX^{\alpha}= a_\alpha L^\alpha, 
\end{aligned}
\label{eq:scalefluctuation}
\end{equation}
$ a_\alpha=a_\alpha(X^\alpha)$ are induced by the micro variation $d(dx^\alpha)$, quantifying the amplitude of fluctuations in the $\alpha$ direction, defined as first-order fluctuating factors. It describes how the magnitude of fluctuations in different modes or directions ``breathes''. Note, $ a_\alpha$ are pre-quantized scale factors before it is promoted to be the quantized operators through the quantization procedure described in section VI.    

Since the introduced scale fluctuating factors \( a_\mu\) are direction-dependent, dimensionless Lorentz-scalar, each component encodes the fluctuation amplitude along its respective direction. The covariant properties of these factors are discussed in detail in Section IV. Because of this, throughout this paper, we adopt the convention as listed in TABLE I for using the Einstein summation.

\begin{table*}[htbp]
\centering
\caption{Summary of Einstein summation conventions}
\label{tab:summation_conventions}
\renewcommand{\arraystretch}{1.5}
\scalebox{1.1}{
\scriptsize
\begin{tabular}{l|l|l}
\hline
\hline
\textbf{Rule} & \textbf{Description} & \textbf{Representative Equations} \\
\hline
(i) & 
\begin{tabular}{@{}l@{}}
Standard Einstein summation applies when an index appears twice (once upper,\\ once lower) in expressions without explicit $a_\mu$ factors and references. 
\end{tabular}
& 
\begin{tabular}{@{}l@{}}
Eqs.~(8), (10), (11), (12), (21), (39),\\(58)$\sim$(66),(68), (69), (71), (75), (76),\\ (77), (84), (86), (89)
\end{tabular}\\
\hline
(ii) &
\begin{tabular}{@{}l@{}}
No summation implied for repeated indices in the same position; for example,\\ $x^\mu = L^\mu\,dx_0^\mu + x_0^\mu$ involves no sum over \(\mu\); reference quantities ($dx_0^\mu$, $p_\mu^0$) never\\ trigger summation.
\end{tabular}
& Eqs.~(1), (3), (4), (9), (22), (26), (40) \\
\hline
(iii) &
\begin{tabular}{@{}l@{}}
In mixed expressions with $a_\mu\;,\; \xi^\mu\; or\;\zeta^\mu$, summation is triggered by other\\ tensorial quantities; the presence of $a_\mu\;,\; \xi^\mu\; or\;\zeta^\mu$ alone does not imply\\ summation over $\mu$. $a_\mu\;,\; \xi^\mu\; or\;\zeta^\mu$ is treated as a direction-dependent scalar\\ coefficient: it participates in the expressions without triggering summation.\\ For example, no summation is implied in
    $
        dX^{\alpha} =  a_{\alpha} L^\alpha, \quad 
        \hat g^{\alpha\beta} = g^{\alpha\beta}\,\frac{ a_\alpha  a_\beta}{ a^2},
    $
    \\where \(dX^\alpha\), \(L^\alpha\), \(\hat g^{\alpha\beta}\), and \(g^{\alpha\beta}\) carry the summation indices. 
\end{tabular}
& 
\begin{tabular}{@{}l@{}}
Eqs.~(5)$\sim$(7), (13), (14), (16)$\sim$(20), \\(23)$\sim$(25), (27)$\sim$(36), (37), (38), (45),\\ (46), (48), (53)$\sim$(57) (59), (72), (73),\\ (85), (87), (88), (90)$\sim$(92) 
\end{tabular}\\
\hline
(iv) & 
\begin{tabular}{@{}l@{}}
When $a_\mu$ are coordinates on (\(\mathcal{M}^{( b)}\), $\tilde{g}$, $ a_\mu$),  they follow standard Einstein\\ summation after reparametrization. 
\end{tabular}
& Eqs.~(74), (78), (79), (80) \\
\hline
(v) & 
\begin{tabular}{@{}l@{}}
Explicit indication when summation does/doesn't apply for remaining cases;\\ for example, $\tilde{g}^{\alpha\beta}=\hat g^{\alpha\beta}\frac{ b_\alpha b_\beta}{ b^2}$ (no sum on $\alpha\,,\beta$) or explicit $\Sigma$ used. 
\end{tabular}
& 
\begin{tabular}{@{}l@{}}
Eqs.~(15), (41)$\sim$(44), (93)$\sim$(97), (101),\\ (103) and all subsequent equations.
\end{tabular}\\
\hline
\hline
\end{tabular}
}
\end{table*}

The most general Lorentz-scalar line element can be written as
\begin{equation}
d\lambda^{2}=g_{\alpha\beta}\,dx^{\alpha}dx^{\beta},
\end{equation}
where $g_{\alpha\beta}$ denotes the metric tensor. The invariant interval provides a natural probe of possible spacetime fluctuations at short distances. As introduced in \cite{weihu}, using the scaling representation introduced above, we obtain an expression for the metric in terms of micro-measurements of spacetime fluctuations:
\begin{equation}
\tau^{2}(s)\,d\lambda_{0}^{2}=g_{\alpha\beta}\,L^{\alpha}(\tau)\,L^{\beta}(\tau)\,dx_{0}^{\alpha}dx_{0}^{\beta}.
\end{equation}
For a static measurement, when $\tau=\tau_{C}$ and $L^{\alpha}=L^{\alpha}_{C}$ (constants), we define
\begin{equation}
\tau_{C}^{2}=g_{\alpha\beta}\,L^{\alpha}_{C}L^{\beta}_{C},
\end{equation}
with the consistent choice $d\lambda_{0}=dx_{0}^{\alpha}=dx_{0}^{\beta}$. This expression reduces to the usual classical Lorentz-scalar line element. For a general (non-static) measurement, one therefore has
\begin{equation}
\tau^2=g_{\alpha\beta}L^{\alpha}L^{\beta},
\end{equation}
again adopting the same consistent choice $d\lambda_{0}=dx_{0}^{\alpha}=dx_{0}^{\beta}$.
With considering Eq. (\ref{eq:scalefluctuation}), one finds
\begin{equation}
dX^2=\hat g_{\alpha\beta}dX^{\alpha}dX^{\beta}, 
\end{equation}
defining a scale-measuring spacetime, where $g_{\alpha\beta}=\Omega^{( a)}_{\alpha\beta} \hat g_{\alpha\beta}$ with $\Omega^{( a)}_{\alpha\beta}=\frac{ a_\alpha a_\beta}{ a^2}$. The definition of $ a$ from proper length (i.e, Lorentz invariant line element $\tau$) is similar to $ a_\alpha$ \cite{weihu}. Defining an inverse metric $\hat g^{\alpha\beta}=g^{\alpha\beta}\frac{ a_\alpha a_\beta}{ a^2}$ to satisfy $\hat{g}^{\alpha\beta} \hat{g}_{\beta\gamma} = \delta^\alpha_\gamma\frac{ a_\alpha}{ a_\gamma}$ when $g^{\alpha\beta} g_{\beta\gamma} = \delta^\alpha_\gamma$.

In this framework, the microscopic fluctuations of the physical metric \(g_{\alpha\beta}\) are captured by promoting local, direction-dependent scale factors \(a_\mu\) to fundamental dynamical variables. Physically, these factors represent anisotropic deformations acting upon the metric \(\hat{g}_{\alpha\beta}\).
In this picture, \(g_{\alpha\beta}\) remains the fundamental field governing classical gravity, while its quantum behavior and Planck-scale structure are effectively described by the dynamics of the scaling sector \((\hat{g}_{\alpha\beta}, a_\alpha)\). This approach provides an alternative to direct metric quantization by focusing on the degrees of freedom most sensitive to quantum geometric fluctuations.
More concretely, the physical metric can be expressed as the result of anisotropic stretching or compression of the elastic substrate metric \(\hat{g}_{\alpha\beta}\), modulated by scale factors \(\Omega^{(a)}_{\alpha\beta}\). The term ``substrate'' reflects the dual role of \(\hat{g}_{\alpha\beta}\): it serves as the geometric substrate on which \(g_{\alpha\beta}\) is constructed via anisotropic scaling, and simultaneously as the dynamical substrate defining the manifold \(\mathcal{M}^{(a)}\) on which the scale factors \(a_\mu\) fluctuate. This composite structure cleanly separates the smooth classical geometry (described by \(\hat{g}\)) from its microscopic fluctuations (encoded in \(a_\mu\)).
Together, \(g_{\alpha\beta}\) and \(\hat{g}_{\alpha\beta}\) form an anisotropic structure, where each \(a_\alpha\) governs local scaling behavior along a coordinate direction. If \(a_\alpha\) is constant and isotropic, the framework reduces to a global conformal transformation; otherwise, it captures directionally varying and position-dependent fluctuations.

In this context, the manifold \((\mathcal{M}, g_{\alpha\beta}, x^\alpha)\) no longer represents a fixed classical geometry. Instead, it is dynamically shaped by microscopic fluctuation effects and the measurement prescription embodied in the scale factors \(a_\alpha\). This perspective is operationally extended to a measurement-sensitive space \((\mathcal{M}, g_{\alpha\beta}, L^\alpha)\), where the scale variables \(L^\alpha\) incorporate both classical smooth contributions and quantum fluctuating components to length measurements.
At a deeper level, the geometry \((\mathcal{M}^{(a)}, \hat{g}_{\alpha\beta}, X^\alpha)\) provides a refined substrate structure that naturally couples to quantum fluctuations. This scaling spacetime encodes first-order local deformations and reflects quantum effects in a directional and localized manner.
This geometric construction is motivated by the recognition that quantum-scale measurement processes---especially those sensitive to spacetime fluctuations---cannot be fully described by smooth classical metrics alone. Instead, fluctuation-responsive geometries built from locally rescaled structures offer a more accurate and physically intuitive framework, bridging microscopic quantum dynamics with spacetime geometry.

Define second-order fluctuation factors
\begin{equation}
\begin{aligned}
& b_{\alpha}=-\frac{d a_{\alpha} }{dX^{\alpha}},
\end{aligned}
\end{equation}
we get
\begin{equation} 
\begin{aligned}
&-da_{\alpha}=b_{\alpha}dX^{\alpha}= b_{\alpha}a_{\alpha}L^\alpha, 
\end{aligned}
\end{equation}
$ b_\alpha a_\alpha$ are introduced as scale dilation to the spacetime micro measurement $dx^\alpha/dx^\alpha_0$, reflecting the spacetime fluctuation amplitude defined by $d a_\alpha$.
Similarly, we have
\begin{equation}
d a^2=\sum_{\alpha\beta}\tilde{g}_{\alpha\beta}d a_{\alpha}d a_{\beta},
\label{eq:manifold_b}
\end{equation}
defining scale-measuring spacetime, where $\hat g_{\alpha\beta}=\Omega^{( b)}_{\alpha\beta} \tilde{g}_{\alpha\beta}$ (no sum on $\alpha\,,\beta$) with $\Omega^{( b)}_{\alpha\beta}=\frac{ b_\alpha b_\beta}{ b^2}$, reflecting metric second-order fluctuation by scale factors $ b_\alpha$ ($ b$). Then defining $\tilde{g}^{\alpha\beta}=\hat g^{\alpha\beta}\frac{ b_\alpha b_\beta}{ b^2}$ (no sum on $\alpha\,,\beta$) to satisfy $\tilde{g}^{\alpha\beta} \tilde{g}_{\beta\gamma} = \delta^\alpha_\gamma\frac{ b_\alpha}{ b_\gamma}\frac{ a_\alpha}{ a_\gamma}$ (no sum on $\alpha\,,\gamma$) when $\hat g^{\alpha\beta} \hat g_{\beta\gamma} = \delta^\alpha_\gamma\frac{ a_\alpha}{ a_\gamma}$. Additionally, $\tilde{g}_{\alpha\beta}=g_{\alpha\beta}\frac{ a^2}{ a_\alpha a_\beta}\frac{ b^2}{ b_\alpha b_\beta}$ (no sum on $\alpha\,,\beta$) and $\tilde{g}^{\alpha\beta}=g^{\alpha\beta}\frac{ a_\alpha a_\beta}{ a^2}\frac{ b_\alpha b_\beta}{ b^2}$ (no sum on $\alpha\,,\beta$). 
$ a_{\alpha}$ are first-order fluctuation factors. $ b_{\alpha}$ are second-order fluctuation factors, appearing in the term of the second-order differential fluctuating factors; they reflect the non-uniformity of scale deformation. 

Metrics \(\hat{g}\) and \(\tilde{g}\) correspond to different hierarchical levels induced by spacetime fluctuations on microscale. The first-order fluctuation geometry \((\mathcal{M}^{( a)}, \hat{g}_{\alpha\beta}, X^\alpha)\) captures the local anisotropic scaling structure generated by the microscopic scaling \( a_\alpha\), encoding directional fluctuations in length measurements. Building upon this, the second-order fluctuation geometry \((\mathcal{M}^{( b)}, \tilde{g}_{\alpha\beta},  a_\alpha)\) characterizes the variation or ``inhomogeneity'' of \( a_\alpha\) itself, as quantified by the second-order scale factors \( b_\alpha\). This additional layer of structure reflects more refined, derivative-level deformation effects in the micro spacetime geometry.

To reflect this, we introduce a hierarchical structure of metrics:
\[
\tilde{g}_{\alpha\beta} \xrightarrow{ b_\alpha} \hat{g}_{\alpha\beta} \xrightarrow{ a_\alpha} g_{\alpha\beta} \implies g_{\alpha\beta} = \tilde{g}_{\alpha\beta} \times \Omega^{( b)} \times \Omega^{( a)},
\]
where:
- \(\tilde{g}_{\alpha\beta}\) captures second-order fluctuation structures, including variation in the fluctuation amplitude itself;
- \(\hat{g}_{\alpha\beta}\) reflects first-order anisotropic scaling effects associated with \( a_\alpha\);
- \(g_{\alpha\beta}\) is the resulting physical geometry that combines both levels of fluctuation into a measurable structure.
Specifically, metric \(g_{\alpha\beta}\) results from two successive anisotropic scale transformations governed by the fluctuation factors \( a_\alpha\) and \( b_\alpha\), which respectively characterize first- and second-order scale deformations. This hierarchical perspective differs from the conventional viewpoint: the classical-physical metric \(g_{\alpha\beta}\) is instead fluctuation-informed and measurement-dependent, bridging scale fluctuation geometry with observable structure in a physically meaningful way.

Then, the first and second-order partial differential operators transformed in scaling form can be introduced by
\begin{equation}
\begin{aligned}
&\frac{\partial}{\partial x^{\nu}}=\frac{dX^{\nu}}{dx^{\nu}}\frac{\partial }{\partial X^{\nu}}=\frac{1}{dx_{0}^{\nu}} a_{\nu}\frac{\partial}{\partial X^{\nu}},
\end{aligned}
\end{equation}
\begin{equation}
\begin{aligned}
\frac{\partial^{2} }{\partial x^{\mu}\partial x^{\nu}}&=\frac{1}{dx_{0}^{\mu}}\frac{1}{dx_{0}^{\nu}}( a_{\mu} a_{\nu}\frac{\partial^{2}}{\partial X^{\mu}\partial X^{\nu}}- a_{\mu} b_{\nu}\frac{\partial  X^\nu}{\partial X^{\mu}}\frac{\partial}{\partial X^{\nu}}),
\end{aligned}
\end{equation}

The connection between differential operators and scaling functions emerges by introducing a rescale factor $\zeta^{\nu}$, which also defines spacetime microelement fluctuations. This incorporates transformation factors $ a_{\nu}$ and $ b_{\nu}$ that modify differential operator behavior under scale transformations. Mapping classical differential operators into the \(X^\nu\) scaling coordinate system via the appropriate scaling functions provides a systematic connection for probing the microscopic (Planck-scale) spacetime structure in an explicitly scale-aware manner.

Using the scale transformation $\bar{X}^{\alpha}=\zeta^{\alpha}X^{\alpha}$ with the rescale factor $\zeta^{\alpha}=\zeta^{\alpha}(X^{\alpha})$, we have $\frac{d\bar{X}^{\alpha}}{dX^{\alpha}}=\xi^\alpha=\zeta^{\alpha}+X^{\alpha}\frac{d\zeta^{\alpha}}{dX^{\alpha}}$ and $\frac{dX^{\alpha}}{d\bar{X}^{\alpha}}=\frac{1}{\xi^\alpha}$. Subsequently, this leads to the derivation of
\begin{equation}
\begin{aligned}
&\frac{\partial}{\partial \bar{X}^{\alpha}}=\frac{dX^{\alpha}}{d\bar{X}^{\alpha}}\frac{\partial}{\partial X^{\alpha}}=\frac{1}{\xi^\alpha}\frac{\partial}{\partial X^{\alpha}};
\end{aligned}
\end{equation}
\begin{equation}
\begin{aligned}
\frac{\partial^{2}}{\partial \bar{X}^{\beta}\partial \bar{X}^{\alpha}}&=\frac{dX^{\beta}}{d\bar{X}^{\beta}}\frac{\partial}{\partial X^{\beta}}(\frac{\partial}{\partial \bar{X}^{\alpha}})=\frac{1}{\xi^{\beta}}\frac{1}{\xi^{\alpha}}\frac{\partial^{2}}{\partial X^{\beta}\partial X^{\alpha}}\\
&-\{\frac{1}{\xi^{\beta}[\xi^{\alpha}]^{2}}\frac{d\xi^{\alpha}}{dX^{\alpha}}\}\frac{\partial X^{\alpha}}{\partial X^{\beta}}\frac{\partial}{\partial X^{\alpha}},
\end{aligned}
\end{equation}
where 
\[
\frac{d\xi^{\alpha}}{dX^{\alpha}}=2\frac{d\zeta^{\alpha}}{dX^{\alpha}} + X^{\alpha} \frac{d^2 \zeta^{\alpha}}{d(X^{\alpha})^2}.
\]
Here, we provide careful revisions to Eq. (13), Eq. (14), and their corresponding inferences presented in Ref. \cite{weihu}.

\subsection{Dual measurement and discrete spacetime}
From $\bar{X}^\alpha=\zeta^\alpha X^\alpha$ with the rescale factor $\zeta^{\alpha}=\zeta^{\alpha}(X^{\alpha})$, and
\begin{equation}
 d\bar{X}^{\alpha}=\xi^\alpha dX^{\alpha}=(\zeta^{\alpha}+X^{\alpha}\frac{d\zeta^{\alpha}}{dX^{\alpha}})dX^{\alpha},
\end{equation}
we have
\begin{equation}
d\bar X^2=\bar g_{\alpha\beta}d\bar X^{\alpha}d\bar X^{\beta}, 
\end{equation}
defining a scale-measuring spacetime, where $\bar g_{\alpha\beta}=\Omega^{(\xi)-1}_{\alpha\beta} \hat g_{\alpha\beta}=\frac{\xi^2}{\xi^\alpha\xi^\beta}\frac{ a^2}{ a_\alpha a_\beta}g_{\alpha\beta}$ with $\Omega^{(\xi)}_{\alpha\beta}=\frac{\xi^\alpha\xi^\beta}{\xi^2}$ and $\bar g^{\alpha\beta}=\hat g^{\alpha\beta}\frac{\xi^\alpha\xi^\beta}{\xi^2}=g^{\alpha\beta}\frac{ a_\alpha a_\beta}{ a^2}\frac{\xi^\alpha\xi^\beta}{\xi^2}$, reflecting metric fluctuation measured by bilinear scale factors $ a_\alpha \xi^\alpha$ ($a\xi$), where $\bar{g}^{\alpha\beta} \bar{g}_{\beta\gamma} = \delta^\alpha_\gamma\frac{ a_\alpha}{ a_\gamma}\frac{ \xi^\alpha}{ \xi^\gamma}$.

A linear mapping is introduced between the physical coordinates \(x^\alpha\) and a set of rescaled variables \(\bar{X}^\alpha\), defined as \(\bar{X}^\alpha = \zeta^\alpha X^\alpha\), where \(\zeta^\alpha\) denotes the rescale factor in each direction. This transformation provides a linearized representation of the original nonlinearly measured spacetime structure, enabling a linearized analysis of microscopic fluctuations. The corresponding linear mapping differential relation,
\begin{equation}
dx^\alpha = d\bar{X}^\alpha \, dx_0^\alpha, 
\label{eq:mapping}
\end{equation}
suggests that the local displacement \(dx^\alpha\) can be viewed as a product of the scaled increment \(d\bar{X}^\alpha\) and a reference length scale \(dx_0^\alpha\). The nonlinear measurement kernel \(L^\alpha(X^\alpha)\) can be effectively mapped onto a linear structure governed by \(\bar{X}^\alpha\).

This linearization aims to simplify the mathematical structure while capturing the essential physical content of the fluctuations. It establishes a dual description in which the complex fluctuation behavior encoded in \(L^\alpha\) can be equivalently analyzed through the evolution of the linearly scaled quantities \(d\bar{X}^\alpha\). As a result, the system's intrinsic dynamical properties can be extracted. This leads to a self-consistent measurement framework in which nonlinear fluctuation behavior is embedded within the evolution of linearly scaled quantities, allowing one to describe complex quantum-scale variations using geometrically intuitive linear dynamics. In this construction, the scaled differential \(d\bar{X}^\alpha\) serves as an effective measure of microscopic length, equivalent in fluctuation content to the nonlinear measure \(L^\alpha\), thereby enabling a dual description of quantum-scale measurements.

The distinction between nonlinear and linear measurements of micro-length $L^{\mu}$ under different scaling coordinates admits an analogy with general relativity: just as the metric in a general coordinate frame can be locally transformed into its Minkowski form in a freely falling frame, the nonlinear fluctuation kernel $L^{\mu}(X^{\mu})$ can be mapped onto a linearized structure described by $d\bar{X}^{\mu}$. Nonlinear \(L^\mu(X^\mu)\) is analogous to the length measurements in a general coordinate frame, while linear \(d\bar{X}^\mu\) is analogous to the measurement in a locally inertial frame. This mapping to a fluctuation-local inertial frame simplifies the analysis of quantum spacetime fluctuations, thereby clarifying the physical motivation for introducing the scaled variables $\bar{X}^{\mu}$ in the context of fluctuation-sensitive geometry.

The associated metric \(\bar{g}_{\alpha\beta} = g_{\alpha\beta}(x^\mu \rightarrow \bar{X}^\mu)\) encodes fluctuation effects within a rescaled coordinate space \((\mathcal{M}^{(\xi)}, \bar{g}, \bar{X}^\alpha)\). Here, the metric incorporates the influence of microscopic fluctuations into the geometry itself, allowing all physical fields to evolve under the combined effects of the elastic substrate structure and the embedded fluctuation scale factors, while preserving general covariance. With linear mapping, the rescale factors are governed by
\begin{equation}
 a_\alpha = 1/\xi^\alpha, \quad \Omega^{\xi}_{\alpha\beta}\Omega^{ a}_{\alpha\beta}=1.
\end{equation}
As a result, consistent mapping relations between the nonlinear and linear representations of length are established. Therefore, the first-order fluctuation mapping is given by
\begin{equation}
\frac{dL^\alpha}{dX^\alpha} = \xi^\alpha\left(\frac{\bar{L}^\alpha}{L^\alpha} - 1\right),
\label{eq:mapping1}
\end{equation}
while the second-order mapping for the fluctuation factors \( b_\alpha\) read
\begin{equation}
 b_\alpha = \frac{1}{(\xi^\alpha)^2} \frac{d\xi^\alpha}{dX^\alpha}.
\end{equation}
These relations provide a bridge between differential operators in the nonlinear and linearized geometries.

Moreover, this construction reveals a dual measurement principle: microscopic length measurements can be equivalently described by either the fluctuation-sensitive quantity \(L^\mu\) or the differential \(d\bar{X}^\mu\), both residing in the rescaled fluctuation spacetime \((\mathcal{M}^{( a)}, \hat{g}, X^\alpha)\). The fluctuation factor \( a_\mu\) thus admits an alternative representation via the inverse scale factor \(1/\xi^\mu\), establishing a one-to-one correspondence between nonlinear and linearized descriptions of quantum geometric fluctuations. This duality forms a consistent measurement theory that allows for the analysis of complex fluctuation dynamics.

The micro distance measurement with $L^{\alpha}(X^{\alpha})$, expressed as $dx^{\alpha}=L^{\alpha}dx^{\alpha}_{0}$. By adding the linear mapping $dx^{\alpha}=d\bar{X}^{\alpha}dx^{\alpha}_{0}=\bar{X}^{\alpha}dx^{\alpha}_{0}-\bar{X}^{\alpha}_0dx^{\alpha}_{0}=d\tilde{x}^{\alpha}-d\tilde{x}^{\alpha}_0$, we obtain $dx_0^\alpha=d\tilde{x}^{\alpha}_0/\bar{X}^{\alpha}_0$ and the following equations:
\begin{equation}
\begin{cases}
&x^{\alpha}=\frac{L^{\alpha}}{\bar{X}^{\alpha}_0}d\tilde{x}^{\alpha}_{0}+x^{\alpha}_{0};\\
&x^{\alpha}=\frac{\bar{X}^{\alpha}}{\bar{X}^{\alpha}_0}d\tilde{x}^{\alpha}_{0}-d\tilde{x}^{\alpha}_{0}+x^{\alpha}_{0},
\end{cases}
\end{equation}
where $d\tilde{x}^{\alpha}_{0}=\bar{X}^{\alpha}_0dx^{\alpha}_{0}=x^{\alpha}_{0}-\tilde{x}^{\alpha}_{0}$, $dx^{\alpha}=L^\alpha dx^{\alpha}_{0}=x^{\alpha}-x^{\alpha}_{0}$, and $d\tilde{x}^{\alpha}=\bar{X}^{\alpha}dx^{\alpha}_{0}=x^{\alpha}-\tilde{x}^{\alpha}_{0}$. $x^{\alpha}$ are the position coordinates relative to reference position coordinates  $x^{\alpha}_{0}$ or $\tilde{x}^{\alpha}_{0}$, we get
\begin{equation}
\bar{X}^\alpha = \zeta^\alpha X^\alpha = \bar{X}^{\alpha}_0 + L^\alpha; \quad L^\alpha=d\bar{X}^\alpha,
\label{eq:mapping2}
\end{equation}
where $\bar{X}^{\alpha}_0$ are reference scales. Substituting into the Eq.~(\ref{eq:mapping1}), we obtain
\begin{equation}
\frac{dL^\alpha}{dX^\alpha} = \zeta^\alpha+X^\alpha\frac{d\zeta^\alpha}{dX^\alpha}=\xi^\alpha,
\end{equation}
and we have $\bar L^\alpha=2L^\alpha$, namely
\begin{equation}
(\zeta^{\alpha})^2-2\zeta^\alpha+\frac{\bar{X}^{\alpha}_0}{X^\alpha}=0.
\label{eq:dual}
\end{equation}
Solved
\begin{equation}
\zeta^\alpha=\zeta^\alpha_\pm = 1 \pm \sqrt{1 - \frac{\bar{X}^{\alpha}_0}{X^\alpha}}=1 \pm\frac{1}{\Upsilon^\alpha},
\end{equation}
with
\begin{equation}
\Upsilon^\alpha=\frac{1}{\sqrt{1 - \frac{\bar{X}^{\alpha}_0}{X^\alpha}}},
\end{equation}
and then
\begin{equation}
\frac{d\zeta^\alpha}{dX^\alpha} = \pm \frac{\bar{X}^{\alpha}_0}{2 \left(X^\alpha\right)^2 \sqrt{1 - \frac{\bar{X}^{\alpha}_0}{X^\alpha}}}=\pm\frac{\Upsilon^\alpha\bar{X}^{\alpha}_0}{2 \left(X^\alpha\right)^2 };
\end{equation}
\begin{equation}
\frac{d^2 \zeta^\alpha}{d(X^\alpha)^2} =\pm \frac{\left(\Upsilon^\alpha\right)^{3}\bar{X}^{\alpha}_0(3\bar{X}^{\alpha}_0-4X^\alpha)}{4(X^\alpha)^4 };
\end{equation}
\begin{equation}
 a_\alpha =\frac{1}{\xi^\alpha} = \left[ 1 \pm \left(\frac{1}{\Upsilon^\alpha} + \frac{\Upsilon^\alpha \bar{X}_0^\alpha}{2 X^\alpha}\right) \right]^{-1}.
\label{eq:hata}
\end{equation}

\begin{figure}[!htb]
    \includegraphics
    [width=0.7\hsize]
      {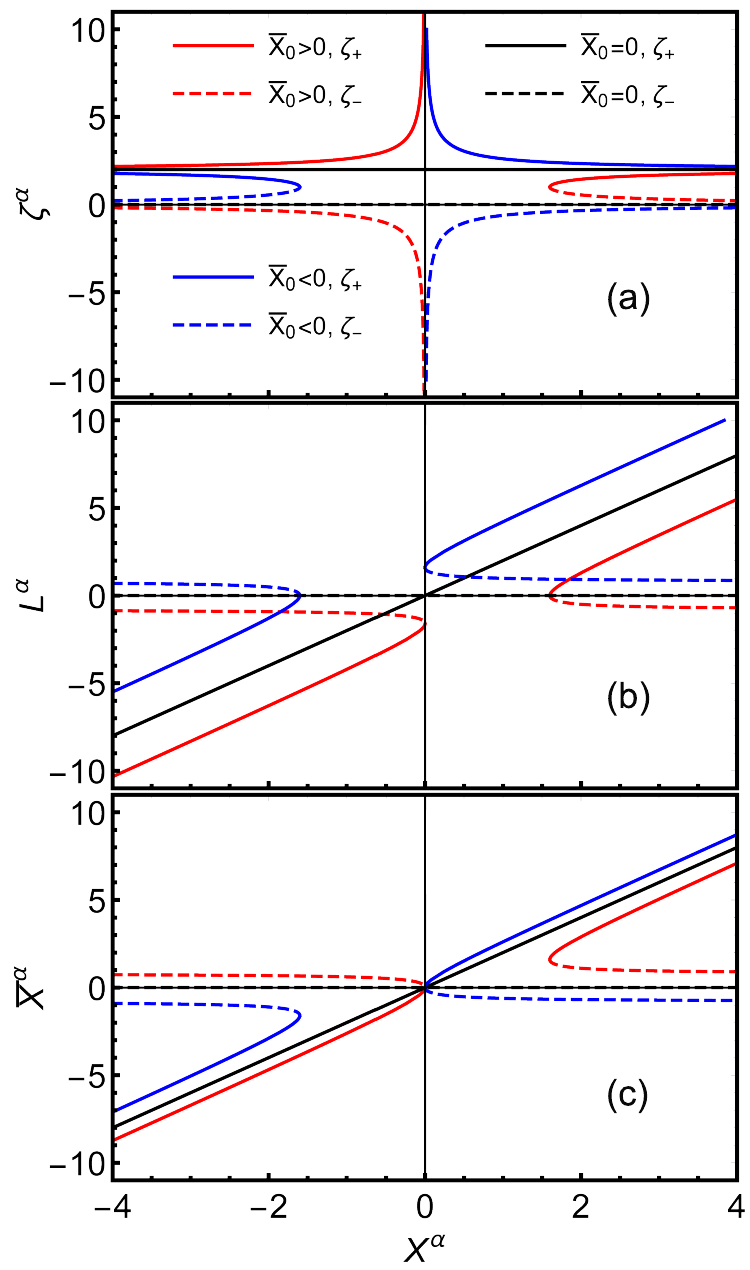}
    \caption{(Color online) Curves of (a) $\zeta^\alpha(X^\alpha)\sim X^\alpha$, (b) $L^\alpha(X^\alpha)\sim X^\alpha$, and (c) $\bar X^\alpha(X^\alpha)\sim X^\alpha$ for $\bar X^\alpha_0=0$ and an arbitrary choice of $|\bar X^\alpha_0|=1.6$.} 
    \label{fig1}
\end{figure}

As shown in Fig.~\ref{fig1}, a nonlinear micro-length measurement occurs when \( |\bar{X}_0^\alpha| > 0 \), whereas \(\bar{X}_0^\alpha = 0\) corresponds to a linear case. In the latter situation, with \(\zeta_-^\alpha = 0\), the micromeasurement reduces to \(L \equiv 0\), describing an absolutely static and continuous spacetime in which ultraviolet divergences arise. For all other configurations, the spacetime becomes fluctuating and discrete, providing a natural ultraviolet cutoff. In this context, the behavior of \(L^\alpha\) is analogous to an order parameter, marking a phase transition between a static, classical continuum \(L^\alpha \equiv 0\) and a dynamic, fluctuating regime with discrete characteristics \(L^\alpha = L^\alpha(X^\alpha)\).

\begin{figure}[!htb]
    \includegraphics
    [width=1.0\hsize]
      {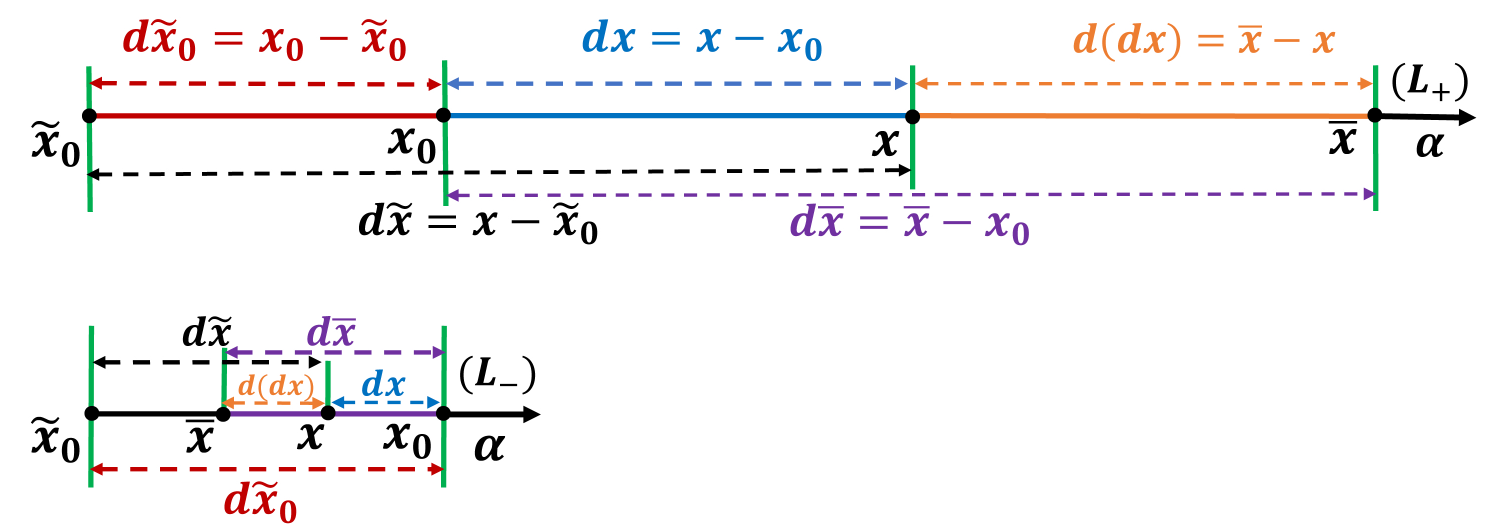}
    \caption{(Color online) The schematic diagram of microscopic spacetime fluctuation measurements when $\bar X^\alpha_0<0$ and $X^\alpha>0$. Where, $\bar X^\alpha_0/2<L_{-}\leq\bar X^\alpha_0$ and $L_{+}\geq\bar X^\alpha_0$. For $\bar X^\alpha_0>0$, the situation is the same but opposite direction.} 
    \label{fig2}
\end{figure}

As depicted in Fig.~\ref{fig2} of the schematic diagram of microscopic spacetime measurements when $\bar X^\alpha_0 < 0$ and $X^\alpha > 0$, the spacetime microstructure is discrete and equidistant ($\bar L^\alpha=2L^\alpha$), bounded by $\bar X^\alpha_0/2 < L_{-} \leq \bar X^\alpha_0$ and $L_{+} \geq \bar X^\alpha_0$. The scaling function $L^\alpha(X^\alpha)$, with $\bar X_0^\alpha$ as the modulating parameter, governs the microscopic length measurement. When $\zeta^\alpha$ takes its negative solution branch and $\bar X^\alpha_0 \to 0$, the system approaches the smooth classical continuum. The existence of dual solutions for the rescaling factor $\zeta^\alpha$ highlights a fundamental duality in the nature of the fluctuating spacetime. By incorporating (i) the Heisenberg uncertainty principle at the Planck scale and (ii) intrinsic spacetime fluctuations, the framework predicts a natural, built-in ultraviolet cutoff, where $L^\alpha \not\equiv 0$.

\begin{figure}[!htb]
    \includegraphics
    [width=1.0\hsize]
      {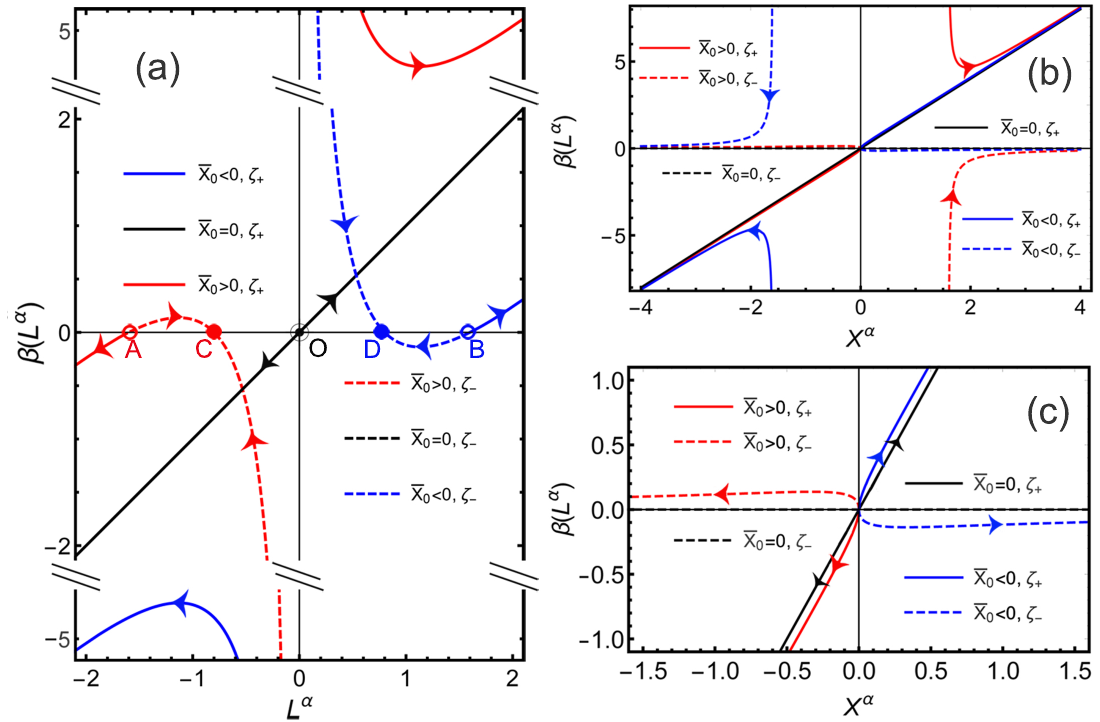}
    \caption{(Color online) Curves of (a) $\beta(L^\alpha(X^\alpha))\sim L^\alpha(X^\alpha)$, (b) $\beta(L^\alpha(X^\alpha))\sim X^\alpha$ and (c) $\beta(L^\alpha(X^\alpha))\sim X^\alpha$ of zoomed scale for $\bar X^\alpha_0=0$ and an arbitrary choice of $|\bar X^\alpha_0|=1.6$.} 
    \label{fig3}
\end{figure}

\begin{figure}[!htb]
    \includegraphics
    [width=0.7\hsize]
      {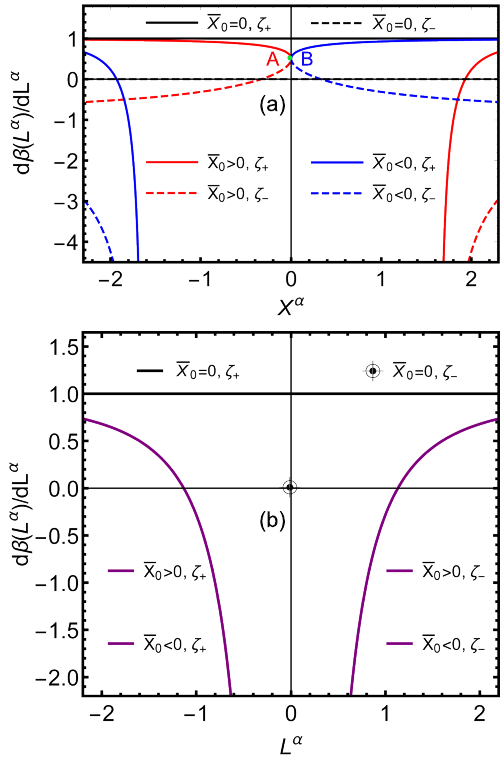}
    \caption{(Color online) Curves of (a) $d\beta(L^\alpha)/dL^\alpha\sim X^\alpha$, (b) $d\beta(L^\alpha)/dL^\alpha\sim L^\alpha$ for $\bar X^\alpha_0=0$ and an arbitrary choice of $|\bar X^\alpha_0|=1.6$.} 
    \label{fig4}
\end{figure}

In theoretical physics, particularly in quantum field theory, the renormalization group (RG) \cite{Stueckelberg, Curtis} provides a systematic framework for analyzing how a physical system behaves under changes in the scale of observation. This general approach applies to any system that exhibits scaling behavior, including those in quantum field theory, statistical mechanics, and other fields where physical quantities depend on scale. The RG flow describes how the parameters of a system evolve as the scale changes, offering valuable insights into phenomena such as phase transitions, critical points, and scale invariance. For spacetime fluctuation measurements, we derive a micro-scale geometry RG equation for $L^\alpha$:
\begin{equation}
\begin{aligned}
\beta(L^\alpha(X^\alpha)) &=X^\alpha \frac{dL^\alpha}{dX^\alpha} =  \zeta^\alpha X^\alpha+(X^\alpha)^2\frac{d\zeta^\alpha}{dX^\alpha}\\
&=  \zeta^\alpha X^\alpha\pm\frac{1}{2}\Upsilon^\alpha\bar X^\alpha_0,
\end{aligned}
\end{equation}

\begin{equation}
\begin{aligned}
\frac{d\beta(L^\alpha)}{dL^\alpha}&=\frac{d\beta/dX^\alpha}{dL^\alpha/dX^\alpha}=\frac{d\beta(\bar X^\alpha)}{d\bar X^\alpha}\\
&=\frac{dL^\alpha/dX^\alpha+2X^\alpha \frac{d\zeta^\alpha}{dX^\alpha}+(X^\alpha)^2\frac{d^2\zeta^\alpha}{d(X^\alpha)^2}}{dL^\alpha/dX^\alpha}\\
&=1+\frac{2X^\alpha d\zeta^\alpha/dX^\alpha}{dL^\alpha/dX^\alpha}+\frac{(X^\alpha)^2d^2\zeta^\alpha/d(X^\alpha)^2}{dL^\alpha/dX^\alpha}\\
&=1 \pm 
\frac{
\displaystyle  \left[ \frac{\bar{X}^\alpha_0}{X^\alpha} \, \Upsilon^\alpha + \frac{3}{4}\left(\frac{\bar{X}^\alpha_0}{X^\alpha}\right)^2(\Upsilon^\alpha)^3  -\frac{\bar{X}^\alpha_0}{X^\alpha} \, (\Upsilon^\alpha)^3 \right]
}{
\displaystyle 1 \pm \left(\frac{1}{\Upsilon^\alpha} + \frac{\bar{X}^\alpha_0}{2X^\alpha} \, \Upsilon^\alpha\right)
},
\end{aligned}
\end{equation}
where $\beta(L^\alpha)$ governs the evolution of scaling functions $L^\alpha(X^\alpha)$. Fixed points occur at $\beta=0$ as shown in FIG. \ref{fig3} and the corresponding $d\beta(L^\alpha)/dL^\alpha$ as shown in FIG. \ref{fig4}. It should be emphasized that the geometry RG $\beta(L^\alpha)$-function essentially characterizes the dynamical behavior of the fluctuation micro-length scale function $L^\alpha$—that is, the dynamics of scale-dependent measurements—enabling the analysis of its evolution and stability.

\begin{figure}[!htb]
    \includegraphics
    [width=0.7\hsize]
      {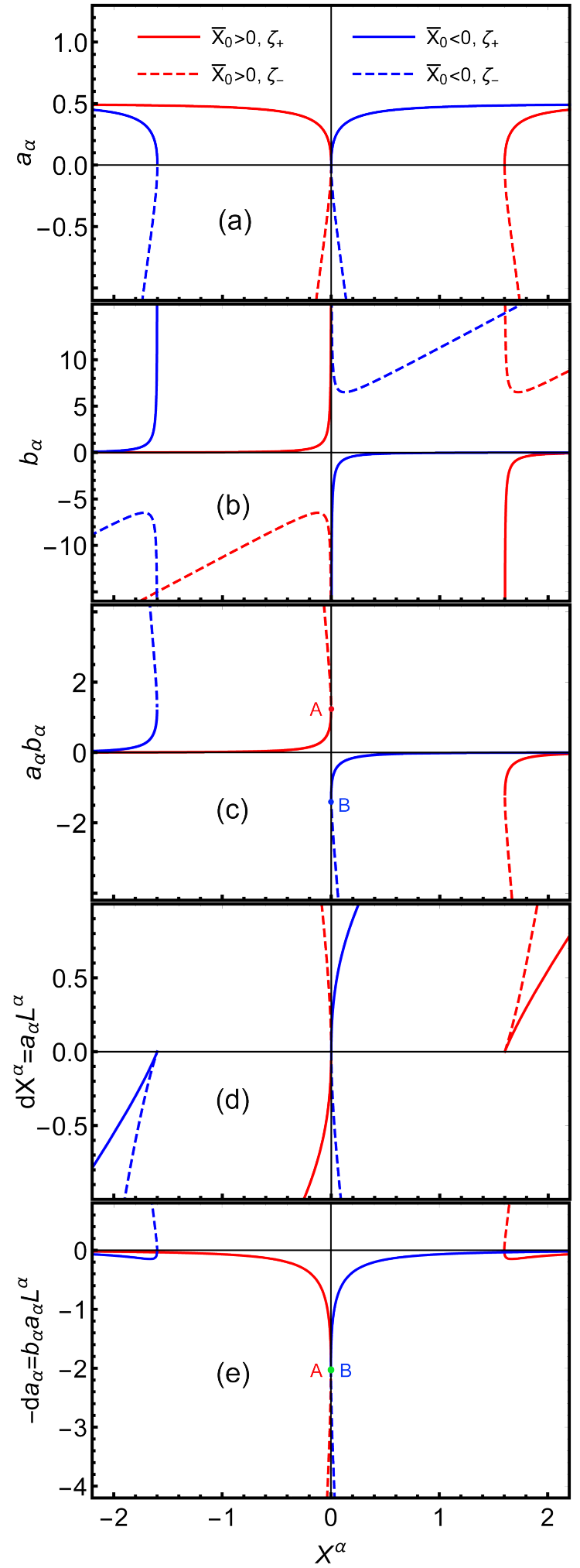}
    \caption{(Color online) Curves of (a) $ a_\alpha\sim X^\alpha$, (b) $ b_\alpha\sim X^\alpha$, (c) $ a_\alpha b_\alpha\sim X^\alpha$, (d) $dX^\alpha\sim X^\alpha$, and (e) $-d a_\alpha\sim X^\alpha$, for an arbitrary choice of $|\bar X^\alpha_0|=1.6$.} 
    \label{fig5}
\end{figure}

\begin{table*}[htbp]
\centering
\caption{Summary of fixed points A, B, C, D, O, and UV/IR endpoints with their limiting values and flow directions.}
\label{tab:fixed_points}
\renewcommand{\arraystretch}{1.5}
\scalebox{1.1}{
\scriptsize
\begin{tabular}{l|l|l|l|l|l|l|l}
\hline
\hline
\centering
\textbf{Attribute} 
    & \textbf{ A: UV} 
    & \textbf{ B: UV} 
    & \textbf{ C: IR} 
    & \textbf{ D: IR} 
    & \textbf{ O: (1) UV; (2) saddle} 
    & \textbf{ UV EP} 
    & \textbf{ IR EP} \\ 
\hline

\(X^\alpha \to\) 
    & \(0^-\) 
    & \(0^+\) 
    & \(-\infty\) 
    & \(+\infty\) 
    & \(0\) 
    & \(\bar X^\alpha_0\) 
    & \(\pm\infty\) \\ 
\hline

\(\bar X^\alpha_0\)  
    & \(>0\) 
    & \(<0\) 
    & \(>0\) 
    & \(<0\) 
    & \(=0\) 
    & \(\neq0\) 
    & \(\neq0\) \\ 
\hline

\(L^\alpha_*\) 
    & \(-\bar X^\alpha_0 \) 
    & \(-\bar X^\alpha_0\) 
    & \(-\tfrac{1}{2}\bar X^\alpha_0 \) 
    & \(-\tfrac{1}{2}\bar X^\alpha_0\) 
    & \(0\) 
    & \(0\) 
    & \(\pm\infty\) \\ 
\hline

\(\beta\) Sign 
    & \(>0\),\,\(X^\alpha<0\)
    & \(<0\),\,\(X^\alpha>0\)
    & \(>0\),\,\(X^\alpha<0\)
    & \(<0\),\,\(X^\alpha>0\)
    & 
    \begin{tabular}{@{}l@{}}
      (1) \(<0\) for \(X^\alpha<0\); \\ 
      \(>0\) for \(X^\alpha>0\) \\ 
      (2) \(\beta=0\) 
    \end{tabular} 
    & \(\to\pm\infty\) 
    & \(\to\pm\infty\) \\ 
\hline

\(d\beta(L^\alpha)/dL^\alpha\)
    & \(1/2\) 
    & \(1/2\) 
    & \(-1\) 
    & \(-1\) 
    & 
    \begin{tabular}{@{}l@{}}
      (1) \(1\) \\ 
      (2) \(0\) 
    \end{tabular} 
    & \(\to-\infty\) 
    & \(\to1\) \\ 
\hline

\( a_\alpha\) 
    & \(0\) 
    & \(0\) 
    & \(-\infty\) 
    & \(-\infty\) 
    & 
    \begin{tabular}{@{}l@{}}
      (1) \(\tfrac{1}{2}\) \\ 
      (2) \(\infty\) 
    \end{tabular} 
    & \(0\) 
    & \(\tfrac{1}{2}\) \\ 
\hline

\( b_\alpha\) 
    & \(\pm\infty\) 
    & \(\pm\infty\) 
    & \(-\infty\) 
    & \(+\infty\) 
    & \(0\) 
    & \(\pm\infty\) 
    & \(0\) \\ 
\hline

\( a_\alpha\, b_\alpha\) 
    & \(+\sqrt{|\bar X^\alpha_0|}\) 
    & \(-\sqrt{|\bar X^\alpha_0|}\) 
    & \(+\infty\) 
    & \(-\infty\) 
    & 
    \begin{tabular}{@{}l@{}}
      (1) \(0\) \\ 
      (2) \(0\)
    \end{tabular} 
    & \(\pm\sqrt{|\bar X^\alpha_0|}\) 
    & \(0\) \\ 
\hline

\( b^2_\alpha/\, a^2_\alpha\) 
    & \(\infty\) 
    & \(\infty\) 
    & \(0\) 
    & \(0\) 
    & 
    \begin{tabular}{@{}l@{}}
      (1) \(0\) \\ 
      (2) \(0\)
    \end{tabular} 
    & \(\infty\) 
    & \(0\) \\ 
\hline

\(dX^\alpha\) 
    & \(0\) 
    & \(0\) 
    & \(+\infty\) 
    & \(-\infty\) 
    & 
    \begin{tabular}{@{}l@{}}
      (1) \(0\) \\ 
      (2) \(\infty \times 0\) 
    \end{tabular} 
    & \(0\) 
    & \(\pm\infty\) \\ 
\hline

\(-\,d a_\alpha\) 
    & \(-\,|\bar X^\alpha_0|^{3/2}\) 
    & \(-\,|\bar X^\alpha_0|^{3/2}\) 
    & \(-\infty\) 
    & \(-\infty\) 
    & 
    \begin{tabular}{@{}l@{}}
      (1) \(0\) \\ 
      (2) \(0 \times \infty \times 0\) 
    \end{tabular} 
    & \(0\) 
    & \(0\) \\ 

\hline
\hline
\end{tabular}
}
\end{table*}

As shown in FIG.~\ref{fig3}, the condition $\beta(L^\alpha) = 0$ defines five fixed points (A, B, C, D, O). The $\beta$-function exhibits two non-trivial zeros at $|X^\alpha| \to 0$ and $|X^\alpha| \to \infty$ for $\bar{X}^\alpha_0 \neq 0$, with $L^\alpha$ converging to finite values $L^\alpha_*$ at these limits. The derivatives of the $\beta$-function take specific universal values:
$d\beta(L^\alpha)/dL^\alpha=1/2$ at $X^\alpha=0$; $d\beta(L^\alpha)/dL^\alpha=1$ at $X^\alpha\to\infty\,\, for\,\,\zeta_+$; $d\beta(L^\alpha)/dL^\alpha=-1$ at $X^\alpha\to\infty\,\,for\,\,\zeta_-$; $d\beta(L^\alpha)/dL^\alpha=1$ for $X^\alpha_0=0\,\,for\,\,\zeta_+=2$; $d\beta(L^\alpha)/dL^\alpha=0$ for $X^\alpha_0=0\,\,for\,\,\zeta_-=0$. 
TABLE II summarizes the fixed points A, B, C, D, O, and UV/IR endpoints with their respective limiting values. These values are universal, being independent of the arbitrary choice of reference scale $X^\alpha_0=1.6$.

A: For $X^\alpha \to 0^-$ with $\bar X^\alpha_0 > 0$, $L^\alpha\to L^\alpha_* = -\bar{X}^\alpha_0 < 0$. Here, $\beta > 0$ in the $X^\alpha < 0$ regime from A to C. This corresponds to a scale ultraviolet (SUV) fixed point with a positive slope of $\beta(L^\alpha)$. The limits yield $ a_\alpha = 0$, $ b_\alpha = \pm\infty$, $ a_\alpha b_\alpha = \sqrt{|\bar X^\alpha_0|}$, $dX^\alpha =0$, and $-d a_\alpha =-|\bar X^\alpha_0|^{3/2}$ at $X^\alpha \to 0$. There is another flow away from the SUV fixed point A to an IR endpoint ($L^\alpha\to-\infty\,\&\,\beta\to-\infty$), at where, $ a_\alpha = 1/2$, $ b_\alpha = 0$, $ a_\alpha b_\alpha = 0$, $dX^\alpha =-\infty$, and $-d a_\alpha =0$ with $X^\alpha \to -\infty$.

B: For $X^\alpha \to 0^+$ with $\bar X^\alpha_0 < 0$, point B represents a SUV fixed point analogous to A, with the specific limitations behavior of all quantities detailed in TABLE II.

C: For $X^\alpha \to -\infty$ with $\bar X^\alpha_0 > 0$, $L^\alpha\to L^\alpha_* = -\bar{X}^\alpha_0/2 < 0$. Here, $\beta > 0$ in the $X^\alpha < 0$ regime attracted to C. This represents a scale infrared (SIR) fixed point with a negative slope of $\beta(L^\alpha)$. The limits yield $ a_\alpha =-\infty$, $ b_\alpha =-\infty$, $ a_\alpha b_\alpha = \infty$, $dX^\alpha =+\infty$, and $-da_\alpha =-\infty$ at $X^\alpha \to \infty$.

D: For $X^\alpha \to +\infty$ with $\bar X^\alpha_0 < 0$, point D represents an SIR fixed point analogous to C, whose specific limiting behavior is cataloged in TABLE II. Since $dL^\alpha/dX^\alpha=\xi^\alpha=1/a_\alpha=0$ and $L^\alpha=L^\alpha_*= -\bar{X}^\alpha_0/2$, the spacetime becomes static $\&$ discrete phase at the SIR stable fixed point (C or D). This stabilization of the microscopic structure corresponds to a phase where spacetime is discrete, consequently breaking the continuous translational symmetry and reducing it to a discrete one.

O: For $X^\alpha \to 0$ with $\bar X^\alpha_0 = 0$, $L^\alpha_* = 0$. When $\zeta^\alpha = 2$, $\beta < 0$ for $X^\alpha < 0$ and $\beta > 0$ for $X^\alpha > 0$, forming a SUV fixed point with a positive slope.
When $\zeta^\alpha = 0$, $\beta = 0$ across all $X^\alpha$, corresponding to unstable fixed points (saddle point) with zero slope ($d\beta/dL=0$). For $\bar{X}^\alpha_0 = 0$ and $\zeta^\alpha = 2$, $L^\alpha=2X^\alpha$, $ a_\alpha = 1/2$, $ b_\alpha =0$, and $ a_\alpha b_\alpha =0$ universally. When $X^\alpha \to 0$, $L^\alpha=0$, $dX^\alpha=0$, $-d a_\alpha=0$. For $\bar{X}^\alpha_0 = 0$ and $\zeta^\alpha = 0$, $L^\alpha=0$, $ a_\alpha = \infty$, $ b_\alpha =0$. This corresponds to an unstable structure, where $dX^\alpha=\infty*0$, $-d a_\alpha=0*\infty*0$, and they are in an indeterminate form. This marks the boundary between classical continuum and fluctuating spacetime; its instability suggests a phase transition critical point.

Endpoints: For $X^\alpha < 0$ with $\bar{X}^\alpha_0 < 0$, the flow originates from the SUV endpoint ($L^\alpha \to 0$, $\beta \to -\infty$) and proceeds either to the SIR endpoint ($L^\alpha \to -\infty$, $\beta \to -\infty$) when $\zeta^\alpha = \zeta^\alpha_+$ (with $\beta < 0$), or to the SIR fixed point C when $\zeta^\alpha = \zeta^\alpha_-$ (with $\beta < 0$). Conversely, for $X^\alpha > 0$ with $\bar{X}^\alpha_0 > 0$, the flow begins at the SUV endpoint ($L^\alpha \to 0$, $\beta \to +\infty$) and evolves either to the SIR endpoint ($L^\alpha \to +\infty$, $\beta \to +\infty$) for $\zeta^\alpha = \zeta^\alpha_+$ (with $\beta > 0$), or to the SIR fixed point D for $\zeta^\alpha = \zeta^\alpha_-$ (with $\beta > 0$). As shown in FIG.~\ref{fig5}, the SUV endpoints are characterized by $a_\alpha = 0$, $b_\alpha = \pm\infty$, $a_\alpha b_\alpha = \pm\sqrt{|\bar{X}^\alpha_0|}$, $dX^\alpha = 0$, and $-da_\alpha = 0$, while the IR endpoints exhibit $a_\alpha = \tfrac{1}{2}$, $b_\alpha = 0$, $a_\alpha b_\alpha = 0$, $dX^\alpha = \pm\infty$, and $-da_\alpha = 0$. The renormalization group flow exhibits distinct behaviors at the endpoints, which are not fixed points since $\beta \ne 0$ in these regimes.

The existence of a UV fixed point ensures a well-defined and predictive high-energy limit for the theory. The renormalization group flow traces the emergence of physical phenomena with the fundamental microscopic description. While traditional renormalization focuses on the running of coupling constants in the action, this work promotes the measurement scales themselves to running parameters, mapped onto geometric scaling functions $ L^{\alpha}(X^{\alpha}) $. Consequently, the $ \beta $-function here describes the scale-dynamical evolution of $ L^{\alpha} $, rather than perturbative corrections to couplings.

As cataloged in TABLE II and FIG. \ref{fig5}, $a_\alpha$ exhibits simple power-law asymptotics. 
As $X^\alpha \to 0$ one finds $ a_\alpha \propto |X^\alpha|^{1/2}$, hence 
$ a_\alpha \to 0$ (SUV fixed points A/B). 
Similarly, approaching the singular SUV endpoints $X^\alpha \to \bar X_0^\alpha$ gives 
$ a_\alpha \propto |X^\alpha - \bar X_0^\alpha|^{1/2}$, so $ a_\alpha$ again vanishes as the square root of 
the distance to $\bar X_0^\alpha$. 
In the large-scale limit $X^\alpha \to \pm\infty$ the two solution branches separate: 
the ``$+$'' branch tends to a constant $ a_\alpha \to \tfrac{1}{2}$ up to 
$\mathcal{O}(1/X^\alpha)$ corrections, while the ``$-$'' branch grows linearly 
$ a_\alpha \propto X^\alpha$ (the IR divergent branch to C/D). 

By anchoring itself in differential measurement and scaling factors, this mathematical structure embeds spacetime fluctuations directly into geometry, offering an intrinsic approach to fluctuation spacetime. The key is a measurement duality that reveals a natural discrete structure, connecting the operation of measurement to the nature of fluctuations, discreteness, and the emergence of classical geometry.

\section{Scaling-Operator Quantization}
\subsection{Deformed commutators}
In parallel, various modifications of the Heisenberg uncertainty principle have been proposed to encode a minimal length into quantum mechanics.  Noncommutative geometry \cite{snyder1947}, generalized uncertainty principles (GUP) \cite{kempf1995}, and doubly special relativity (DSR) \cite{magueijo2002} all predict deformations of the canonical commutation relations that become significant near \(l_p\).  Anisotropic scaling frameworks, such as Ho\v{r}ava--Lifshitz gravity \cite{horava2009}, further suggest that space and time may scale differently at high energies, leading to novel dispersion relations and potential observational consequences.

In this work, we introduce an approach in which the usual position and momentum operators are endowed with local scaling functions and rescaling factors to capture spacetime fluctuations at the microscale. By defining dimensionless scaling operators and computing their commutators, we reveal how a fluctuation-dependent scaled Planck constant deforms the canonical algebra. This deformation naturally leads to a generalized scale fluctuation uncertainty relation. 

First, we define the rescaled momentum operator. Let
\begin{equation}
\hat{p}_\alpha =-i\,\hbar\,\frac{\partial}{\partial x^\alpha}= -i\, r_{\hbar}\, a_{\alpha}\,\frac{\partial}{\partial X^\alpha}\,p_\alpha^0,
\label{eq:p}
\end{equation}
where $r_{\hbar}$ is a dimensionless normalization Planck constant, defined as $r_\hbar=\frac{\hbar}{\hbar_0}=\frac{\hbar}{dx_0^\alpha p^0_\alpha}$ (no sum on $\alpha$), with $p_\alpha^0$ is a fixed reference momentum. 
And then
\begin{equation}
\hat{p}^\mu\hat{p}_\mu=\eta^{\mu\nu}\hat{p}_\mu\hat{p}_\nu=(p^0)^2\eta^{\mu\nu}r^2_{\hbar} a_{\mu}( a_{\nu}\frac{\partial^{2}}{\partial X^{\mu}\partial X^{\nu}}- b_{\nu}\delta^\nu_\mu\frac{\partial}{\partial X^{\nu}}),
\end{equation}
where the consistency selection $p^0_\alpha=p^0$ of momentum reference is a global gauge.
Taking the expectation value in a spacetime-fluctuation state $|\Phi\rangle$, allows us to identify a dimensionless mass scale $r_m$ by
\begin{equation}
p^2=r^2_{m} r^2_{c}(p^0)^2=\langle\Phi|\hat{p}^\mu\hat{p}_\mu|\Phi\rangle,
\end{equation}
where $r_{c}$ and $r_m$ are dimensionless, defined relative to fixed reference constants, respectively \cite{weihu}.

Next, we define the rescaled position operator via
\begin{equation}
\hat{x}^\alpha = \hat L^\alpha(X^\alpha)\,dx^\alpha_0 + \hat x^\alpha_0,
\label{eq:x}
\end{equation}
where \( \hat L^\alpha(X) \) is the local scaling function, \( dx^\alpha_0 \) is the reference length, and \( \hat x^\alpha_0 \) is a fixed reference coordinate. The nontrivial commutator then arises from the product of \( \hat L^\alpha(X) \, dx^\alpha_0 \) with the differential operator in \( \hat{p}_\alpha \).
The operator commutator is defined as
\begin{equation}
[\hat{x}^\alpha,\hat{p}_\alpha] = \hat{x}^\alpha\hat{p}_\alpha - \hat{p}_\alpha\hat{x}^\alpha,\, \text{no\, sum\,on\,\,$\alpha$ (same\,to\, following)}
\end{equation}
Acting on a test function $f(X^\alpha)$, one finds
\begin{equation}
\begin{aligned}
\hat{x}^\alpha\hat{p}_\alpha
&=\Bigl[\hat L^\alpha\,dx^\alpha_0 + \hat x^\alpha_0\Bigr]\left[-i\,r_\hbar\,p_\alpha^0\, a_\alpha\,f'\right] \\
&=-i\,r_\hbar\,p_\alpha^0\,\Bigl(\hat L^\alpha\,dx^\alpha_0 + \hat x^\alpha_0\Bigr) a_\alpha\,f'.
\end{aligned}
\end{equation}
and
\begin{equation}
\begin{aligned}
&\hat{p}_\alpha\hat{x}^\alpha f(X^\alpha)=-i\,r_\hbar\,p_\alpha^0\, a_\alpha\,\partial_\alpha\Bigl[\bigl(\hat L^\alpha\,dx^\alpha_0 + \hat x^\alpha_0\bigr)f\Bigr] \\
&=-i\,r_\hbar\,p_\alpha^0\, a_\alpha\left[ \hat L'^\alpha\,dx^\alpha_0\,f + \Bigl(\hat L^\alpha\,dx^\alpha_0 + \hat x^\alpha_0\Bigr)f' \right],
\end{aligned}
\end{equation}
where $\hat L'^\alpha=\frac{d\hat L^\alpha}{dX^\alpha}$ and $f'=\frac{\partial f}{\partial X^\alpha}$
Subtracting to find the commutator
\begin{equation}
\begin{aligned}
\bigl[\hat{x}^\alpha,\hat{p}_\alpha\bigr]f&=\hat{x}^\alpha\hat{p}_\alpha f-\hat{p}_\alpha\hat{x}^\alpha f\\
&= i\,r_\hbar\,p_\alpha^0\, a_\alpha\,\hat L'^\alpha\,dx^\alpha_0\,f.
\end{aligned}
\end{equation}
So that the operator-valued commutator is
\begin{equation}
[\hat{x}^\alpha,\hat{p}_\alpha] = i\,\hbar\, a_\alpha\,\hat L'^\alpha,
\end{equation}
In conventional quantum mechanics, one has \([\hat{x}, \hat{p}] = i\hbar\). Here, the appearance of the scaling \( a_{\alpha}\) and the derivative \(\hat L^{\alpha}\) implies a scaled Planck constant, which is also expressed into (due to Eq.~(\ref{eq:firstorder})):
\begin{equation}
  \hat \hbar_{\text{s}} = \hbar a_\alpha\,\hat L'^\alpha=\hbar\,(\frac{\hat {\bar{L}}^\alpha}{\hat L^\alpha}-1),
\end{equation}
A vanishing commutator occurs in the limit where $\hat {\bar{L}}^\alpha\to \hat L^\alpha$. This describes a regime where the local spacetime fluctuations are suppressed. In this limit, the quantum uncertainty from the non-commutativity vanishes, and the formalism reduces to a classical, commuting geometry. This demonstrates how a classical limit can arise from the fluctuating framework.

Using the generalized commutator in the uncertainty principle,
\begin{equation}
\Delta x\,\Delta p \ge \frac{1}{2}\left|\langle[\hat{x},\hat{p}]\rangle\right|.
\end{equation}
We have its expected value is
\begin{equation}
\langle[\hat{x},\hat{p}]\rangle = i\,r_\hbar\,p_\alpha^0\,dx^\alpha_0\,\langle a_\alpha\,\hat L'^\alpha\rangle=\langle i\hat \hbar_{s}\rangle.
\end{equation}
Finally, the generalized uncertainty relation becomes
\begin{equation}
\Delta x\,\Delta p \ge \frac{1}{2}\Bigl|\langle i\hat \hbar_{s}\rangle\Bigr|.
\end{equation}
The linear mapping condition can be extended to $\hat{\bar{L}}^\alpha=2\hat L^\alpha$, makes
\begin{equation}
  \hat \hbar_{s} = \hbar\,(\frac{\hat{\bar{L}}^\alpha}{\hat L^\alpha}-1)\to\hbar;
\end{equation}
recovering the usual constant. And the spacetime is quantized equidistantly with a minimum interval $\bar X_0^\alpha$, as shown in FIG. \ref{fig2}.

The deformed commutation relation and the generalized uncertainty principle, induced by the scaling functions \( \hat L^\alpha \) and the factors \(  a_\alpha \), result in a non-uniform quantum structure with local scaling and microscopic fluctuations, replacing the standard lower bound \( \hbar/2 \).

In any coordinate, the generalized angular momentum (or Lorentz rotation generator) is defined as
\begin{equation}
\hat{J}^\alpha_\beta = \hat{x}^\alpha \hat{p}_\beta - \hat{x}^\beta \hat{p}_\alpha.
\end{equation}
and the commutation relations are
\begin{equation}
\begin{aligned}
\left[\hat{J}^{\alpha}_{\ \beta}, \hat{J}^{\gamma}_{\ \delta}\right] = -i\hbar \left( \delta^{\alpha}_{\delta} \hat{J}^{\gamma}_{\ \beta} - \delta^{\alpha}_{\gamma} \hat{J}^{\delta}_{\ \beta} + \delta^{\gamma}_{\beta} \hat{J}^{\alpha}_{\ \delta} - \delta^{\delta}_{\beta} \hat{J}^{\alpha}_{\ \gamma} \right).
\end{aligned}
\label{eq:Lorentzalgebra}
\end{equation}
Substitute Eq.~(\ref{eq:p}) and Eq.~(\ref{eq:x}) into it and perform some simplifications. First, write the original expression as
\begin{equation}
\begin{aligned}
\hat{J}^\alpha_\beta &= \left(\hat L^\alpha dx^\alpha_0 + \hat x^\alpha_0\right) \left( -i \, r_\hbar p^0_\beta\,  a_\beta \partial_{X^\beta}  \right) \\
&-  \left(\hat L^\beta dx^\beta_0 + \hat x^\beta_0\right) \left( -i \, r_\hbar p^0_\alpha\, a_\alpha \partial_{X^\alpha}  \right)\\
&=-i \, \hbar(\mathcal{\hat D}^\alpha_\beta+ w\mathcal{\hat E}_{\alpha\beta}),
\end{aligned}
\end{equation}
where the references are gauged and $w=\frac{x_0}{dx_0}=\bar X_0\frac{x_0}{d\tilde{x}_0}=\bar X_0$ with gauging $x_0=d\tilde{x}_0$ as showing FIG. \ref{fig2},
and
\begin{equation}
\begin{aligned}
\mathcal{\hat D}^\alpha_\beta &= \hat L^\alpha  a_\beta \partial_{X^\beta} - \hat L^\beta  a_\alpha \partial_{X^\alpha},
\end{aligned}
\end{equation}
and 
\begin{equation}
\begin{aligned}
\mathcal{\hat E}_{\alpha\beta} &=  a_\beta \partial_{X^\beta} -  a_\alpha \partial_{X^\alpha},
\end{aligned}
\end{equation} 
and then
\begin{equation}
\begin{aligned}
&-\hbar^{-2}\left[\hat{J}^{\alpha}_{\ \beta}, \hat{J}^{\gamma}_{\ \delta}\right]\\
&=\left[\hat{\mathcal{D}}^\alpha_\beta, \hat{\mathcal{D}}^\gamma_\delta\right] + w\left[\hat{\mathcal{D}}^\alpha_\beta, \hat{\mathcal{E}}_{\gamma\delta}\right]+ w\left[\hat{\mathcal{E}}_{\alpha\beta}, \hat{\mathcal{D}}^\gamma_\delta\right] + w^2\left[\hat{\mathcal{E}}_{\alpha\beta}, \hat{\mathcal{E}}_{\gamma\delta}\right]
\end{aligned}
\end{equation}
\begin{equation}
\begin{aligned}
&-\hbar^{-2}\left[-i\hbar \left( \delta^{\alpha}_{\delta} \hat{J}^{\gamma}_{\ \beta} - \delta^{\alpha}_{\gamma} \hat{J}^{\delta}_{\ \beta} + \delta^{\gamma}_{\beta} \hat{J}^{\alpha}_{\ \delta} - \delta^{\delta}_{\beta} \hat{J}^{\alpha}_{\ \gamma} \right)\right]\\
&=\mathcal{\hat S}^{(\mathcal{\hat D})\alpha\gamma}{}_{\beta\delta}+w\mathcal{\hat S}^{(\mathcal{\hat E})\alpha\gamma}{}_{\beta\delta}
\end{aligned}
\end{equation}
with defining
\[
\mathcal{\hat S}^{(\mathcal{\hat D})\alpha\gamma}{}_{\beta\delta}=\delta_{\delta}^{\alpha}\hat{\mathcal{D}}^{\gamma}{}_{\beta}-\delta_{\gamma}^{\alpha}\hat{\mathcal{D}}^{\delta}{}_{\beta}+\delta_{\beta}^{\gamma}\hat{\mathcal{D}}^{\alpha}{}_{\delta}-\delta_{\beta}^{\delta}\hat{\mathcal{D}}^{\alpha}{}_{\gamma}
\]
and 
\[
\mathcal{\hat S}^{(\mathcal{\hat E})\alpha\gamma}{}_{\beta\delta}=\delta_{\delta}^{\alpha}\hat{\mathcal{E}}^{\gamma}{}_{\beta}-\delta_{\gamma}^{\alpha}\hat{\mathcal{E}}^{\delta}{}_{\beta}+\delta_{\beta}^{\gamma}\hat{\mathcal{E}}^{\alpha}{}_{\delta}-\delta_{\beta}^{\delta}\hat{\mathcal{E}}^{\alpha}{}_{\gamma}.
\]

Define $\hat \Delta$ as the difference between the left-hand side and the right-hand side of Eq.~(\ref{eq:Lorentzalgebra}).
If $\partial_{i}a_{\mu}=0$ and $\partial_{i}\hat{L}^{\mu}=0$ (i.e., $a$ and $\hat{L}$ are scale-independent constants), then all derivative terms appearing in $[\mathcal{\hat D},\mathcal{\hat D}]$, $[\mathcal{\hat D},\mathcal{\hat E}]$, and $[\mathcal{\hat E},\mathcal{\hat E}]$ vanish. In that case, the commutators of $\mathcal{\hat D}$ and $\mathcal{\hat E}$ can be expressed in the usual form of structure constants multiplying $\mathcal{\hat D}$ or $\mathcal{\hat E}$, so $\hat \Delta=0$: the standard Lorentz algebra is recovered and the algebra closes. Conversely, if $\hat{L}$ or $a$ depend on scale, extra derivative terms $\partial\hat{L}$ and $\partial a$ appear; these modify the left-hand side of $[\hat{J},\hat{J}]$ so that it is no longer given solely by structure constants times $\hat{J}$, but contains additional terms. Physically, those extra terms can be interpreted as corrections to the rotation/boost generator algebra induced by local scaling (microscopic) fluctuations, i.e. as deformations of the standard Lorentz closure.

This section generalizes the canonical commutation relations by introducing locally scaled position and momentum operators, where the scaling functions $\hat L^\alpha(X)$ and factors $ a_\alpha$ encode microscopic spacetime fluctuations. The resulting commutator acquires a scaled Planck constant $\hat\hbar_{s}$, leading to a generalized uncertainty principle whose lower bound depends on local fluctuation properties. In the fluctuation scale-free limit ($\hat{\bar L}^\alpha = \hat L^\alpha$), $\hat\hbar_{s} \to 0$ and the theory reduces to classical mechanics. A linear scaling restores the standard $\hbar$ and yields a discretized spacetime with a minimal length. The generalized angular momentum operators correct the Lorentz algebra with fluctuation-dependent embeddings. This picture suggests that the local deformations of spacetime, encoded in the factors $a_\mu$, dynamically govern the effective strength of quantum non-commutativity and uncertainty.

\subsection{Locally scaled Klein-Gordon and Dirac Equations} 
In Ref. \cite{weihu}, key physical differential equations such as the Klein-Gordon \cite{Klein, Gordon} and the Dirac equations \cite{Dirac} are transformed into scaling form, thereby encoding the effects of microscopic spacetime fluctuations (governed by the factors $a_\mu$ and $b_\mu$ ) on the propagation of fields. The scaling Klein-Gordon equation is given by
\begin{footnotesize}
\begin{equation}
\begin{aligned}
&\left[\eta^{\mu\nu} a_{\mu}\left(  a_{\nu}\frac{\partial^{2}}{\partial X^{\mu}\partial X^{\nu}}- b_{\nu}\frac{\partial X^{\nu}}{\partial X^{\mu}}\frac{\partial}{\partial X^{\nu}}\right)-\left( \frac{r_{m} r_{c}}{r_{\hbar}} \right)^{2}\right]\phi=0,
\end{aligned}
\end{equation}
\end{footnotesize}
where, $\phi$ is now regarded as a scaled field $\phi(X^\sigma)$ defined on the scale manifold. Its dynamics are influenced by the local spacetime fluctuations encoded in the scaling factors appearing in the equation. This scaled field is projected onto spacetime coordinates via the measurement relation $x^{\sigma} = x^{\sigma}_{0} + L^{\sigma}(X^{\sigma})dx^{\sigma}_{0}$.

The scaling covariant Euler-Lagrange equation is modified to be
\begin{equation}
\frac{\partial \mathcal{L}}{\partial \phi} -  a_\mu\frac{\partial}{\partial X^\mu} \left( \frac{\partial \mathcal{L}}{\partial ( a_\mu\frac{\partial \phi}{\partial X^\mu})} \right) = 0 
\end{equation}
Thus, the corresponding scaling Lagrangian density is given by
\begin{equation}
\mathcal{L} = 
\frac{1}{2} \, \eta^{\mu\nu}  a_{\mu}  a_{\nu} \left( \frac{\partial \phi}{\partial X^\mu} \, \frac{\partial \phi}{\partial X^\nu} \right) 
+ \frac{1}{2} \left( \frac{r_{m} r_{c}}{r_{\hbar}} \right)^2 \phi^2
\end{equation}
and the corresponding action is 
\begin{equation}
S=\int \mathcal{L}\frac{1}{d^4x_0}d^4x=\int \mathcal{L}\frac{1}{\prod_{\mu=0}^{D-1} a_\mu}d^4X.
\end{equation}

Given the scaling Lagrangian density, the conjugate momentum is defined as:
\begin{equation}
\pi(x) = \frac{\partial \mathcal{L}}{\partial( a_{0}\partial_0 \phi)}=\eta^{00} a_{0}\partial_0 \phi.
\end{equation}
The Hamiltonian density is defined as:
\begin{equation}
\mathcal{H} = \pi \,  a_{0}\partial_0 \phi - \mathcal{L}.
\end{equation}
Then, we have
\begin{equation}
\mathcal{H} =\frac{1}{2} \Bigl[\eta^{00}( a_{0}\partial_0\phi)^2- \eta^{ij} a_{i} a_{j}\partial_i\phi\partial_j\phi-\left( \frac{r_{m} r_{c}}{r_{\hbar}} \right)^2\phi^2\Bigr].
\end{equation}

The Scaling Dirac equation is
\begin{equation}
\begin{aligned}
&[i\hat{\gamma}^{\mu} a_{\mu}\frac{\partial}{\partial X^{\mu}}-\frac{r_{m} r_{c}}{r_{\hbar}}]\psi=0
\end{aligned}
\end{equation}
with Einstein notation. Here, $\psi=\psi(x^{\sigma}\rightarrow x^{\sigma}_{0}+L^{\sigma}(X^{\sigma})dx^{\sigma}_{0})$. The corresponding scaling Lagrangian density is given by
\begin{equation}
\mathcal{L} = 
\bar{\psi} \left( i \hat{\gamma}^\mu  a_\mu \frac{\partial}{\partial X^\mu} 
- \frac{r_m r_c}{r_\hbar} \right) \psi
\end{equation}

The key idea of this part is to view spacetime at the smallest scales as a locally rescaled version of the smooth geometry, so that every derivative and metric component in the usual Klein-Gordon and Dirac equations is modified by direction-dependent scale factors. These factors encode how nonlinear fluctuations stretch or compress distances and times in each coordinate direction. When one constructs the Lagrangian, each kinetic term carries the same scale factors, and the corresponding conjugate momenta and Hamiltonian densities reflect how local distortions of spacetime alter energy densities. In the Dirac case, the gamma matrices are coupled to the rescaled geometry so that the spinor field dynamics incorporate microscale spacetime fluctuations. 

\section{Covariant Geometry and Vierbein Structure}
\subsection{Lorentz and general covariance}
Under a Lorentz transformation \( \Lambda^{\mu}{}_{\nu} \), spacetime coordinates and partial derivatives transform according to the standard rules of special relativity \cite{Wald1984, Weinberg1972}:
\[ 
x'^{\mu} = \Lambda^{\mu}{}_{\nu} x^{\nu}, \quad \partial'_{\mu} = \frac{\partial}{\partial x'^{\mu}} = (\Lambda^{-1})^{\nu}{}_{\mu} \partial_{\nu}.
\]

In the present framework, we adopt a fundamental postulate: all inertial observers employ the same fixed reference length scale for measurements. This is implemented through the invariant reference length condition:
\begin{equation}
d\lambda_0 = dx_0^\alpha = dx_0^\beta = dx_0'^\alpha = dx_0'^\beta,
\label{eq:reference}
\end{equation}
which ensures a consistent operational foundation across different reference frames.

The physical coordinates are expressed in terms of scaling functions as
\[
x^\mu = L^\mu(X^\mu) dx^\mu_0 + x^\mu_0.
\]
Under Lorentz transformation, the coordinates transform as
\[
x'^\mu = \Lambda^{\mu}{}_{\nu} x^\nu = \Lambda^{\mu}{}_{\nu} L^\nu dx^\nu_0 + \Lambda^{\mu}{}_{\nu} x^\nu_0.
\]
Physically, \( x_0^{\mu} \) is just another set of coordinate components, representing the position of the `origin'' or reference point'' in spacetime. resulting in the reference point \( x_0^\mu \) transforms as a vector, \( x'^\mu_0 = \Lambda^{\mu}{}_{\nu} x^\nu_0 \), and invoking the invariant reference length condition, we obtain the transformation law for scaling functions:
\begin{equation}
L'^\mu = \Lambda^{\mu}{}_{\nu} L^\nu.
\end{equation}
Thus, the scaling functions \( L^\mu \) transform as Lorentz vectors.

We now examine the transformation properties of the scale factors \( a_\mu \). Recall the definition of scaled derivatives:
\[
\frac{\partial}{\partial x^\mu} = a_\mu \frac{\partial}{\partial X^\mu} \frac{1}{dx_0^\mu}.
\]
Under Lorentz transformations, the left-hand side transforms covariantly:
\[
\frac{\partial}{\partial x'^\mu} = (\Lambda^{-1})^\nu{}_\mu \frac{\partial}{\partial x^\nu}.
\]
Using the invariant reference length condition, the scaled derivative relation in the primed frame becomes
\[
a'_\mu \frac{\partial}{\partial X'^\mu} \frac{1}{dx_0^\mu} = (\Lambda^{-1})^\nu{}_\mu a_\nu \frac{\partial}{\partial X^\nu} \frac{1}{dx_0^\nu}.
\]
To proceed, we postulate that the scale coordinates \( X^\mu \) transform as Lorentz vectors:
\begin{equation}
X'^\mu = \Lambda^{\mu}{}_{\nu} X^\nu, \quad \frac{\partial}{\partial X'^\mu} = (\Lambda^{-1})^\nu{}_\mu \frac{\partial}{\partial X^\nu}.
\end{equation}
This assumption is natural given the vectorial character of \( L^\mu \) and the relation \( dX^\mu = a_\mu L^\mu \). Comparing both sides of the transformed derivative relation, we deduce the invariance of scale factors:
\begin{equation}
a'_\mu = a_\mu.
\end{equation}
That is, the scale factors \( a_\mu \) remain invariant under Lorentz transformations. Consequently, the second-order factors \( b_\mu = -d a_\mu / dX^\mu \) transform as covectors:
\begin{equation}
b'_\mu = (\Lambda^{-1})^\nu{}_\mu b_\nu.
\end{equation}

To extend Lorentz invariance to the scaled geometry, we introduce scaled Lorentz transformations that preserve the hierarchical metric structures. For the first-order scaled geometry, we define:
\begin{equation}
\Xi^\alpha{}_\mu = \frac{\partial X^\alpha}{\partial X'^\mu} = \frac{\partial x^\alpha}{\partial x'^\mu} \frac{a_\alpha}{a'_\mu} = \Lambda^\alpha{}_\mu \frac{a_\alpha}{a_\mu}, 
\end{equation}
where we have used the invariance \( a'_\mu = a_\mu \). For Minkowski background \( g_{\alpha\beta} = \eta_{\alpha\beta} \), the scaled metric \( \hat{g}_{\alpha\beta} = \eta_{\alpha\beta} \frac{a^2}{a_\alpha a_\beta} \) transforms as:
\begin{equation}
\begin{aligned}
\hat{g}'_{\mu\nu} &= \hat{g}_{\alpha\beta} \Xi^\alpha{}_\mu \Xi^\beta{}_\nu = \eta_{\alpha\beta} \frac{a^2}{a_\alpha a_\beta} \left( \Lambda^\alpha{}_\mu \frac{a_\alpha}{a_\mu} \right) \left( \Lambda^\beta{}_\nu \frac{a_\beta}{a_\nu} \right)\\
&= \frac{a^2}{a_\mu a_\nu} \eta_{\mu\nu} = \hat{g}_{\mu\nu}.
\end{aligned}
\end{equation}
Thus, \( \Xi \) preserves the first-order scaled metric \( \hat{g} \).

The standard Lorentz transformation \( \Lambda \) preserves the Minkowski metric \( \eta \), while the scaled transformations \( \Xi \) preserve the first-order metrics \( \hat{g} \). The directional scale factors \( a_\mu \) remain invariant under Lorentz transformations, carrying intrinsic fluctuation spectra that cannot be removed by coordinate transformations. The geometry and physical observables are fundamentally rooted in these spectral properties.

The principle of general covariance requires form-invariance of physical laws under general coordinate transformations \cite{Misner1973}. The metric tensor transforms as:
\[
g'_{\mu\nu} = \frac{\partial x^\alpha}{\partial x'^\mu} \frac{\partial x^\beta}{\partial x'^\nu} g_{\alpha\beta}.
\]
We now verify the general covariance of the scaled metrics. For \( \hat{g}_{\mu\nu} = g_{\mu\nu} \frac{a^2}{a_\mu a_\nu} \), we have:
\[
\hat{g}'_{\mu\nu} = g'_{\mu\nu} \frac{a^2}{a'_\mu a'_\nu} = \left( \frac{\partial x^\alpha}{\partial x'^\mu} \frac{\partial x^\beta}{\partial x'^\nu} g_{\alpha\beta} \right) \frac{a^2}{a_\mu a_\nu}.
\]
Based on Eq. (\ref{eq:defination}) and (\ref{eq:reference}), one finds the scale coordinates transform \( \frac{\partial x^\alpha}{\partial x'^\mu} = \frac{\partial X^\alpha}{\partial X'^\mu} \frac{a_\mu}{a_\alpha} \), we obtain:
\[
\hat{g}'_{\mu\nu} = \frac{\partial X^\alpha}{\partial X'^\mu} \frac{\partial X^\beta}{\partial X'^\nu} \left( g_{\alpha\beta} \frac{a^2}{a_\alpha a_\beta} \right) = \frac{\partial X^\alpha}{\partial X'^\mu} \frac{\partial X^\beta}{\partial X'^\nu} \hat{g}_{\alpha\beta}.
\]
Thus, \( \hat{g}_{\mu\nu} \) transforms as a tensor under general coordinate transformations.

For the second-order metric \( \tilde{g}_{\mu\nu} = \hat{g}_{\mu\nu} \frac{b^2}{b_\mu b_\nu} \), we have:
\begin{equation}
\begin{aligned}
\tilde{g}'_{\mu\nu}&=\frac{ b^2}{ b'_\mu b'_\nu}\hat{g}'_{\mu\nu}=\frac{ b^2}{ b'_\mu b'_\nu} \left(\frac{\partial X^\alpha}{\partial X'^\mu}\frac{\partial X^\beta}{\partial X'^\nu}\hat g_{\alpha\beta} \right)\\
&=\frac{ b^2}{ b'_\mu b'_\nu}\frac{\partial  a_\alpha}{\partial  a'_\mu}\frac{\partial  a_\beta}{\partial  a'_\nu}\frac{ b'_\mu b'_\nu}{ b_\alpha b_\beta}\hat g_{\alpha\beta}\\
&=\frac{ b^2}{ b_\alpha b_\beta}\frac{\partial  a_\alpha}{\partial  a'_\mu}\frac{\partial  a_\beta}{\partial  a'_\nu}\hat g_{\alpha\beta}\\
&=\frac{\partial  a_\alpha}{\partial  a_\mu}\frac{\partial  a_\beta}{\partial  a_\nu}\tilde{g}_{\alpha\beta}\\
&=\delta^\alpha_\mu\delta^\beta_\nu\tilde{g}_{\alpha\beta}\\
&=\tilde{g}_{\mu\nu},
\end{aligned}
\end{equation}
where, the amplitude coordinates \( a_\mu \) are scalars under general coordinate transformations, \( \frac{\partial a_\alpha}{\partial a'_\mu} = \delta^\alpha_\mu \).
Hence, \( \tilde{g}_{\mu\nu} \) is invariant under general coordinate transformations.

In this scale-based framework, the coordinate \( X^\mu \) represents a fluctuation-sensitive rescaled coordinate system defined via \( dX^\mu = a_\mu L^\mu \), where \( a_\mu \) encodes quantum fluctuation amplitudes along each direction. Although \( a_\mu \) is not a conventional tensor, the composite operator \( a_\mu \partial / \partial X^\mu \) transforms covariantly, and the anisotropic metric \( \hat{g}_{\mu\nu} = [\Omega^{(a)}_{\mu\nu}]^{-1} g_{\mu\nu} \) maintains proper transformation properties under general coordinate changes. This defines a first-order scale geometry $(\mathcal{M}^{( a)}, \hat g, X^\mu)$, encoding local scale effects induced by spacetime fluctuations. 

Practically, \( a_\mu\) influences field propagation, dispersion relations, and integration measures. Upon quantization, the spectral properties and self-adjointness of these operators govern the mathematical consistency of constructs such as the micro-area operator and the associated state counting. In summary, \( a_\mu\) is interpreted as a collection of direction-labeled, dimensionless scalar factors whose anisotropic amplitudes embody the fundamental degrees of freedom encoding the intrinsic local structure and dynamics of microscopic spacetime fluctuations.

The second-order factors \( b_\mu = -d a_\mu / dX^\mu \) capture non-uniformities in scale deformations, leading to the second-order metric \( \tilde{g}_{\mu\nu} = [\Omega^{(b)}_{\mu\nu}]^{-1} \hat{g}_{\mu\nu} \), which remains invariant under general coordinate transformation. Together, \( a_\mu \) and \( b_\mu \) define a hierarchical, fluctuation-driven geometry with generalized covariance at each level, realized through successive anisotropic conformal transformations.

Operationally, \( a_\mu \) exhibits dual characteristics: in mixed tensor-scale expressions, it functions as a direction-dependent multiplicative coefficient without triggering index contraction; when treated as coordinates on \( \mathcal{M}^{(b)} \), it follows standard differential geometry rules after reparameterization. The composite operator \( a_\mu \partial_{X^\mu} \) and scaled constructions ensure that field equations and geometric objects maintain general covariance.

Concerning covariance, we make a summary as following: the reference lengths \( dx_0^\alpha \) and \( dx_0'^\alpha \), reference momenta \( p^0_\mu \) and \( p'^0_\mu \) are the same across different reference frames, respectively. The scale factors \( a_\mu, \zeta_\mu, \xi_\mu \) are dimensionless, direction-dependent invariants under coordinate transformations. The micromeasures \( L^\mu \), second-order factors \( b_\mu \), and metric \( \hat{g} \) are Lorentz-covariant.  The metric \( \tilde{g} \) is a generalized invariant, and \( \hat{g} \) is generally covariant.

\subsection{The connection of $ a_\alpha$ to the vierbein $e^m_{\ \mu}$}
To establish a rigorous connection between the scale factors \( a_\alpha \) and the standard vierbein formalism, we begin by recalling the fundamental role of the vierbein in curved spacetime geometry. At every point \( x \) in the curved spacetime \( (M, g_{\mu\nu}) \), we select a set of local orthonormal bases \( \{e_m(x)\} \), where Latin indices \( m,n = 0,1,2,3 \) represent local Minkowski (Lorentz) indices, and Greek indices \( \mu,\nu = 0,1,2,3 \) represent coordinate indices.

This set of bases can be written as
\[
e_m(x) = e_m^{\ \mu}(x) \,\partial_\mu, 
\quad\text{or}\quad 
e^m(x) = e^m_{\ \mu}(x) \, dx^\mu ,
\]
satisfying the orthogonality relations
\[
e_m^{\ \mu} e^n_{\ \mu} = \delta_m^n, 
\quad
e^m_{\ \mu} e_m^{\ \nu} = \delta_\mu^\nu .
\]

The vierbein relates the metric \( g_{\mu\nu}(x) \) to the local Minkowski metric \( \eta_{mn} = \mathrm{diag}(-1,+1,+1,+1,\,...) \) via
\[
g_{\mu\nu}(x) = e^m_{\ \mu}(x)\, e^n_{\ \nu}(x)\, \eta_{mn}.
\]
Thus, the vierbein can be viewed as the ``square root'' of the metric \cite{DeFelice1990,Eguchi1980}.

Due to the equivalence principle, physical laws are naturally expressed in Minkowski space. After introducing the vierbein, we can find at each point in curved space a local inertial coordinate system where the metric measured by a free-falling observer is \(\eta_{mn}\) \cite{Kannenberg2016,Yepez2011}. 

The vierbein has additional degrees of freedom: if a metric \( g_{\mu\nu} \) is given, the vierbein can be locally Lorentz transformed without changing \( g_{\mu\nu} \). Specifically, if \( \Lambda^m_{\ n}(x) \in SO(1,3) \), then
\[
e'^m_{\ \mu}(x) = \Lambda^m_{\ n}(x) \, e^n_{\ \mu}(x) 
\quad\Rightarrow\quad 
g_{\mu\nu} = e'^m_{\ \mu} e'^n_{\ \nu} \,\eta_{mn}.
\]

To connect this formalism with our scaling framework, we recall the relation between the physical metric and the scaled metric: $g_{\alpha\beta}=\hat g_{\alpha\beta}\frac{ a_\alpha a_\beta}{ a^2}$. We now present a minimal embedding of the anisotropic scaling factors into the vierbein formalism.
To illustrate the embedding in a simple case, we consider a scale-local inertial frame where the substrate metric is flat, \( \hat{g}_{\alpha\beta}= \eta_{\alpha\beta} \). In this frame, the anisotropic scaling can be embedded into the vierbein as follows:
\[
g_{\mu\nu} = e^m_{\ \mu} e^n_{\ \nu} \,\eta_{mn},
\]
with \(\hat g_{\mu\nu}=\eta_{\mu\nu}\) and \(\eta_{\mu\nu} = \delta^m_{\ \mu}\delta^n_{\ \nu}\eta_{mn}\), by setting
\[
e^m_{\ \mu} = \frac{ a_\mu}{ a} \,\delta^m_{\ \mu}, 
\qquad 
e_n^{\ \mu} = \frac{ a}{ a_\mu} \,\delta_n^{\ \mu}.
\]
To restore the complete local Lorentz symmetry while retaining directional scaling, we introduce a local scaling Lorentz rotation through \( \Xi^m_{\ n}(X) \):
\[
E^m_{\ \mu} = \Xi^m_{\ n} e^n_{\ \mu} 
= \Xi^m_{\ n} \frac{ a_\mu}{ a} \,\delta^n_{\ \mu}.
\]
This extended vierbein preserves both directional stretching effects and local Lorentz invariance. It reproduces the anisotropic conformal scaling presented in this work while embedding it within the standard vierbein framework, which facilitates the introduction of spin fields and the writing of covariant derivatives using spin connections. From the transformation property \( \Xi^\alpha_{\ \mu} = \frac{a_\alpha}{a_\mu} \Lambda^\alpha_{\ \mu} \) (no sum), we derive the consistency condition:
\[
\frac{1}{a_m} e^m_{\ \mu} = \Lambda^m_{\ n} \frac{1}{a_n} e^n_{\ \mu}.
\]

We now extend this construction to the second-order scaled geometry. From the relations: \( \hat g_{\alpha\beta} = \eta_{\alpha\beta} \) and \(\tilde{g}_{\alpha\beta} = \eta_{\alpha\beta} \,\frac{ b^2}{ b_\alpha b_\beta}\), define
\[
\tilde{g}_{\alpha\beta} = \tilde{e}^m_{\ \alpha} \tilde{e}^n_{\ \beta} \,\eta_{mn},
\]
with orthogonality
\[
\tilde{e}_m^{\ \mu} \tilde{e}^n_{\ \mu} = \delta_m^n, 
\quad
\tilde{e}^m_{\ \mu} \tilde{e}_m^{\ \nu} = \delta_\mu^\nu,
\]
and
\[
\tilde{e}^m_{\ \alpha} = \frac{ b}{ b_\alpha} \,\delta^m_{\ \alpha}.
\]
Restoring the complete scaling vierbein in metric \(\hat g\) gives
\[
\tilde{E}^m_{\ \mu} = \Xi^m_{\ n} \tilde{e}^n_{\ \mu} 
= \Xi^m_{\ n} \frac{ b}{ b_\mu} \,\delta^n_{\ \mu}
= \Lambda^m_{\ n} \frac{ a_m}{ a_n} \frac{ b}{ b_\mu} \,\delta^n_{\ \mu},
\]
which preserves both directional stretching and local Lorentz symmetry.

$a_\alpha$ is a scale position-dependent or dynamical quantity in the case of spacetime fluctuations considered in this work and its derivatives contribute nontrivially to the spin connection. This affects the covariant derivatives of spinor and vector fields, potentially leading to observable anisotropic gravitational effects. The preservation of local Lorentz invariance via the matrix $\Xi^m_{\ n}$ ensures that the formalism remains consistent with standard vierbein theory, even when such anisotropies are present.  The dependence of the spin connection on derivatives of $a_\alpha$ modifies the covariant derivative operator for fermions. This results in a non-trivial, anisotropic coupling between spacetime fluctuations and spinor fields, such as modifications to dispersion relations.

\section{First- and Second-Order Geodesic Equations in Scaling Manifolds} 
Quantum gravity suggests that spacetime at microscopic scales exhibits complex fluctuations, challenging classical geometric descriptions. To systematically capture these scale-dependent spacetime fluctuations, inspired by the geometric description of classical trajectories, we propose a geometric analogue for the evolution of scale fluctuations. We hypothesize that the locally ``shortest'' or ``least'' scale path of the scale degrees of freedom themselves can be described by a geodesic principle on an appropriately defined abstract scale manifold. This provides a variational principle for the dynamics of fluctuations, analogous to the least-action principle for particle motion. This manifold, equipped with a metric encoding scale fluctuations, provides a geometric framework to study the evolution of scale degrees of freedom. Furthermore, by introducing a second-order manifold for fluctuation amplitudes, one can describe how the internal structure of these fluctuations evolves and interacts, revealing richer nonlinear dynamics beyond the first-order scale space. The first- and second-order geodesic equations derived from variational principles thus generalize classical geodesic geometry to describe spacetime fluctuation in both scale coordinates \(X^\alpha\) and fluctuation amplitude spaces $a_\alpha$. This thought offers a natural and intuitive geometric language to investigate quantum spacetime hierarchical and nonlinear structure, bridging classical geometric methods with quantum gravitational phenomena.

Let \(\mathcal{M}^{( a)}\) be the scale space manifold, with scale coordinates \(X^\alpha\), describing different scale degrees of freedom. Introduce $\hat g_{\alpha\beta}(X)$ as a smooth metric tensor on \(\mathcal{M}^{( a)}\). Take the following action with the affine parameter \(\mathcal{X}\)
\begin{equation}
S[X] = \frac{1}{2} \int \hat g_{\alpha\beta}(X) \frac{dX^\alpha}{d\mathcal{X}} \frac{dX^\beta}{d\mathcal{X}} d\mathcal{X}.
\end{equation}
After the variation \(X^\alpha(\mathcal{X}) \rightarrow X^\alpha(\mathcal{X}) + \delta X^\alpha(\mathcal{X})\) (with fixed endpoints), the geodesic equations can be obtained, ensuring torsion-free and metric compatibility $\nabla_\rho \hat g_{\alpha\beta}=0$. Therefore, the first-order equations govern the dynamics of scaling coordinates $X^\alpha$ in fluctuating spacetime:
\begin{equation}
\frac{\mathrm{d}^2 X^\alpha}{\mathrm{d}\mathcal{X}^2} + \hat{\Gamma}^\alpha_{\mu\nu} \frac{\mathrm{d}X^\mu}{\mathrm{d}\mathcal{X}} \frac{\mathrm{d}X^\nu}{\mathrm{d}\mathcal{X}} = 0,
\label{eq:firstgeodesic}
\end{equation}
where
\begin{equation}
\hat{\Gamma}^{\alpha}_{\sigma\gamma}=\frac{1}{2}\hat g^{\alpha\mu}(\frac{\partial \hat g_{\gamma\mu}}{\partial X^{\sigma}}+\frac{\partial\hat g_{\sigma\mu}}{\partial X^{\gamma}}-\frac{\partial\hat g_{\sigma\gamma}}{\partial X^{\mu}})
\end{equation}
is defined as a First-order Scaling Levi-Civita Connection (FSLCC).

These equations are the geodesic equations on the scale manifold \((\mathcal{M}^{( a)}, \hat g)\), describing the evolution of scale coordinates, under the extremal scale variation, which we interpret as representing the most natural or least fluctuating evolution of the scale degrees of freedom themselves, depicting a free evolution path on the metric $\hat g_{\alpha\beta}$. 
The geodesic equations on the scale manifold prescribe the free evolution of the scale degrees of freedom determined solely by the geometry of the scale manifold.  In a local scaling-inertial frame (i.e. in a neighborhood where $\hat\Gamma^\alpha{}_{\mu\nu}=0$), the connection contributes no coordinate-dependent inertial term and the scale coordinates evolve linearly with the affine parameter, corresponding to a locally inertial, uncoupled (or free) scale flow. This indicates that, under this coordinate description, there is no nonlinear scale coupling introduced by the connection.  By contrast, nonzero connection components \(\hat\Gamma^\alpha_{\mu\nu}\neq0\) indicates that the local basis on the scale manifold varies with scale coordinates, affected by effective, force-like contributions that drive departures from linear scale evolution and generate nonlinear coupling. As discussed next section VI, it can be interpreted as effective feedback from the amplitude distribution or microstructure onto the scale flow, leading to nonlinear interactions and coupling between scales.  Crucially, however, the intrinsic and observable content of such departures is encoded not in the connection alone but in curvature and in the geodesic-deviation equation: a nonvanishing Riemann tensor \(\hat R^\alpha_{\beta\gamma\delta}\) signals irreversible inter-scale ``tidal'' effects, controls the separation or focusing of nearby scale flows, and thus provides a quantitative measure of the strength and stability of inter-scale coupling, whereas a nonzero \(\hat\Gamma\) with vanishing curvature can be removed by a coordinate transformation and therefore reflects a choice of parametrization rather than an intrinsic physical interaction.

The physical picture of the first-order scale manifold may be imagined as an elastic substrate, characterizing the fundamental stretching and contraction of spacetime scales in different directions. In this sense, the coordinates $X^\alpha$ describe how the elastic foundation is deformed at the most basic level. The amplitudes $a_\mu$ represent the internal fluctuation intensity defined on this substrate, quantifying how strongly the substrate fluctuates along each direction, thus providing a possible description of the microstructure.

To investigate the interactions between the fluctuation amplitudes \( a_\alpha\), we can define the amplitude coordinates \( a_\alpha\) and the metric \(\tilde{g}_{\alpha\beta}\) on the manifold \(\mathcal{M}^{( b)}\). Clarifying that $a_\alpha$ are coordinates on $\mathcal{M}^{(b)}$ representing the configuration space of fluctuation amplitudes. Here, the amplitude manifold $\mathcal{M}^{(b)}$ is not an extension of spacetime, but rather the configuration space or field space of the fluctuation amplitudes $a_{\alpha}$. A point in $\mathcal{M}^{(b)}$ corresponds to a specific configuration of these amplitudes across spacetime. A geodesic on $\mathcal{M}^{(b)}$, parameterized by $\sigma$, then describes a continuous transformation from one such global configuration to another. 

Note that two distinct types of coordinate transformations are introduced. The first is the diffeomorphisms of the external spacetime (spacetime diffeomorphisms / Lorentz transformations), under which \( a_{\alpha}(X) \) are proven to be a set of scalar variables (i.e., invariant under external spacetime scale coordinate transformations). The second is the internal reparametrization on the amplitude manifold \( M^{(b)} \), whose transformation rule is the standard one for internal coordinates. To avoid confusion, we denote: external spacetime indices with Greek letters \(\alpha, \beta, \mu, \nu, \ldots \) and amplitude (internal) manifold indices with capital Latin letters \( I, J, K, \ldots \). With this distinction, reparametrize $a_\mu$ as $a_I=a_I(a)$ to be the internal coordinates, satisfying $a'_I(X')=a_I(X)$, $a^I=\tilde{g}^{IJ}a_J$, and $\tilde{g}'_{IJ}(a')=\frac{\partial a^M}{\partial a'^I}\frac{\partial a^K}{\partial a'^J}\tilde{g}_{MK}(a)$. It follows that the geodesic equation on the amplitude space possesses the correct covariance under both types of transformations – invariant under external spacetime transformations because $a_I$ are scalars, and covariant under internal reparametrizations because the equation is tensorial in form.

Thus, we have the following action
\begin{equation}
\tilde{S}[a] = \frac{1}{2} \int \tilde{g}_{IJ}(a) \frac{d a^I}{d\sigma} \frac{d a^J}{d\sigma} d\sigma,
\label{eq:action_a}
\end{equation}
where $\sigma$ is the affine parameter.
Therefore, the second-order equations govern the dynamics of fluctuation amplitudes
\begin{equation}
\frac{\mathrm{d}^2  a^I}{\mathrm{d} \sigma^2} + \tilde{\Gamma}^I_{JK} \frac{\mathrm{d} a^I}{\mathrm{d} \sigma} \frac{\mathrm{d} a^J}{\mathrm{d} \sigma} = 0,
\label{eq:geodesic}
\end{equation}
where
\begin{equation}
\tilde{\Gamma}^I_{JK}=\frac{1}{2}\tilde{g}^{IM}(\frac{\partial \tilde{g}_{JM}}{\partial  a^{K}}+\frac{\partial\tilde{g}_{KM}}{\partial  a^{J}}-\frac{\partial\tilde{g}_{JK}}{\partial  a^M})
\label{eq:SSLCC}
\end{equation}
is defined as a Second-order Scaling Levi-Civita Connection (SSLCC). Here, $ a_{\alpha}$ is regarded as a coordinate system of a multi-dimensional parameter space, used to describe the ``internal state'' of fluctuation amplitudes or the second-order hierarchy of scale geometry. This approach is similar to that in physics, where "phase space" or the "internal degrees of freedom space" is treated. Although the coordinates themselves are dimensionless scalars, a metric and connection can still be defined to characterize the geometric properties of the fluctuation amplitude space. This represents a conceptual extension of scale space: from scale coordinates $X^{\alpha}$ to fluctuation amplitude coordinates $ a_{\alpha}$, forming a hierarchical geometry.
When the connection $\tilde{\Gamma}^I_{JK}$ vanishes, the amplitude configurations evolve independently and linearly, corresponding to uncoupled “breathing modes” on the elastic substrate. By contrast, when $\tilde{\Gamma}^I_{JK}\neq 0$, it induces nonlinear coupling among different configurations. 

In this framework, the geodesic equations specify the natural, least-distorted trajectory, capturing how the internal states of microscope spacetime fluctuations propagate, couple, and evolve within a hierarchical geometric structure, which can be summarized as follows: the first-order scale manifold $\mathcal{M}^{(a)}$ provides the substrate frame ($X^\alpha$) for scaling, while the second-order amplitude manifold $\mathcal{M}^{(b)}$ describes the configurational degrees of freedom ($a^I$) of the fluctuations themselves.

Although each component \( a_\alpha\) carries an index, it is fundamentally a set of dimensionless Lorentz scalar functions, each representing the amplitude of quantum fluctuations along a particular direction. It does not transform as a conventional vector field but rather as multiple scalar fields labeled by direction. Consequently, it is mathematically consistent to treat \( a_\alpha\) as coordinates on the manifold \(\mathcal{M}^{( b)}\) that describes the internal state or geometry of fluctuation amplitudes. This construction extends the geometric framework of scale space by introducing a higher-level manifold where the metric and connection capture the evolution of fluctuation inhomogeneities. Clarifying this distinction avoids confusion and highlights that these coordinates serve as parameters for a hierarchical, scale-dependent geometric structure rather than conventional spacetime vectors.

By explicitly introducing affine parameters and applying action-based variational principles, while ensuring metric compatibility of the connections for both the first-order (scale $\hat g$) and second-order (amplitude $\tilde g$) geodesic equations, which provides an effective description of the scale degrees of freedom ($X^\alpha$) and fluctuation amplitude ($ a_\alpha$) of microscope spacetime geometry. Extending classical geodesic concepts to scale space geodesics, it provides a natural geometric language and an intuitive principle of least action for investigating microscopic scale fluctuations.
The derivation of these generalized geodesic equations establishes a complete geometric formalism for spacetime fluctuations. While their full physical interpretation is a subject of ongoing research, they provide a self-consistent, covariant framework for hypothesizing about the dynamics of the microscopic spacetime structure. The first-order equations (\ref{eq:firstgeodesic}) might govern the evolution of preferred scale coordinates in the presence of fluctuations, while the second-order equations (\ref{eq:geodesic}) could constrain the allowed profiles of the fluctuation amplitudes $a_{\alpha}$ themselves. Future work must focus on solving these equations in specific contexts and identifying potential observable signatures.

Within this scaling-based framework, the metrics $\hat{g}$, $g$, and $\tilde{g}$ form a symbiotic, hierarchical structure. The substrate metric $\hat{g}_{\alpha\beta}$ defines the elastic geometry of the scale manifold $\mathcal{M}^{(a)}$. The physical metric $g_{\alpha\beta}$ is dynamically shaped by anisotropically stretching on $\hat{g}$ through the scale factors $a_{\alpha}$. These factors $a_{\alpha}$ are, in turn, the coordinates of the amplitude manifold $\mathcal{M}^{(b)}$, whose intrinsic geometry is described by $\tilde{g}_{IJ}$. This creates a closed, co-dependent loop: the dynamics on the amplitude manifold (governed by $\tilde{g}$) drive the configuration of $a_{\alpha}$, which then deforms the substrate $\hat{g}$ into the physical metric $g$. Consequently, no single metric exists in isolation; their meanings and dynamics are mutually defined, where spacetime and its fluctuations co-evolve.

\section{Scaling quantization of the first-order fluctuating spacetime}
\subsection{The scaling Lagrangian of the first-order fluctuation}
The scaling expression for the Christoffel symbols is given by \cite{weihu} 
\[
\Gamma^{\alpha}_{\sigma\gamma}=\frac{1}{2}g^{\alpha\mu}\Bigl( a_{\sigma}\frac{\partial g_{\gamma\mu}}{\partial X^{\sigma}}+ a_{\gamma}\frac{\partial g_{\sigma\mu}}{\partial X^{\gamma}}- a_{\mu}\frac{\partial g_{\sigma\gamma}}{\partial X^{\mu}}\Bigr),
\]
and the scaling Riemann curvature tensor is
\[
R^{\rho}_{\sigma\mu\nu}= a_{\mu}\frac{\partial}{\partial X^{\mu}}\Gamma^{\rho}_{\nu\sigma}- a_{\nu}\frac{\partial}{\partial X^{\nu}}\Gamma^{\rho}_{\mu\sigma}+\Gamma^{\rho}_{\mu\lambda}\Gamma^{\lambda}_{\nu\sigma}
-\Gamma^{\rho}_{\nu\lambda}\Gamma^{\lambda}_{\mu\sigma},
\]
where $R_{\mu\nu} = R^\sigma_{\mu\sigma \nu}$ and $R=g^{\mu\nu}R_{\mu\nu}$. 
From the above scaling description, we get the scaling Einstein-Hilbert Lagrangian
\begin{equation}
\begin{aligned}
\mathcal{L}_{EH}=\frac{1}{2\kappa c}R\sqrt{-g}\frac{1}{dx_0^2},
\end{aligned}
\end{equation}
where $\kappa=\frac{8\pi G}{c^4}=\frac{8\pi l_p^2}{\hbar c}=\frac{8\pi r_{l_p}^2d\lambda^2_0}{\hbar c}$ is the Einstein gravitational constant with defining  $r_{l_p}=\frac{l_p}{d\lambda_0}$, the ratio of the Planck length to the chosen reference length $d\lambda_0$. and $R\sqrt{-g}$ is dimensionless with dimension $\frac{1}{dx_0^2}$ being stripped out. The Einstein-Hilbert action is given by
\begin{equation}
\begin{aligned} 
S_{\text{EH}} &= \int \mathcal{L}_{\text{EH}} d^4x = \frac{1}{2\kappa c} \frac{1}{dx_0^2}\int R\sqrt{-g} d^4x\\
&=\frac{dx_0^2}{2\kappa c} \int R\sqrt{-g} \frac{1}{\prod_\mu^4 a_\mu}d^4X\\
&= \frac{\hbar}{2\hat\kappa} \int R\sqrt{-g} \frac{1}{\prod_\mu^4 a_\mu}d^4X\\
&=\hbar \int \mathcal{L}_{\text{scale}} d^4X
\end{aligned}
\end{equation}
with defining dimensionless scaled Lagrangian 
\begin{equation}
\mathcal{L}_{\text{scale}}=\frac{1}{2\hat\kappa} R\sqrt{-g} \frac{1}{\prod_\mu^4 a_\mu}
\label{eq:scaledLagrangian}
\end{equation}
where $dX^\mu= a_\mu L^\mu$ and $\frac{1}{\hat\kappa}=\frac{dx_0^2}{\kappa \hbar c}=\frac{dx_0^2}{d\lambda_0^2}\frac{1}{8\pi r_{l_p}^2}=\frac{1}{8\pi r_{l_p}^2}$ with the consistent reference gauge Eq. (\ref{eq:reference}), resulting in $\hat\kappa=8\pi r_{l_p}^2$ being dimensionless. The action has the unit of $\hbar$. When generalizing to a $D$-dimensional spacetime, the dimensionless coupling transforms as $\hat{\kappa} \rightarrow \hat{\kappa}_D=8\pi r_{l_p}^{D-2}$, where the gravitational constant in $D$-dimensions is defined by $G_D=\frac{l_p^{D-2}c^3}{\hbar}$.

By applying the variational principle to this action (varying the metric \( g_{\mu\nu} \)), one can derive the vacuum scaling Einstein field equations：
\begin{equation}
\begin{aligned}
G_{\mu\nu}=R_{\mu\nu}-\frac{1}{2}g_{\mu\nu}R=0.
\end{aligned}
\label{eq:Einstein}
\end{equation}
Since  \( g_{\mu\nu} \) is a functional of \( a_\mu\), these equations equivalently impose dynamical constraints on the scale factors.
For the first-order fluctuation, from $g_{\alpha\beta}=\hat g_{\alpha\beta}\frac{ a_\alpha a_\beta}{ a^2 }$ and $g^{\alpha\beta}=\hat g^{\alpha\beta}\frac{ a^2}{ a_\alpha a_\beta}$, remind that $ a_\mu= a_\mu(X^\mu)$ and then $\partial_{\nu} a_\mu=\partial_{X^\nu} a_\mu=- b_\mu$ if $\mu=\nu$ (=0, if $\mu\neq\nu$), thus,
%\begin{widetext}
\begin{equation}
\begin{aligned} 
\Gamma^{\alpha}_{\sigma\gamma}&=\frac{1}{2}\hat g^{\alpha\mu}\frac{ a^2}{ a_\alpha a_\mu}[ a_{\sigma}\frac{\partial }{\partial X^{\sigma}}(\hat g_{\gamma\mu} \frac{ a_\gamma a_\mu}{ a^2})\\
&+ a_{\gamma}\frac{\partial}{\partial X^{\gamma}}(\hat g_{\sigma\mu} \frac{ a_\sigma a_\mu}{ a^2})- a_{\mu}\frac{\partial}{\partial X^{\mu}}(\hat g_{\sigma\gamma} \frac{ a_\sigma a_\gamma}{ a^2})]\\
&=\frac{1}{2}\hat g^{\alpha\mu}\frac{ a_{\sigma} a_\gamma}{ a_\alpha}(\frac{\partial \hat g_{\gamma\mu}}{\partial X^{\sigma}}+\frac{\partial\hat g_{\sigma\mu}}{\partial X^{\gamma}}-\frac{\partial\hat g_{\sigma\gamma}}{\partial X^{\mu}})\\
&+\frac{1}{2}\hat g^{\alpha\mu}\frac{1}{ a_\alpha a_\mu}(\hat g_{\gamma\mu} a_\sigma a_\gamma\partial_\sigma a_\mu+\hat g_{\gamma\mu} a_\sigma a_\mu\partial_\sigma a_\gamma\\
&+\hat g_{\sigma\mu} a_\gamma a_\mu\partial_\gamma a_\sigma+\hat g_{\sigma\mu} a_\gamma a_\sigma\partial_\gamma a_\mu-\hat g_{\sigma\gamma} a_{\mu} a_{\gamma}\partial_\mu a_\sigma\\&-\hat g_{\sigma\gamma} a_{\mu} a_{\sigma}\partial_\mu a_\gamma)=\frac{ a_{\sigma} a_\gamma}{ a_\alpha}\hat{\Gamma}^{\alpha}_{\sigma\gamma}-\frac{1}{2}\hat g^{\alpha\mu}\frac{1}{ a_\alpha a_\mu}\\
&\times(\hat g_{\gamma\mu} a_\mu a_\gamma b_\mu+\hat g_{\gamma\mu} a_\gamma a_\mu b_\gamma+\hat g_{\sigma\mu} a_\sigma a_\mu b_\sigma\\
&+\hat g_{\sigma\mu} a_\mu a_\sigma b_\mu-\hat g_{\mu\gamma} a_{\mu} a_{\gamma} b_\mu-\hat g_{\sigma\mu} a_{\mu} a_{\sigma} b_\mu)\\
&=\frac{ a_{\sigma} a_\gamma}{ a_\alpha}\hat{\Gamma}^{\alpha}_{\sigma\gamma}-\frac{1}{2} \hat{g}^{\alpha\mu} \frac{1}{ a_\alpha  a_\mu} (\hat{g}_{\gamma\mu}  a_\gamma  a_\mu  b_\gamma +\hat{g}_{\sigma\mu}  a_\sigma  a_\mu  b_\sigma )\\
&=\frac{ a_{\sigma} a_\gamma}{ a_\alpha}\hat{\Gamma}^{\alpha}_{\sigma\gamma}-\frac{1}{2} ( \delta^\alpha_\gamma  b_\gamma +\delta^\alpha_\sigma  b_\sigma ) \\
&=U^{\alpha}_{\ \sigma\gamma}-V^{\alpha}_{\ \sigma\gamma},
\end{aligned}
\end{equation}
%\end{widetext}
where $U^{\alpha}_{\ \sigma\gamma}=\frac{ a_{\sigma} a_\gamma}{ a_\alpha}\hat{\Gamma}^{\alpha}_{\sigma\gamma}$ and $V^{\alpha}_{\ \sigma\gamma}=\frac{1}{2} ( \delta^\alpha{}_\gamma  b_\gamma +\delta^\alpha{}_\sigma  b_\sigma )$. $V^{\alpha}_{\ \sigma\gamma}=0$ if $\alpha\neq\sigma$ and $\alpha\neq\gamma$; $V^{\alpha}_{\ \sigma\gamma}= b_\alpha$ if $\alpha=\sigma=\gamma$;  $V^{\alpha}_{\ \sigma\gamma}=0$ if $\alpha\neq\sigma$ and $\sigma=\gamma$; $V^{\alpha}_{\ \sigma\gamma}=\frac{1}{2} b_\alpha$ if $\alpha=\sigma$ and $\alpha\neq\gamma$ or $\alpha=\gamma$ and $\alpha\neq\sigma$. The total connection $\Gamma_{\sigma\gamma}^{\alpha}$ is decomposed into two distinct parts: $U_{\sigma\gamma}^{\alpha}$ and $V_{\sigma\gamma}^{\alpha}$. The first, $U_{\sigma\gamma}^{\alpha}$, is taken to be the scaling Levi-Civita connection associated with metric \(\hat g_{\mu\nu}\); it is metric-compatible and torsion-free, and thus preserves the local rotational structure while encoding the direction-dependent scaling through its dependence on \(\hat g_{\mu\nu}\). The second, $V_{\sigma\gamma}^{\alpha}$, encodes an induced anisotropic correction; its components vanish except when the upper index coincides with one of the lower indices (i.e. \(\alpha=\sigma\) or \(\alpha=\gamma\)), and the nonzero components are proportional to \(b_{\alpha}\).

A First-order Scaling Riemann Curvature Tensor (FSRCT) on the manifold $\mathcal{M}^{a}$ is introduced as
\begin{equation}
\begin{aligned} 
\hat R^{\rho}_{\sigma\mu\nu}=\frac{\partial}{\partial X^{\mu}}\hat{\Gamma}^{\rho}_{\nu\sigma}-\frac{\partial}{\partial X^{\nu}}\hat{\Gamma}^{\rho}_{\mu\sigma}+\hat{\Gamma}^{\rho}_{\mu\lambda}\hat{\Gamma}^{\lambda}_{\nu\sigma}
-\hat{\Gamma}^{\rho}_{\nu\lambda}\hat{\Gamma}^{\lambda}_{\mu\sigma},
\end{aligned}
\end{equation}
and thus, 
\begin{equation}
R^{\rho}_{\sigma\mu\nu} = \frac{ a_\mu  a_\nu  a_\sigma}{ a_\rho} \hat{R}_{\sigma\mu\nu}^{\rho} + \mathcal{\hat R}_{\sigma\mu\nu}^{\rho},
\label{eq:REHtotal}
\end{equation}
where
%\begin{widetext}
\begin{equation}
\begin{aligned} 
&\mathcal{\hat R}_{\sigma\mu\nu}^{\rho}= \underbrace{ a_\mu \hat{\Gamma}^\rho_{\nu\sigma} \partial_\mu \left( \frac{ a_\nu  a_\sigma}{ a_\rho} \right) -  a_\nu \hat{\Gamma}^\rho_{\mu\sigma} \partial_\nu \left( \frac{ a_\mu  a_\sigma}{ a_\rho} \right)}_{\text{mixed }  \partial a \times \hat{\Gamma}}\\
&- \underbrace{\left( U^\rho{}_{\mu\lambda} V^\lambda{}_{\nu\sigma} + V^\rho{}_{\mu\lambda} U^\lambda{}_{\nu\sigma} \right) - (\mu \leftrightarrow \nu)}_{\text{mixed } \hat{\Gamma} \text{-} b \text{ terms}} \\
& -\frac{1}{2}  a_{\mu} \delta^\rho_{\sigma} \dfrac{\partial  b_{\sigma}}{\partial X^{\mu}} 
- \frac{1}{2}  a_{\mu} \delta^\rho_{\nu} \dfrac{\partial  b_{\nu}}{\partial X^{\mu}} \\
&+ \frac{1}{2}  a_{\nu} \delta^\rho_{\sigma} \dfrac{\partial  b_{\sigma}}{\partial X^{\nu}} 
+ \frac{1}{2}  a_{\nu} \delta^\rho_{\mu} \dfrac{\partial  b_{\mu}}{\partial X^{\nu}}\\ 
&+ \frac{1}{4} 
    \delta^\rho_{\nu}  b_{\rho}  b_{\nu} 
    - \frac{1}{4}\delta^\rho_{\mu}  b_{\rho}  b_{\mu} 
    + \frac{1}{4}\delta^\rho_{\mu}  b_{\mu}  b_{\sigma} \\
    &- \frac{1}{4}\delta^\rho_{\nu}  b_{\nu}  b_{\sigma} 
    + \delta^\rho_{\mu} \frac{1}{4} b_{\mu}  b_{\nu} 
    - \delta^\rho_{\nu} \frac{1}{4} b_{\mu}  b_{\nu} .
\end{aligned}
\label{eq:REH}
\end{equation}
%\end{widetext}
The first term in $R^{\rho}_{\sigma\mu\nu}$ is the elastic substrate curvature stretched by the scale factors; the second term, $\mathcal{\hat R}$, contains contributions from $\partial a\times\hat\Gamma$ and mixed $ b\times\hat\Gamma$ terms, as well as terms in $ \partial b$ and $ b^{2}$, and genuinely reflects the intrinsic curvature induced by the scaling fluctuation.

Thus, a First-order Scaling Einstein Tensor (FSET) can be expressed as
\begin{equation}
\begin{aligned} 
\hat G_{\mu\nu}=\hat R_{\mu\nu}-\frac{1}{2}\hat g_{\mu\nu}\hat R,
\end{aligned}
\end{equation}
where $\hat R_{\mu\nu}=\hat R^{\rho}_{\mu\rho\nu}$ and $\hat R=\hat g^{\mu\nu}\hat R_{\mu\nu}$.
Resulting in
\begin{equation}
\begin{aligned} 
R_{\mu\nu}&= a_\mu a_\nu\hat R_{\mu\nu}+\mathcal{\hat R}_{\mu\nu};\\
R&= a^2 \hat R+a^2\mathcal{\hat R};\\
G_{\mu\nu}&= a_\mu a_\nu\hat G_{\mu\nu}+\mathcal{\hat G}_{\mu\nu};
\end{aligned}
\end{equation}
where $\mathcal{\hat R}_{\mu\nu}=\mathcal{\hat R}_{\mu\rho\nu}^{\rho}$, $\mathcal{\hat R}=\hat g^{\mu\nu}\frac{1}{ a_\mu a_\nu}\mathcal{\hat R}_{\mu\nu}$, and $\mathcal{\hat G}_{\mu\nu}=\mathcal{\hat R}_{\mu\nu}-\frac{1}{2}\hat g_{\mu\nu}\mathcal{\hat R}$.

Thus, the action is converted from $S[g_{\mu\nu}]$ to $S[ a_\mu,\hat{g}_{\mu\nu}]$, allowing for direct variation with respect to $ a_\mu$ (and $\hat{g}_{\mu\nu}$ when necessary). 
The expression of the metric shifts all physical information to the factor spectrum $a_\mu$ and combinations of $ \hat g_{\mu\nu}$. Since the action, measure, connection, curvature, and even the Einstein tensor can be algebraically expressed in terms of $ a_{\mu}$, $ b_{\mu}$ (and the Levi-Civita part of $\hat g_{\mu\nu}$), the physical field equations ultimately reduce to the equations of $ a_{\mu}$. In physical terms, the metric $g_{\mu\nu}$ is an object determined by microscopic degrees of freedom, arising under the algebra and constraints of $ a_{\mu}$.
This construction preserves manifest covariance while making the metric $g_{\mu\nu}$ a scaling-determined object. Covariance is thus realized at the level of physical, measurement-defined geometry. The variational principle is here formulated so that the Euler-Lagrange equations determine the physical metric $g_{\mu\nu}$ through $ a_{\mu}$ and $\hat g_{\mu\nu}$, which satisfy the constraints Eq. (\ref{eq:Einstein}).

To clarify the structure, it is instructive to evaluate the explicit form of the connection in a simplified setting. In particular, by adopting a scale inertial or a scale local inertial frame, one can consider the contributions arising solely from the scaling-induced terms. Hence, in the simple case where $\hat{\Gamma}^{\alpha}_{\sigma\gamma}=0$ (e.g., $\hat g=\eta$ or $\delta$ metric) for scale measurement, one finds
\begin{equation}
\begin{aligned} 
&\Gamma^{\alpha}_{\sigma\gamma} =-\frac{1}{2}  \left[ \delta^\alpha{}_\gamma  b_\gamma +\delta^\alpha{}_\sigma  b_\sigma \right]; 
\end{aligned}
\end{equation}
and then
%\begin{widetext}
\begin{equation}
\begin{aligned} 
R^{\rho}_{\sigma\mu\nu}&= -\frac{1}{2}  a_{\mu} \delta^\rho_{\sigma} \dfrac{\partial  b_{\sigma}}{\partial X^{\mu}} 
- \frac{1}{2}  a_{\mu} \delta^\rho_{\nu} \dfrac{\partial  b_{\nu}}{\partial X^{\mu}} \\
&+ \frac{1}{2}  a_{\nu} \delta^\rho_{\sigma} \dfrac{\partial  b_{\sigma}}{\partial X^{\nu}} 
+ \frac{1}{2}  a_{\nu} \delta^\rho_{\mu} \dfrac{\partial  b_{\mu}}{\partial X^{\nu}} \\
&+ \frac{1}{4}  
    \delta^\rho_{\nu}  b_{\rho}  b_{\nu} 
    - \frac{1}{4}\delta^\rho_{\mu}  b_{\rho}  b_{\mu} 
    + \frac{1}{4}\delta^\rho_{\mu}  b_{\mu}  b_{\sigma} \\
    &- \frac{1}{4}\delta^\rho_{\nu}  b_{\nu}  b_{\sigma} 
    + \delta^\rho_{\mu} \frac{1}{4} b_{\mu}  b_{\nu} 
    - \delta^\rho_{\nu} \frac{1}{4} b_{\mu}  b_{\nu} .
\end{aligned}
\end{equation}
%\end{widetext}
In this representation, the usual concept of a local inertial frame--where one chooses coordinates that eliminate the connection terms and make it flat--is extended by introducing direction-dependent scale factors. A scale local inertial frame is then defined as the coordinate system that makes the first-order scaling connection terms vanish, and the corresponding local basis explicitly includes the direction-dependent factors $ a_\mu$ so that spacetime fluctuations are woven into a flat elastic substrate. If, however, the scale factors happen to be the same constant in every direction, its effect reduces to an ordinary uniform rescaling, and the metric $\hat g$ becomes globally proportional to the one $g$. Therefore, the scale local inertial frame coincides with the classical local inertial frame of general relativity. 

For a scale-curved metric spacetime, the Lagrangian acquires additional structural complexity.  
Starting from Eq.~(\ref{eq:REHtotal},\ref{eq:REH},), one finds not only the scaled $\hat R$ and pure quadratic forms $b^2$ contributions, but also explicit derivative terms $\partial b$ and coupling terms involving the affine connection --- in particular derivative couplings of the form $(\partial a)\times\hat\Gamma$ as well as mixed $\hat\Gamma$--$ b$ interactions, where $\hat\Gamma$ appears as an interacting geometric object and is not an explicit functional of $ a_\mu$ or $ b_\mu$. The Lagrangian for a scale-curved metric spacetime and its quantization will be derived in subsequent works.

By setting $\sigma \rightarrow \mu, \mu \rightarrow \rho$  and summing over $\rho$, we get
%\begin{widetext}
\begin{equation}
\begin{aligned} 
R_{\mu\nu} &= R^\rho_{~\mu\rho\nu} = -\frac{1}{2}  a_{\rho} \delta^\rho_{\mu} \dfrac{\partial  b_{\mu}}{\partial X^{\rho}} 
- \frac{1}{2}  a_{\rho} \delta^\rho_{\nu} \dfrac{\partial  b_{\nu}}{\partial X^{\rho}} \\
&+ \frac{1}{2}  a_{\nu} \delta^\rho_{\mu} \dfrac{\partial  b_{\mu}}{\partial X^{\nu}} 
+ \frac{1}{2}  a_{\nu} \delta^\rho_{\rho} \dfrac{\partial  b_{\rho}}{\partial X^{\nu}} \\
&+ \frac{1}{4} 
    \delta^\rho_{\nu}  b_{\rho}  b_{\nu} 
    - \frac{1}{4}\delta^\rho_{\rho}  b_{\rho}  b_{\rho} 
    + \frac{1}{4}\delta^\rho_{\rho}  b_{\rho}  b_{\mu} \\
    &- \frac{1}{4}\delta^\rho_{\nu}  b_{\nu}  b_{\mu} 
    + \delta^\rho_{\rho} \frac{1}{4} b_{\rho}  b_{\nu}  
    - \delta^\rho_{\nu} \frac{1}{4} b_{\rho}  b_{\nu}  
\\
&=-\frac{1}{2}  a_{\mu} \dfrac{\partial  b_{\mu}}{\partial X^{\mu}} 
+ \frac{1}{2}  a_{\nu}  \dfrac{\partial  b_{\mu}}{\partial X^{\mu}}\delta_{\mu\nu}\\
&+ \frac{1}{4} \left( b_{\mu}\sum_{\rho}  b_{\rho}
    - \sum_{\rho}  b_{\rho}^2
    -  b_{\nu}  b_{\mu} 
    +  b_{\nu}\sum_{\rho} b_{\rho}   
\right)\\
&=-\frac{1}{2}  a_{\mu} \dfrac{\partial  b_{\mu}}{\partial X^{\mu}} 
+ \frac{1}{2}  a_{\nu}  \dfrac{\partial  b_{\mu}}{\partial X^{\mu}}\delta_{\mu\nu}\\
&+ \frac{1}{4} \left[- B_{bb} + ( b_{\mu}+ b_{\nu})B_{b}  
    -  b_{\nu}  b_{\mu}\right]
\end{aligned}
\end{equation}
%\end{widetext}
with defining  $ B_b=\sum_{\rho}  b_\rho$ and $B_{bb}=\sum_{\rho}  b_\rho^2 $.
\begin{equation}
\begin{aligned} 
R&=g^{\mu\nu}R_{\mu\nu}=\eta^{\mu\nu}\frac{ a^2}{ a_\mu a_\nu}R_{\mu\nu}\\
&=-\frac{1}{4}\sum_{\mu} \eta^{\mu\mu} \frac{ a^2}{ a_\mu^2}  \left( 
     B_{bb} 
    - 2 b_{\mu}B_{b}  
    +  b_{\mu}^2 
    \right).
\end{aligned}
\end{equation}
Thus, we have the reduced scaling vacuum Einstein tensor
%\begin{widetext}
\begin{equation}
\begin{aligned}
G_{\mu\nu}&=R_{\mu\nu}-\frac{1}{2}g_{\mu\nu}R\\
&=-\frac{1}{2}  a_{\mu} \dfrac{\partial  b_{\mu}}{\partial X^{\mu}} + \frac{1}{2}  a_{\nu}  \dfrac{\partial  b_{\mu}}{\partial X^{\mu}}\delta_{\mu\nu}\\
&+ \frac{1}{4} \left[- B_{bb} + ( b_{\mu}+ b_{\nu})B_{b} -  b_{\nu}  b_{\mu}\right]\\
&+\frac{1}{2}g_{\mu\nu}\frac{1}{4}\sum_{\mu} \eta^{\mu\mu} \frac{ a^2}{ a_\mu^2}   \left( B^{\rho\neq\mu}_{bb} -  b_{\mu}B^{\rho\neq\mu}_{b} \right)\\
&=-\frac{1}{2}  a_{\mu} \dfrac{\partial  b_{\mu}}{\partial X^{\mu}} + \frac{1}{2}  a_{\nu}  \dfrac{\partial  b_{\mu}}{\partial X^{\mu}}\delta_{\mu\nu}\\
&+ \frac{1}{4} \left[- B_{bb} + ( b_{\mu}+ b_{\nu})B_{b} -  b_{\nu}  b_{\mu}\right]\\
&+\frac{1}{8}\eta_{\mu\nu} a_\mu a_\nu\sum_{\mu} \eta^{\mu\mu} \frac{1}{ a_\mu^2}\left( B^{\rho\neq\mu}_{bb} -  b_{\mu}B^{\rho\neq\mu}_{b} \right).
\end{aligned}
\end{equation}
%\end{widetext}
In such case of  \(\hat g_{\mu\nu}=\eta_{\mu\nu}\) with \(\det\eta=-1\), the determine of $g_{\mu\nu}$ is
\begin{equation}
\begin{aligned} 
g&=det(g)=det(\eta_{\mu\nu}\frac{ a_\mu a_\nu}{ a^2})=-\frac{\prod_\mu^D a_\mu^2}{ a^{2D}}
\end{aligned} 
\end{equation}
The dimensionless scaling Lagrangian $\mathcal{L}_{\text{scale}}$ purely containing scale fluctuations becomes 
%\begin{widetext}
\begin{equation}
\begin{aligned} 
\mathcal{L}_{\text{scale}}&=\frac{1}{2\hat \kappa}R\sqrt{-g}\frac{1}{\prod_\mu^D a_\mu}=\frac{1}{2\hat \kappa}\frac{1}{\prod_\mu^D a_\mu}\frac{\Pi_\alpha^D a_\alpha}{ a^{D}}\\
&\times\sum_{\mu} \eta^{\mu\mu} \frac{ a^{2}}{ a_\mu^2} \left[ - \frac{1}{4} \left( 
     B_{bb} 
    - 2 b_{\mu}B_{b}  
    +  b_{\mu}^2 
    \right) \right]\\
    &=-\frac{1}{8\hat \kappa}\frac{1}{ a^{D-2}}\sum_{\mu} \eta^{\mu\mu} \frac{1}{ a_\mu^4}   \left( 
      a_\mu^2B_{bb} 
    - 2 a_\mu^2 b_{\mu}B_{b}  
    +  a_\mu^2 b_{\mu}^2 
    \right) \\
    &=-\mathcal{G}\sum_{\mu} \eta^{\mu\mu} \frac{1}{ a_\mu^2}   \left( 
     B^{\rho\neq\mu}_{bb} 
    - 2 b_{\mu}B^{\rho\neq\mu}_{b}   
    \right),
\end{aligned}
\label{eq:EHL}
\end{equation}
%\end{widetext}
where a dimensionless gravitational constant is defined to be
\begin{equation}
\mathcal{G}=\frac{1}{8\hat k_D} \frac{1}{ a^{D-2}}=\frac{1}{64\pi } \frac{1}{r^{D-2}_{l_p} a^{D-2}}.
\label{eq:dimensionlessG}
\end{equation}
The scaling Lagrangian \(\mathcal{L}_{\mathrm{scale}}\) encodes the kinetic terms of the anisotropic fluctuation modes of the scale factors. Here \(B_{bb}^{\rho\neq\mu}\) denotes the sum of squared magnitudes of the scale derivatives over all directions except the \(\mu\)-direction; it therefore represents the local density of fluctuation energy (i.e., the intensity of scale fluctuations) in the transverse directions. Similarly, \(B_{b}^{\rho\neq\mu}\) denotes the sum of the (first-power) magnitudes of the scale derivatives over all directions except \(\mu\), and characterizes the anisotropy of the fluctuations. The coupling term \(-\,b_{\mu}\,B_{b}^{\rho\neq\mu}\) couples the derivative amplitude in the \(\mu\)-direction to the anisotropy in the transverse directions; physically, it represents a nonlinear interaction between fluctuation components (for example, an energy-transfer or interference term between fluctuations along \(b_{\mu}\) and fluctuations in the other directions). In other words, this term quantifies how fluctuations in the \(\mu\)-direction influence, and are influenced by, anisotropic fluctuations in the orthogonal directions.

The Lagrangian presented here represents a generalized Einstein-Hilbert action formulated in an anisotropic, scale-dependent spacetime geometry. Unlike classical general relativity, where curvature is encoded in the smooth metric tensor \( g_{\mu\nu} \), this framework strengthens local fluctuation factors \(  a_\mu \) (first-order dilation) and \(  b_\mu \) (second-order non-uniformity) to encode the fluctuation effects on the curvature. These factors characterize the direction-dependent amplitude and variation of microscopic spacetime fluctuations, respectively.

Under isotropic limit, \( a_\alpha =  a_{iso} \ \forall \,\alpha\); \( b_\alpha =  b_{iso} = -\partial_\alpha  a_{iso}\) (equal in all directions). Therefore,
\[
B_b^{\rho \neq \mu} = (D-1) b_{iso}, \quad B_{bb}^{\rho \neq \mu} = (D-1) b_{iso}^2
\]
Substitute into the Lagrangian:
\begin{equation}
\begin{aligned} 
\mathcal{L}_{\text{iso}} &= -\mathcal{G}\frac{1}{ a_{iso}^2} \sum_\mu \eta^{\mu\mu} \left[ (D-1) b_{iso}^2 - 2(D-1) b_{iso}^2 \right]\\
&= \mathcal{G}\frac{1}{ a_{iso}^2} (D-1) b_{iso}^2 (\text{tr}\, \eta)
\end{aligned}
\end{equation}
For Minkowski space \(\eta^{\mu\mu} = \{-1, 1, 1, 1\}\), we have \(\text{tr}(\eta) = -1 + (D-1) = D-2\). The final result is:
\begin{equation}
\begin{aligned} 
\mathcal{L}_{\text{iso}} &= \mathcal{G} (D-1)(D-2)\frac{b_{iso}^2}{ a_{iso}^2} \\
&= \mathcal{G}(D-1)(D-2) (d\ln a_{iso}/dX^\mu)^2.
\end{aligned}
\end{equation}
In the isotropic limit, where every directional scale factor is equal and the directional derivatives coincide, the full scaling Lagrangian reduces to a single effective kinetic term for the overall scale field. Substituting the isotropic values shows that the Lagrangian is proportional to the square of the gradient of the logarithmic scale factor. The overall coefficient of this term is proportional to the dimensionless coupling \(\mathcal{G}\) and carries a purely geometric dimension-dependent factor \((D-1)(D-2)\), the appearance of \(D-2\) arising from the trace of the Minkowski metric. 

\subsection{Linear representation and diagonalization}
We rewrite Eq. (~\ref{eq:EHL}) to make the matrix/quadratic structure more explicit:
\begin{equation}
\begin{aligned} 
\mathcal{L}_{\text{scale}}&= \mathcal{G}[ (D-1)\sum_\mu \eta^{\mu\mu} \frac{1}{ a_\mu^2}  b_\mu^2\\
&-\sum_{\mu<\nu}(\eta^{\mu\mu} \frac{1}{ a_\mu^2}+\eta^{\nu\nu} \frac{1}{ a_\nu^2})\,( b_\mu- b_\nu)^2 \; ]\\
&=\mathcal{G}\{ (D-1)\sum_\mu \eta^{\mu\mu} \frac{1}{ a_\mu^4} v_\mu^2\\
&+\sum_{\mu<\nu}(\eta^{\mu\mu} \frac{1}{ a_\mu^2}+\eta^{\nu\nu} \frac{1}{ a_\nu^2})\,(\frac{1}{ a_\mu}v_\mu-\frac{1}{ a_\nu}v_\nu)^2 \; \}
\end{aligned}
\label{eq:Lab}
\end{equation}
where 
\[
v_\mu=- a_\mu b_\mu= a_\mu\frac{d a_\mu}{dX^\mu}.
\]
This is a quadratic form involving diagonal and adjacent differences, suitable for expression in matrix form:
\begin{equation}
\mathcal{L}_{\text{scale}} = \mathcal{G}\, \mathbf{v}^T M \mathbf{v},
\end{equation}
where 
\[
\mathbf{v}= (v_0, v_1, \ldots, v_{D-1})^T,
\]
and the elements of the matrix \(M\) are given by:
\begin{equation}
\begin{aligned} 
M_{\mu \mu} &= (D - 1) \eta^{\mu\mu} \frac{1}{ a_\mu^4} + \sum_{\nu \neq \mu} (\eta^{\mu\mu} \frac{1}{ a_\mu^4} + \eta^{\nu\nu} \frac{1}{ a^2_\mu a_\nu^2}),\\ 
M_{\mu \nu} &= -(\eta^{\mu\mu} \frac{1}{ a_\mu^3a_\nu} + \eta^{\nu\nu} \frac{1}{ a_\mu a_\nu^3}).
\end{aligned}
\end{equation}
Since matrix $M$ is real symmetric, it admits an orthogonal diagonalization with real eigenvalues, thereby enabling a spectral decomposition. Such a decomposition is crucial for the subsequent canonical quantization, as it allows the quadratic form in the Lagrangian to be expressed in terms of independent modes, which are identified as the physical degrees of freedom. A necessary and sufficient condition for $M$ to be invertible is that all its eigenvalues are non-vanishing, i.e., $\det(M) \neq 0$. However, zero eigenvalues may appear if certain parameters $ a_\mu$ approach $\infty$, or if they satisfy special algebraic relations that impose constraints. Under the Minkowski convention $(\eta_{00} = -1,\ \eta_{ij} = +1)$, the time component introduces a natural negative sign in the expressions, so $M$ is not necessarily positive definite. Whether it is positive definite depends on the relative magnitudes of the $ a_\mu$: if the positive contributions from the spatial components dominate over the negative temporal contribution, eigenvalues may become positive.

We compute the conjugate momentum of $ a_\mu$,
\begin{equation}
\begin{aligned} 
\pi_\mu = \frac{\partial\mathcal{L}_{\text{scale}}}{\partial v_\mu} = 2 \mathcal{G} (M v)_{\mu} = 2 \mathcal{G} \sum_{\nu} M_{\mu \nu} v_{\nu}. 
\end{aligned}
\label{eq:piab}
\end{equation}
With the definition of $\boldsymbol{\pi} = (\pi_{0}, \pi_{1}, \dots, \pi_{D-1})^{T}$ and from
\[
\sum_{\mu} \pi_{\mu} v_{\mu} = \boldsymbol{\pi}^T \mathbf{v} = (2\mathcal{G} M \mathbf{v})^T \mathbf{v} = 2\mathcal{G} \mathbf{v}^T M \mathbf{v},
\]
the Hamiltonian density is given by
\begin{equation}
\begin{aligned} 
\mathcal{H}_{ab} = \sum_\mu \pi_\mu v_\mu - \mathcal{L}_{\text{scale}} = \mathcal{G} \mathbf{v}^T M \mathbf{v}.
\end{aligned}
\label{eq:Hab}
\end{equation}
If to express the Hamiltonian in terms of \(\boldsymbol{\pi}\) (instead of \(\mathbf{v}\)), one can solve for  
\[ \mathbf{v} = \frac{1}{2\mathcal{G}} M^{-1} \boldsymbol{\pi}= \frac{1}{2\mathcal{G}} M^{-1} \boldsymbol{\pi}, \]
and substitute to obtain (as long as \(M\) is invertible)
\begin{equation}
\mathcal{H}_{ab} = \frac{1}{4\mathcal{G}} \boldsymbol{\pi}^{T} M^{-1} \boldsymbol{\pi}.
\end{equation}
From
\[
\frac{\partial\mathcal{L}_{\text{scale}}}{\partial a_{\mu}}
=\mathcal{G}\Big(\mathbf v^{T}\frac{\partial M}{\partial a_{\mu}}\mathbf v
+2\,(M\mathbf v)_{\mu}\,\partial_{\mu} a_{\mu}\Big),
\]
and
\[
 a_{\mu}\frac{\partial}{\partial X^{\mu}}\!\Big(\frac{\partial\mathcal{L}_{\text{scale}}}{\partial v_{\mu}}\Big)
=2\mathcal{G}\, a_{\mu}\,\partial_{\mu}\big[(M\mathbf v)_{\mu}\big].
\]
Substituting these into the variation form
\[
0=\frac{\partial\mathcal{L}_{\text{scale}}}{\partial a_{\mu}}
- a_{\mu}\frac{\partial}{\partial X^{\mu}}\!\Big(\frac{\partial\mathcal{L}_{\text{scale}}}{\partial v_{\mu}}\Big)
\]
and cancelling the common factor \(\mathcal{G}\) yields the direct component the Euler-Lagrange equations (for each component label \(\mu\)):
\begin{equation}
0 \;=\; \mathbf v^{T}\frac{\partial M}{\partial a_{\mu}}\mathbf v
\;+\;2\,(M\mathbf v)_{\mu}\,\partial_{\mu} a_{\mu}
\;-\;2\, a_{\mu}\,\partial_{\mu}\big[(M\mathbf v)_{\mu}\big].
\end{equation}

Assume isotropic \( a_\mu= a_{iso}\) for all \(\mu\). The matrix elements simplify and depend only on the metric signs \(\eta^{\mu\mu}\). Define the trace of the diagonal metric signs \(T_\eta=\sum_\mu\eta^{\mu\mu}\). One finds
\[
M_{\mu\mu}=\frac{(2D-3)\eta^{\mu\mu}+T_\eta}{ a_{iso}^4},\,
M_{\mu\nu}=-\frac{\eta^{\mu\mu}+\eta^{\nu\nu}}{ a_{iso}^4}\,(\mu\neq\nu).
\]
In the isotropic limit, the matrix $M$ simplifies significantly. Its spectrum can be obtained in closed form for both Euclidean and Minkowski spacetimes, as detailed below. The Lorentzian signature naturally leads to a decoupling between the temporal and spatial sectors.

For Euclidean signature (\(\eta^{\mu\mu}=+1\) for all \(\mu\)) \(T_\eta=D\) and
\[
M_{\mu\mu}=\frac{3(D-1)}{ a_{iso}^4},\quad M_{\mu\nu}=-\frac{2}{ a_{iso}^4}\,(\mu\ne\nu).
\]
Such a constant-diagonal / constant-off-diagonal matrix has one eigenvector proportional to the all-ones vector \(\mathbf{1}=(1,\dots,1)^T\) and \(D-1\) eigenvectors orthogonal to \(\mathbf{1}\). The corresponding eigenvalues are
\begin{equation}\label{eq:euclid-eigs}
\begin{aligned}
\lambda_{\parallel}&=\frac{D-1}{ a_{iso}^4}\quad(\text{multiplicity }1),\\
\lambda_{\perp}&=\frac{3D-1}{ a_{iso}^4}\quad(\text{multiplicity }D-1)\;.
\end{aligned}
\end{equation}
An elementary verification is given by applying \(M\) to \(\mathbf{1}\) and to any vector orthogonal to \(\mathbf{1}\).

For the conventional Minkowski choice (\(\eta={\rm diag}(-1,+1,\dots,+1)\), time index \(\mu=0\)), one has \(T_\eta=-1+(D-1)=D-2\). Substituting into the isotropic expressions yields a decoupling between the time direction and the spatial \((D-1)\)-block:
\[
M_{00}=-\frac{D-1}{ a_{iso}^4},\quad M_{0i}=0\; (i\ge1),
\]
\[
M_{ii}=\frac{3D-5}{ a_{iso}^4},\quad M_{ij}=-\frac{2}{ a_{iso}^4}\; (i\ne j,\; i,j\ge1).
\]
Hence, the spectrum splits into a time eigenvalue and the spectrum of the spatial \((D-1)\times(D-1)\) constant-type matrix. One obtains
\begin{equation}\label{eq:mink-eigs}
\begin{aligned}
\lambda_{\rm time}&=-\frac{D-1}{ a_{iso}^4}\;(\text{multiplicity }1),\\
\lambda_{\rm space}^{(\parallel)}&=\frac{D-1}{ a_{iso}^4}\;(\text{multiplicity }1),\\
\lambda_{\rm space}^{(\perp)}&=\frac{3(D-1)}{ a_{iso}^4}\;(\text{multiplicity }D-2)\;.
\end{aligned}
\end{equation}
The presence of a negative eigenvalue $\lambda_{\text{time}}<0$ in the Minkowski case indicates a potential ghost mode for the isotropic situation. For the theory to be physically viable, this mode must be eliminated. We will restrict the physical Hilbert space $I_{\text{phys}}$ to exclude such negative-norm states.

If considering the anisotropic specialization of the general matrix \(M\) in which the metric is Minkowskian, the time-scale is distinct from the spatial scale, and the spatial directions remain isotropic. Concretely we set
\[
 a_0= a_t,\quad  a_i= a_s\quad(i=1,\dots,D-1),
\quad \epsilon=\frac{ a_t^2}{ a_s^2}>0,
\]
and we seek necessary and sufficient conditions on \(\epsilon\) (and \(D\)) so that \(M\) is positive definite (all eigenvalues \(>0\)).  We keep the dependence on \( a_s\) explicit where convenient.

With the above specialization, the independent components of \(M\) reduce to a time–time entry, a single spatial diagonal entry (same for each spatial direction), a common time–space coupling, and a uniform spatial off-diagonal entry. After straightforward substitution and algebra one obtains
\begin{align}
M_{00} &= (D-1)\!\left(-\frac{2}{ a_t^{4}}+\frac{1}{ a_t^{2} a_s^{2}}\right),
\\[6pt]
M_{ii} &= \frac{3D-4}{ a_s^{4}}-\frac{1}{ a_t^{2} a_s^{2}}\quad(i=1,\dots,D-1),
\\[6pt]
M_{0i} &= \frac{1/ a_t^{2}-1/ a_s^{2}}{ a_t a_s},
\\[6pt]
M_{ij} &= -\frac{2}{ a_s^{4}}\quad(i\neq j).
\end{align}
Because the spatial submatrix has the constant-diagonal, constant-off-diagonal form, the spatial sector decomposes into the one-dimensional ``all-ones'' direction and its \((D-2)\)-dimensional orthogonal complement. This yields an explicit and simple spectral decomposition.

For transverse spatial modes (multiplicity \(D-2\)) and any vector with spatial components summing to zero (and vanishing time component) one finds the repeated eigenvalue
\begin{equation}\label{eq:lambda_perp}
\lambda_{\perp} \;=\; \frac{3D-2}{ a_s^{4}}-\frac{1}{ a_t^{2} a_s^{2}}
=\frac{1}{ a_s^{4}}\!\left(3D-2-\frac{1}{\epsilon}\right).
\end{equation}
 
The remaining two eigenvalues are those of the \(2\times2\) matrix acting on the subspace spanned by the time basis vector and the spatial ``all-ones'' vector. Extracting the common factor \((D-1)/ a_s^{4}\) yields the dimensionless \(2\times2\) matrix
\begin{equation}\label{eq:S-def}
S=\begin{pmatrix}
-\dfrac{2}{\epsilon^{2}}+\dfrac{1}{\epsilon} & \dfrac{1/\epsilon-1}{\epsilon^{1/2}}\\[8pt]
\dfrac{1/\epsilon-1}{\epsilon^{1/2}} & D-\dfrac{1}{\epsilon}
\end{pmatrix},
\end{equation}
so that the two eigenvalues in the original normalization are
\begin{equation}\label{eq:lambda_pm}
\lambda_{\pm}= \frac{D-1}{ a_s^{4}}\left[\frac{S_{11}+S_{22}}{2}\pm
\sqrt{\Big(\frac{S_{11}-S_{22}}{2}\Big)^2+S_{12}^2}\,\right].
\end{equation}

Since \(M\) is symmetric, \(M\) is positive definite iff all eigenvalues \(\lambda_{\perp}\) and \(\lambda_{\pm}\) are positive. Because \(\lambda_{\pm}\) are the eigenvalues of the symmetric \(2\times2\) matrix \((D-1)S/ a_s^{4}\), positivity of \(\lambda_{\pm}\) is equivalent to positivity of \(S\). Thus the complete positivity condition is equivalent to
\[
\lambda_{\perp}>0\quad\text{and}\quad S\succ 0\,\, \text{(positive definite)}.
\]

For the \(2\times2\) matrix \(S\) a convenient and standard criterion is
\[
S_{11}>0\quad\text{and}\quad \det S>0.
\]
From Eq. \eqref{eq:S-def} these give the inequalities
\begin{align}
S_{11}>0 &\iff -\frac{2}{\epsilon^{2}}+\frac{1}{\epsilon}>0
\Longleftrightarrow \epsilon>2,\label{eq:S11-cond}\\[6pt]
\det S>0 &\iff P(\epsilon)= (D-1)\epsilon^{2}+(1-2D)\epsilon+1>0.\label{eq:detS-cond}
\end{align}
The transverse eigenvalue condition \(\lambda_{\perp}>0\) from \eqref{eq:lambda_perp} reads
\begin{equation}\label{eq:perp-cond}
3D-2-\frac{1}{\epsilon}>0\,\Longleftrightarrow\, \epsilon>\frac{1}{3D-2},
\end{equation}
which is automatically satisfied for any \(\epsilon\) fulfilling Eq. \eqref{eq:S11-cond} and Eq. \eqref{eq:detS-cond} for \(D\ge2\).

Because \(P(2)=-1<0\) and \(P(\epsilon)\to+\infty\) as \(\epsilon\to\infty\), the quadratic \(P(\epsilon)\) has two real roots and positivity of \(P\) for large \(\epsilon\) defines a threshold. Denote by \(\epsilon_{*}\) the larger positive root of \(P(\epsilon)=0\). Solving the quadratic yields the explicit threshold
\begin{equation}\label{eq:r-star}
\epsilon_{*} \;=\; \frac{2D-1+\sqrt{4D^{2}-8D+5}}{2(D-1)}.
\end{equation}
When $D=2$, $\epsilon_*=\frac{3+\sqrt{5}}{2}=g_1^2$ with $g_1=\frac{1+\sqrt{5}}{2}\simeq1.618034$ being the golden section.
One checks that \(\epsilon_{*}>2\) for all \(D\ge2\); Hence, the single compact necessary and sufficient condition for positive definiteness is
\begin{equation}\label{eq:final-cond}
\;\epsilon=\frac{ a_t^2}{ a_s^2} > \epsilon_{*}.
\end{equation}
Under this condition the two eigenvalues \(\lambda_{\pm}\) are strictly positive and \(\lambda_{\perp}>0\) as well.
For four spacetime dimensions (\(D=4\)) the threshold evaluates to
$\epsilon_{*}=\frac{7+\sqrt{37}}{6}\approx 2.18046$, so positivity requires \( a_t^{2}>\,2.18046\, a_s^{2}\). An interesting and non-trivial consistency check is that Eq. (\ref{eq:final-cond}) $(|\frac{a_t}{a_s}|\ge g_1;\, D\ge2)$ aligns with the micro-length measuring ratio
$\frac{L^0}{L^1}=\frac{r_{l^0}}{r_{l^1}}=\gamma_{ltz}^2=\frac{1}{2}(1+\sqrt{1+4[\varsigma^{1}]^{2}})\ge g_1$ with $\varsigma^{1}\ge1$ \cite{weihu}. Since $\gamma_{ltz}^2>1$ ensures compliance with the speed-of-light limit, the microscopic scale factors $a_\mu$ associated with $L^\mu$ meet the same physical constraint. Consequently, the requirement $\epsilon_*>2$, which ensures that $M$ is positive definite, is consistent with avoiding obvious violations of the speed-of-light limit and therefore with causality.

Requiring the matrix $M$ to be positive definite ensures that the quadratic form entering the action remains bounded from below. In physical terms, this guarantees that all fluctuation modes contribute with non–negative energy, thereby excluding unphysical instabilities or ghostlike excitations. The spectral conditions derived above show that only in this regime can the fluctuation basis be consistently diagonalized and quantized, leading to a well-defined Hilbert space and vacuum structure. Hence, the positivity of $M$ provides the necessary criterion for the stability and consistency of the spacetime microstructure quantization framework.

The eigenvalues of the fluctuation matrix $M$ scale as inverse fourth powers of the scale factors, schematically $\lambda \propto C/ a_{iso}^4$ (or $1/( a^2_t a^2_s)$ with $C$ a geometry- and signature-dependent prefactor). Hence the asymptotic behavior of any eigenvalue is governed entirely by the limit of $ a_\mu(X^\mu)$ (see Eq. (\ref{eq:hata}), TABLE II, and Fig. \ref{fig5}): if $ a_\mu\to 0$ (as at the UV fixed points A/B) then $\lambda\to +\infty$; if $ a_\mu$ approaches a finite nonzero constant (e.g.\ $ a_\mu\to 1/2$ at a regular IR endpoint) then $\lambda$ remains finite and $\mathcal{O}(1)$; and if $ a\to\infty$ (in certain IR flow branches, C/D) then $\lambda\to 0$.

\subsection{Mode decomposition and canonical quantization}
The canonical quantization of general relativity faces well-known challenges, particularly due to the constraint structure inherent in metric-based formulations. In Einstein's theory, the spacetime geometry is encoded in the metric tensor $g_{\mu\nu}$, and its diffeomorphism invariance gives rise to four first-class constraints—namely, the Hamiltonian constraint ($\mathcal{H} \approx 0$) and three momentum constraints ($\mathcal{H}_i \approx 0$), which restrict the physical phase space evolution \cite{Dirac1964, rovelli2004}. These constraints introduce notable complications in the quantization process:(1) Mathematical structure: The constraints form a non-trivial algebra, requiring Dirac's quantization procedure. Physical states must satisfy $\hat{\mathcal{H}}|\Psi\rangle = 0$ (the Wheeler--DeWitt equation) \cite{DeWitt1967}, though obtaining explicit solutions remains challenging due to operator-ordering ambiguities and non-renormalizability \cite{tHooft1974}. (2) Problem of time: The Hamiltonian constraint $\mathcal{H} \approx 0$ suggests a lack of explicit time evolution at the quantum level, raising questions about the emergence of classical time \cite{Kuchar2011}. (3) Issue with observables: Physical observables must commute with all constraints, but constructing local, gauge-invariant operators from the metric $g_{\mu\nu}$ is conceptually and technically difficult \cite{Mottola1995, Marolf1999}. These challenges have motivated the development of alternative approaches to quantum gravity, including string theory \cite{Polchinski1998}, loop quantum gravity \cite{rovelli2004,Ashtekar2004}, and asymptotically safe gravity \cite{Addazi2022,Niedermaier2006}, among others.

This persistent difficulty motivates alternative strategies in which the quantum geometry may emerge from simpler degrees of freedom rather than direct metric quantization. In current canonical quantization, by following Eqs.~(\ref{eq:Lab}, \ref{eq:piab}, \ref{eq:Hab}), we promote the scale fluctuation $a_{\mu}$ and its conjugate momentum $\pi_{\mu}$ to operators. In this approach, we do not attempt a direct canonical quantization of the metric $g_{\mu\nu}$; instead, we quantize the scale degrees of freedom $a_{\mu}$. It is noteworthy that the following quantization procedure is applicable to scaling measurement spacetimes characterized by both scale-flat and scale-curved metrics, while the Lagrangian in the case of a scale-curved metric spacetime is more complex. In the following, we will focus on the scale-flat metric $\hat g=\eta$.

In the following of this section we present a self-contained derivation that (i) diagonalizes the quadratic kinetic form introduced in the main text, (ii) identifies and treats the constraints that arise when the kinetic operator is degenerate, and (iii) promotes the scale factors $a_\mu$ into operators $\hat a_\mu(X^\mu)$, expresses them and their conjugates $\hat\pi_\mu(X^\mu)$ in terms of bosonic creation/annihilation operators. The presentation supplies the normalization conventions that guarantee canonical commutation relations, and indicates how to treat zero- and negative-eigenvalue (ghost) modes. 

Since \(M\) is real symmetric, there exists an orthogonal matrix \(S\) such that
\[
S^{\!T} M S=\operatorname{diag}(\lambda_{0},\lambda_{1},\dots,\lambda_{D-1}),\quad S^{\!T}S=I,
\]
where $S=\bigl(s^{(0)},s^{(1)},\dots,s^{(D-1)}\bigr)$ is the orthogonal matrix whose columns \( s^{(i)} \) are the (normalized) eigenvectors of \( M \).
Define the modal coordinates $\mathbf{ q}$ by \(\mathbf{a}=S \mathbf{ q}\) and the corresponding modal momenta $\mathbf{p}$ by \(\boldsymbol{\pi}=S\mathbf{p}\), where
\[
\mathbf{ a} = ( a_{0},  a_{1}, \dots,  a_{D-1})^{T}.
\]
From \(\boldsymbol{\pi}=2\mathcal{G} M \mathbf{v}\) we obtain 
\begin{equation}
\mathcal{H}=\frac{1}{4\mathcal{G}}\sum_{i=0}^{D-1}\frac{p_i^{2}}{\lambda_i}.
\end{equation}
This is a clear form: the ``coefficient'' for each mode \(i\) is determined by \(\lambda_i\). 

In the spectral basis, we introduce a set of canonical modal coordinates and their conjugate momenta $(\hat q_i,\hat p_i)$ \cite{FetterWalecka2003} (these are standard independent degrees of freedom that satisfy the canonical commutation relations)
\[
[\hat q_i, \hat p_j] = ir_\hbar \delta_{ij}, \quad [\hat q_i, \hat q_j] = [\hat p_i, \hat p_j] = 0.
\]
Linear transformation from modal variables to the original ones
\begin{equation}
\pi_{\mu}=\sum_{i}s_{\mu}^{(i)} p_{i},\qquad
 a_{\mu}=\sum_{i}s_{\mu}^{(i)} q_{i}.
\end{equation}
Note: $\mathbf{s}^{(i)}$ denotes the $i$-th column of $S$; its $\mu$-th component is $S_{\mu i}$ (equivalently written $s^{(i)}_\mu$).
For the $i$-th mode, introduce positive modal mass parameters $m_i > 0$ and a reference frequency (or scale) $\omega_i > 0$--physically, these can be determined by a complete Hamiltonian with kinetic term and a potential term. If a physical quadratic potential is present, it determines the physical modal frequency.
In the absence of an explicit potential, one may temporarily introduce a positive reference frequency as a mathematical regulator, but its numerical value is not physical until linked to a specific stabilizing kernel.

The algebraic relation  \(\boldsymbol{\pi}=2\mathcal{G} M \mathbf{v}\) implies in modal form
\begin{equation}
p_i=m_i\,v_i,\, (i\in I_{\mathrm{phys}});
\quad
p_i=0,\, (i\in I_{\mathrm{null}}),
\end{equation}
where, partition the spectrum into the physical, nonzero eigenvalues
\[
I_{\mathrm{phys}}=\{i\mid \lambda_i>0\}
\]
and the \emph{null} (zero-eigenvalue) set $I_{\mathrm{null}}$.  The null modes generate algebraic constraints that are treated in the next subsection.
To diagonalize the kinetic term and cast the modal Hamiltonian into standard form, we define the effective modal mass $m_i$ for each mode via:
\begin{equation}
m_i = 2\mathcal{G}\lambda_i,
\end{equation}
obtained from the kinetic kernel eigenvalue.
This quantity characterizes the inertia or resistance to change of the spacetime fluctuation mode labeled by $i$.

Introduce the canonical creation/annihilation operators $\hat{d}_i^\dagger$ and $\hat{d}_i$  \cite{Peskin}:
\begin{equation}
\hat{d}_i = \sqrt{\frac{m_i\omega_i}{2r_\hbar}} \hat{q}_i + i \frac{1}{\sqrt{2r_\hbar m_i\omega_i}} \hat{p}_i,
\end{equation}
with $\hat p_i=\partial/\partial \hat q_i$ and
\[
[\hat q_i,\hat p_i]=\delta_{ij}; \quad
[\hat{d}_i, \hat{d}_j^\dagger] = \delta_{ij}, \quad [\hat{d}_i, \hat{d}_j] = 0.
\]
Solve for $\hat{q}_i, \hat{p}_i$ in reverse (to substitute back into the original operators)
\begin{equation}
\begin{aligned}
\hat{q}_i &= \sqrt{\frac{r_\hbar}{2m_i\omega_i}} (\hat{d}_i + \hat{d}_i^\dagger), \\
\hat{p}_i &= -i\sqrt{\frac{r_\hbar m_i\omega_i}{2}} (\hat{d}_i - \hat{d}_i^\dagger).
\end{aligned}
\end{equation}
Finally, express the original operators as sums of creation/annihilation operators
\begin{equation}
\begin{aligned}
\hat{a}_{\mu} &= \sum_i s_{\mu}^{(i)} \sqrt{\frac{r_\hbar}{2m_i\omega_i}} (\hat{d}_i + \hat{d}_i^\dagger),\\
\hat{\pi}^{\mu} &= \sum_i s_{\mu}^{(i)} \left[ -i\sqrt{\frac{r_\hbar m_i\omega_i}{2}} (\hat{d}_i - \hat{d}_i^\dagger) \right].
\end{aligned}
\end{equation}
This is the standard expansion for discrete modes: each original variable is a linear combination of creation/annihilation operators for each mode, with coefficients determined by the mode functions $s_{\mu}^{(i)}$ and the normalization scales $m_i, \omega_i$.

The introduction of \( m_i \) and \( \omega_i \) serves distinct mathematical and physical purposes. The parameter \( m_i \), or modal mass, is defined to transform the original Hamiltonian into the standard canonical form \( p_i^2 / (2m_i) + \cdots \) for each mode. Physically, this represents an effective mass scale for the mode in the kinetic term, where \( 2G \) is the Lagrangian/Hamiltonian coupling constant and \( \lambda_i \) is the eigenvalue of the matrix \( M \).
The parameter \( \omega_i \), or reference frequency, should be naturally determined by the potential energy coefficient and \( m_i \) (e.g., \( \omega_i^2 = k_i / m_i \)) when a potential energy term is present, allowing the Hamiltonian to be diagonalized via the operators \( \hat d_i \) into the form
$\omega_i (\hat{d}_i^\dagger \hat{d}_i + 1/2)$. In the absence of an explicit potential, \( \omega_i \) functions merely as a reference scale; while different choices alter the zero-point energy and operator decomposition weighting, they do not affect the fundamental commutation structure of the modes, with the value that diagonalizes the Hamiltonian being the preferred choice.

Starting from the quadratic Lagrangian density given in the text and adding appropriate orthonormalization conditions \cite{BirrellDavies,Peskin}. The canonical momenta are related by the algebraic relation for a continuous situation, namely
\begin{equation}
\pi_\mu(X) \;=\; 2\mathcal{G} \int dY\; M_{\mu\nu}(X,Y)\,v_\nu(Y).
\end{equation}
Eliminating $v$ yields the reduced Hamiltonian functional (valid if $M^{-1}$ exists):
\begin{equation}
\mathcal{H} \;=\; \frac{1}{4\mathcal{G}}\!\int\!dX\,dY\; \pi_\mu(X)\,[M^{-1}]_{\mu\nu}(X,Y)\,\pi_\nu(Y)\;.
\end{equation}
$M$ is a symmetric kernel on the index $\mu$ and the $X$-space and solve the spectral problem
\begin{equation}\label{eq:M-spectrum}
\int dY\; M_{\mu\nu}(X,Y)\,\Phi_{\nu;i}(Y) \;=\; \lambda_i\,\Phi_{\mu;i}(X).
\end{equation}
whose functional-analytic foundations (spectral expansions of symmetric kernels) are standard \cite{ReedSimon}. Here $i$ enumerates orthogonal mode families (e.g., polarizations), and the eigenfunctions are orthonormal
\begin{equation}
\sum_\mu\int dX\;\Phi^*_{\mu;i}(X)\Phi_{\mu;j}(X)=\delta_{ij}
\end{equation}
and complete
\begin{equation}
\sum_{i\in I} \Phi_{\mu;i}^{*}(X)\,\Phi_{\nu;i}(Y)
= \delta_{\mu\nu}\delta(X-Y).
\end{equation}
The spectral expansion of $M$ is then
\begin{equation}
M_{\mu\nu}(X,Y)=\sum_i \lambda_i\,\Phi_{\mu;i}^{*}(X)\,\Phi_{\nu;i}(Y),
\end{equation}
\begin{equation}
[M^{-1}]_{\mu\nu}(X,Y)=\sum_{i\in I_{\rm phys}}\frac{1}{\lambda_i}\,\Phi_{\mu;i}(X)\Phi^*_{\nu;i}(Y).
\end{equation}

The $M$-matrix serves as the fundamental kinetic kernel that encodes the anisotropic coupling between spacetime fluctuation modes along different coordinate directions.
Physically, the eigenvalues $\lambda_i$ represent the effective inertia or modal mass of each fluctuation mode, quantifying how strongly quantum fluctuations resist amplitude changes. The eigenfunctions $\Phi_{\mu;i}(X)$ describe the spatial profile of these fluctuation modes within the scale manifold $\mathcal{M}^{(a)}$, mapping how microscopic spacetime deformations propagate through different scale coordinates.
If $\lambda_i<0$, then $m_i<0$ and the corresponding mode is a ghost (instability) that must be removed or treated specially.

The integral eigenproblem uses position-dependent eigenfunctions $\Phi_{\mu;i}(X)$ to diagonalize the symmetric kernel $M_{\mu\nu}(X,Y)$ in function space, with orthonormality defined by integrals and completeness as a resolution of the identity. The matrix form treats $M$ as a finite real symmetric matrix and diagonalizes it with an orthogonal matrix $S$ whose columns are discrete eigenvectors $s^{(i)}$. These notations are equivalent when the composite index $(\mu,X)$ is flattened or the continuous domain is discretized; care must be taken with quadrature weights, normalization conventions, and the treatment of degenerate or zero modes. They yield identical modal amplitudes $q_i,p_i$ and modal masses when mapped consistently.

Define modal scale coordinates and conjugate momentum as
\begin{equation}
\begin{aligned}
q_i&=\sum_\mu\int dX\;\Phi^*_{\mu;i}(X)\,a_\mu(X),\\
p_i&=\sum_\mu\int dX\;\Phi^*_{\mu;i}(X)\,\pi_\mu(X),
\end{aligned}
\end{equation}
Project the canonical scale variables onto the spectral basis by the modal projections
\begin{equation}
\begin{aligned}
a_\mu(X)=\sum_{i\in I_{phys}} \Phi_{\mu;i}(X)\, q_i,\\
\pi_\mu(X)=\sum_{i\in I_{phys}} \Phi_{\mu;i}(X)\,p_i.
\end{aligned}
\end{equation}

Each null mode $i$ produces a primary constraint
\begin{equation}
\phi_\alpha= p_\alpha\approx 0.
\end{equation}
Typically, these constraints are first-class (they generate shifts of $q_\alpha$).  To quantize one may either (a) fix a gauge (for example, $q_\alpha=0$) and construct Dirac brackets, or (b) perform a direct reduction by eliminating the null-mode sector entirely.  Both approaches lead to a reduced canonical algebra on the physical subspace; the theory of constrained Hamiltonian systems and Dirac brackets is standard \cite{Henneaux}:
\begin{equation}
\{\hat q_i,\hat p_j\}_D = \delta_{ij}\,;\; i,j\in I_{\mathrm{phys}},
\end{equation}
where $\{\cdot,\cdot\}_D$ denotes the Dirac bracket.  Quantization promotes these Dirac brackets to commutators, $\{\cdot,\cdot\}_D\mapsto(1/ir_\hbar)[\cdot,\cdot]$. In generic cases, the primary constraints Poisson-commute among themselves and hence are first-class; however, this conclusion assumes no further structure functions arise from the full constraint algebra and must be checked in each concrete model.

Restricting sums to the physical spectrum $I_{\mathrm{phys}}$, the reconstruction formulas that express the original operators in terms of creation/annihilation operators read
\begin{equation}\label{eq:afield}
\begin{aligned} 
&\hat a_\mu(X)
= \sum_{i\in I_{\mathrm{phys}}}
\sqrt{\frac{r_\hbar}{2\,m_i\,\omega_i}}\;\\
&\quad\qquad\times\big[\,\Phi_{\mu;i}(X)\,\hat d_{i} + \Phi_{\mu;i}^*(X)\,\hat d_{i}^\dagger\big],\\
&\hat\pi_\mu(X)
= \sum_{i\in I_{\mathrm{phys}}}
\bigg(-\,i\sqrt{\frac{r_\hbar\,m_i\,\omega_i}{2}}\bigg)\\
&\quad\qquad\times\big[\,\Phi_{\mu;i}(X)\,\hat d_{i} - \Phi_{\mu;i}^{*}(X)\,\hat d_{i}^\dagger\big].
\end{aligned} 
\end{equation}
One may verify by direct substitution, using orthonormality of $\Phi$ and the commutators of $\hat b$, that the canonical commutation relations are obtained as
\begin{equation}
\big[\hat a_\mu(X),\ \hat\pi_\nu(Y)\big] \;=\; ir_\hbar\,\delta_{\mu\nu}\,\delta^{(d)}(X-Y),
\label{eq:commutation}
\end{equation}
and all other commutators vanish.

For Zero modes ($\lambda_i=0$): these are constrained directions. After gauge-fixing or reduction, they are absent from the physical expansions (\ref{eq:afield}). If one does not eliminate them explicitly, physical states must satisfy the Dirac constraint operators $\hat p_i|{\Psi_{\mathrm{phys}}}\rangle=0$ for all null labels $i$.

For Negative modes ($\lambda_i<0$): these produce negative modal masses $m_i<0$. Such modes render the quadratic form indefinite and typically correspond to ghosts (instabilities). A physically acceptable quantization requires either that these modes are gauge (and removed by constraints) or that parameters are adjusted so they do not appear in the physical spectrum.  Retaining them in a naive Fock construction yields a Hamiltonian unbounded from below and is not physically acceptable.

The Hamiltonian, in the situations of only the non-zero modes, is written as
\begin{equation}
\hat{\mathcal{H}} = \frac{1}{4\mathcal{G}} \sum_{i \in I_{\mathrm{phys}}} \; \frac{\hat{p}_{i}^2}{\lambda_i} = \sum_{i \in I_{\mathrm{phys}}} \; \frac{\hat{p}_{i}^2}{2 m_i}.
\end{equation}
Substituting \( \hat{p}_{i} = -i \sqrt{\frac{r_\hbar m_i \omega_i}{2}} \left( \hat{d}_{i} - \hat{d}_{i}^\dagger \right) \), the Hamiltonian will contain three types of terms: \( \hat{d}_{i}^\dagger\hat{d}_{i} \), \( \hat{d}_{i}^{\dagger2} \), and \( \hat{d}_{i}^2 \).
Thus, the Hamiltonian can be given by:
\begin{equation}
\hat{\mathcal{H}} = \sum_{i \in I_{\mathrm{phys}}} \frac{r_\hbar \omega_i}{4} \left(  -2\hat{d}_{i}^\dagger \hat{d}_{i} + \hat{d}_{i}^{\dagger2} + \hat{d}_{i}^2 -1\right),
\end{equation}
where  \( -2\hat{d}_{i}^\dagger \hat{d}_{i} \) is the usual number operator term; \( \hat{d}_{i}^{\dagger2} \) and \( \hat{d}_{i}^2 \) are the non-diagonal terms associated with pair creation and annihilation. This Hamiltonian is a pure kinetic form and lacks an explicit potential term, indicating that the spacetime fluctuation modes are non-interacting and unstable at this level of description (source-free vacuum spacetime). Nevertheless, the kinetic energy term itself gives rise to effective parameters (specifically, the modal masses $m_{i}$ and frequencies $\omega_{i}$) through the process of mode decomposition and quantization. These parameters originate from the geometric structure encoded in the scale factors $a_{\mu}$ and $b_{\mu}$. Thus, in this formulation, the energy stems entirely from the kinetic contributions of the fluctuations. These modes correspond to the fundamental quantum fluctuations of spacetime. The structure of this Hamiltonian is consistent with a central premise of this work: that the microscopic structure of spacetime can be described as a collection of quantum fluctuation modes, whose dynamics resemble those of a free bosonic field. A more complete characterization of the physical ground state and excitation spectrum, however, would require the introduction of a stabilizing potential term.

\subsection{A potential term for a full quadratic Lagrangian}
On introducing the modal frequency $\omega_i$, the single-mode Hamiltonian is
\begin{equation}
\begin{aligned} 
\hat H_{i}&=\frac{\hat p_{i}^2}{2m_i}+\frac12 m_i\omega_i^2\hat q_{i}^2\\
&=r_\hbar\omega_i\Big(\hat d_{i}^\dagger\hat d_{i}+\tfrac12\Big),
\end{aligned} 
\label{eq:hamiltonian}
\end{equation}
with $\omega_i^2>0$.  In the limiting case $\omega_i\to0$ (pure kinetic/free case), the required squeeze diverges and the spectrum becomes continuous; this limiting behavior signals that a small regularizing potential (finite $\omega_i$) is needed to define normal modes properly.

From here on, we work exclusively on the reduced (physical) sector and omit the subscript ``phys'' where no ambiguity arises.

Equations (\ref{eq:afield}) provide the explicit, canonical expansion of the operators $\hat a_\mu(X)$ and $\hat\pi_\mu(X)$ in terms of bosonic creation and annihilation operators on the reduced physical Fock space. The construction is independent of whether the spectrum is discrete or continuous, making the role of constraints manifest, and provides a clear pathway for both analytical and numerical implementations. Choosing frequencies $\omega_i$ when a physical restoring kernel $K$ is available, which simplifies the algebra and ensures that the modal operators $\hat d_{i}$ diagonalize the modal Hamiltonians without further Bogoliubov squeezing.

The introduction of the finite modal frequency $\omega_i$ and the subsequent construction of a full harmonic oscillator Hamiltonian are necessary steps to ensure the mathematical consistency of the quantization procedure and to obtain a physically viable theory with a stable ground state. The purely kinetic Hamiltonian $\hat{H} \propto \hat{p}^2$, which arises from the fluctuation modes, is pathologically deficient for canonical quantization: it possesses a continuous energy spectrum, lacks a unique, stable ground state (vacuum), and its attempted diagonalization leads to divergent Bogoliubov transformations. These issues prevent the definition of a Fock space of particle states and undermine unitarity. Introducing the potential energy term $\frac{1}{2} m_i\omega_i^2 \hat{q}_{i}^2$ rectifies this by providing the essential restoring force that regularizes the theory. This transforms the Hamiltonian into a non-degenerate, quadratic form that can be properly diagonalized, yielding a discrete spectrum and a well-defined vacuum, which are fundamental prerequisites for any subsequent physical interpretation and application.

The diagonalized Hamiltonian $\hat H_{i}$ carries profound physical meaning. It recasts the picture of spacetime microstructure from one of free, featureless fluctuations to a collection of quantum harmonic oscillators, each with a characteristic frequency $\omega_i$ intrinsic to the geometry itself. Consequently, the energy of each mode becomes quantized in discrete units of $r_\hbar\omega_i$, and the operators $\hat{d}_{i}^\dagger$ and $\hat{d}_{i}$ acquire the clear physical interpretation of creating and annihilating discrete elementary excitations, or quanta, of spacetime. The constant term $\frac{1}{2}r_\hbar\omega_i$ represents the zero-point energy of each mode, a signature of irreducible quantum vacuum fluctuations. Most importantly, this discrete energy level structure is the very foundation that enables the statistical mechanical counting of quantum states, which is essential for deriving the Bekenstein-Hawking entropy.

To generate the modal restoring term \(\tfrac12\,m_i\,\omega_i^2\,\hat q_{i}^2\), one adds a quadratic restoring kernel directly on the original scaling \( a_\mu\).  Introduce a symmetric kernel matrix \(K_{\mu\nu}(X,Y)\) and the potential
\begin{equation}
\Delta H_{\mathrm{pot}} \;=\; \tfrac12\!\int dX\,dY\; a_\mu(X)\,K_{\mu\nu}(X,Y)\, a_\nu(Y).
\end{equation}
Projecting \(\Delta H_{\mathrm{pot}}\) onto the modal basis \(\{\Phi_{\mu;i}(X)\}\) used in the main text yields the modal kernel
\[
K_{ij}\;=\; \iint dX\,dY\; \Phi^{*}_{\mu;i}(X)\,K_{\mu\nu}(X,Y)\,\Phi_{\nu;j}(Y),
\]
and hence
\begin{equation}
\Delta H_{\mathrm{pot}} \;=\; \tfrac12\sum_{i,j}\; q_i\,K_{ij}\, q_j.
\end{equation}
A sufficient and explicit choice that yields uncoupled modal potentials is to take \(K\) diagonal in the \(M\)-eigenbasis:
\[
K_{\mu\nu}(X,Y)
\;=\; \sum_i\!\; m_i\,\omega_i^2\; \Phi_{\mu;i}(X)\,\Phi^{*}_{\nu;i}(Y).
\]
Using the modal normalization and \(m_i=2\mathcal G\,\lambda_i\) (as defined in the text), one then obtains
\begin{equation}
K_{ij} \;=\; m_i\,\omega_i^2\;\delta_{ij}=k_i\delta_{ij},
\end{equation}
and therefore
\begin{equation}
\Delta H_{\mathrm{pot}} \;=\; \tfrac12\sum_i\; m_i\,\omega_i^2\, q_i^2,
\end{equation}
as required.

A simple local example is \(K_{\mu\nu}(X,Y)=\varrho^2\,\delta_{\mu\nu}\,\delta(X-Y)\), which projects to \(K_{ii}=\varrho^2\) and hence gives
\begin{equation}
\omega_i^2 \;=\; \frac{\varrho^2}{m_i}.
\end{equation}
Note that \(K\) must be chosen self-adjoint and so that \(m_i\omega_i^2\ge0\) for physical oscillatory modes; genuine gauge/constraint directions (zero modes of \(M\)) should be treated by constraint reduction (Dirac/BRST).

The squared frequency is a stiffness-to-mass ratio, so larger stiffness or smaller mass raises the frequency and vice versa. They differ, however, in physical origin and infrared/ultraviolet behavior. With a constant \(\varrho^2\), a mode-independent (or local) ``mass'' term typically produces a nonzero gap (\(\omega\) finite as \(a_{\mu}\to \, constant\)) and behaves like an optical mode in condensed-matter language; it therefore contributes a modewise zero-point energy \(\tfrac12r_\hbar\omega\) that is insensitive to long-wavelength kinematics. Practically, stability requires the stiffness and \(m_i\) to be positive; a negative numerator signals a linear instability (imaginary \(\omega\)). In sum, $\omega$ is a natural frequency scale, corresponding to the ``elastic response" of spacetime microstructures; $K_{ij}$ are not only the ``harmonic oscillator stiffness" scale constant for each mode, but also equivalent to the modal projection of the quadratic self-interaction kernel of the original scaling $a_\mu$; $m_i$ is the ``mass" scale or inertia intensity of the modal.

To provide modal restoring forces, one adds a symmetric restoring kernel acting on the original scale operators \(a_\mu\). The full quadratic Lagrangian then reads
\begin{equation}
\begin{aligned}
\mathcal L_{ab}&=\mathcal{G}\,\mathbf v^{T} M \mathbf v \;-\; \tfrac12\,\mathbf a^{T} K \mathbf a\\
\end{aligned}
\end{equation}
Projecting onto the \(M\)-eigenbasis \(\{\Phi_{\mu;i}\}\) yields the Hamiltonian $\hat{\mathcal{H}}=\sum_{i}\, \hat H_{i}$.

When the scale factors become constant, the construction reduces smoothly to a classical description, providing a direct correspondence with standard low-energy geometry. Namely, $a_\mu = \text{constant}$, it signifies a complete absence of spacetime fluctuations--the static regime described in Section II. In this limit, the derivative $b_\mu = -da_\mu/dX^\mu$ vanishes identically. Consequently, the kinetic term in the Lagrangian, which is built from $b_\mu$ (or $v_\mu$), disappears.
Simultaneously, the potential energy term, which is quadratic in $a_\mu$, reduces to a constant value, contributing only a fixed offset to the energy density without any dynamical content. The Hamiltonian, therefore, loses all its quantum-mechanical, operator-valued character. This state describes a rigid, non-fluctuating spacetime where the scaled metric $\hat{g}_{\mu\nu}$ is simply a constant conformal rescaling of the physical metric $g_{\mu\nu}$. Thus, the formalism smoothly reverts to a classical, deterministic geometric description, as required for correspondence with classical general relativity in the low-energy, macroscopic limit. 

From the most fundamental perspective, the potential energy term itself is a quadratic representation of spacetime fluctuations, described by the scale factors $a_\mu$. The origin of this term lies in the intrinsic quantum fluctuations of spacetime. It describes the self-interaction of these spacetime fluctuations, where the kernel $K$ is an inherent property of spacetime geometry, characterizing its intrinsic stiffness in resisting metric fluctuations. This perspective suggests a possible mechanism for the origin of the cosmological constant: it could emerge as a residual vacuum energy density when the fundamental fluctuations are suppressed, leading to a classical spacetime geometry.

This framework could, in principle, be constrained or tested by future high-precision observations, such as measurements of primordial gravitational waves, searches for vacuum birefringence, or comparisons of ultra-precise atomic clocks. Primordial gravitational waves, a relic from the early universe, could carry imprints of these fundamental spacetime fluctuations. Furthermore, the fluctuating metric may manifest as vacuum birefringence, altering the propagation of light over cosmic distances. Finally, on laboratory scales, ultra-precise atomic clocks could detect minuscule variations in the flow of time, providing a direct probe of the metric's microscopic dynamics.

\subsection{Hilbert space and operator domains of $\hat a_\mu$ and $\hat \pi_\mu$}
In the present framework, the scale operator $\hat a_\mu(X)$ and its conjugate momentum $\hat \pi_\mu(X)$ are defined as operator-valued distributions acting on a Fock space constructed from the physical eigenmodes of the kinetic kernel. Formally, one introduces a complete spectral decomposition of the symmetric kernel $M_{\mu\nu}(X,Y)$, and projects the fields onto the corresponding modal coordinates $q_i$ and $p_i$. Each non-vanishing eigenvalue $\lambda_i$ defines a physical mode with effective mass $m_i=2\mathcal{G}\lambda_i$, which can be quantized as a harmonic oscillator by introducing creation and annihilation operators $\hat d_{i}^\dagger$ and $\hat d_{i}$. The Hilbert space $\mathcal{H}^{space}_{\mathrm{phys}}$ is therefore the bosonic Fock space
\[
\mathcal{H}^{space}_{\mathrm{phys}} \;=\; \mathcal{F}\!\left(\bigoplus_{(i)\in I_{\mathrm{phys}}} \mathbb{C}\right),
\]
where $I_{\mathrm{phys}}$ denotes the set of modes with $\lambda_i\neq 0$, after removing null modes (primary constraints) and discarding potential ghost modes with $\lambda_i<0$.

The operators $\hat a_\mu(X)$ and $\hat \pi_\mu(X)$ only acquire meaning after smearing with test functions $f_\mu(X), g_\mu(X)\in C_c^\infty$, yielding
\[
\hat a(f)=\int dX\,f_\mu(X)\hat a_\mu(X), \; 
\hat \pi(g)=\int dX\,g_\mu(X)\hat \pi_\mu(X),
\]
which are densely defined, unbounded operators on the Fock space. Their canonical commutation relations hold on the dense subspace $\mathcal{D}_{\text{fin}}$, which is a standard choice for the operator domain, spanned by finite linear combinations of finite-particle states 
\[
\mathcal D_{\mathrm{fin}}=\operatorname{span}\{d_{i_1}^\dagger \cdots d_{i_n}^\dagger |0\rangle\},
\]
on which all canonical commutation relations Eq.\ref{eq:commutation} are satisfied.

We assume that for all retained modes $(i)\in I_{\text{phys}}$ with positive $m_i$, the modal coordinates $(q_i,p_i)$ are essentially self-adjoint on $\mathcal{D}_{\text{fin}}$. Under this assumption, they extend uniquely to self-adjoint operators on $\mathcal{H}_{\text{phys}}^{space}$, which is necessary for their interpretation as physical observables.

\subsection{Zero-point energy}
Zero-point energy denotes the irreducible ground-state energy of quantized degrees of freedom. In nonrelativistic quantum mechanics, it arises from the spectrum of the harmonic oscillator; in quantum field theory, it is the sum over vacuum fluctuations of all field modes. This notion has broad consequences: in condensed matter, it underlies stability and phonon dynamics, while in cosmology it is central to the cosmological-constant problem (the enormous mismatch between naive vacuum-energy estimates and the tiny observed value driving cosmic acceleration \cite{Weinberg1989}). More recent approaches that quantize spacetime microstructure suggest that zero-point fluctuations may be intrinsic to geometry, with renormalization-group flows providing a physical mechanism to regulate ultraviolet divergences and to generate effective cutoffs \cite{rovelli2004}.

Within this framework, the zero-point energy of the quantized scale modes takes the familiar harmonic-oscillator form, guaranteeing vacuum stability and the existence of a Fock-space construction. Explicitly,
\[
E^{\rm UV}_0=\frac{r_\hbar}{2}\sum_{i=1}^D \omega_i,
\]
with modal frequencies
\[
\omega_i^2=\frac{k_i}{m_i}=\frac{1}{2\mathcal{G}}\frac{k_i}{\lambda_i}=\frac{1}{2\mathcal{G}}\,k_i\,a_{\rm s}^4\;C_i(D,\epsilon),
\]
where (\ref{eq:lambda_perp}) and (\ref{eq:lambda_pm}) justify the scaling \(\propto C_i(D,\epsilon)a_{\rm s}^4\) with $C_i(D,\epsilon)$ corresponding to the coefficients as the function of spacetime dimension $D$ and $\epsilon$.
From
\begin{align}
\sqrt{2\mathcal{G}}
&= \sqrt{ \frac{2}{64\pi\,r_{l_p}^2}\,\frac{1}{a^{\,D-2}} }= \frac{1}{r_{l_p}\,a^{(D-2)/2}}\frac{1}{\sqrt{32\pi}}.
\end{align}
Hence
\[
E^{\rm UV}_0=2\sqrt{2\pi}\; r_{\hbar}\; r_{l_p}\; a^{(D-2)/2}a_{\rm s}^2\sum_{i=1}^D\sqrt{k_i}\;C_i(D,\epsilon).
\]
If one adopts the Planck units where $ d\lambda_{0}=l_{p} $ and the momentum unit $ p^{0}=p_{p} $ (the Planck momentum), it follows that $ r_{lp}=1 $ and $ r_{\hbar}=1 $, respectively.

This derivation shows that the quantized scale modes yield a discrete spectrum and a stable vacuum: as \(a_{\rm iso}\to 0\) the zero-point energy vanishes, while finite \(a_{\rm iso}\) produces vacuum energy controlled by the geometric stiffness and the mode structure. Without the oscillator structure, the Hamiltonian would admit a continuous spectrum and an unstable vacuum.

The analogy with phonons is manifest: high-frequency modes correspond to rigid geometric fluctuations, low-frequency modes to soft fluctuations, together forming a ``stretchable spacetime lattice'' with intrinsic vacuum energy. The geometric renormalization-group (RG) flow of the scale functions \(L^\alpha(X^\alpha)\), encoded by the beta function \(\beta(L^\alpha)\) and the fixed points listed in TABLE~I, constrains the modal spectrum and thus the ultraviolet zero-point energy. 
TABLE~I identifies ultraviolet-like fixed points (A and B) at small scales and scale ultraviolet endpoint at finite scale where \(a_\alpha\to 0\), and scale infrared endpoints where \(a_\alpha\to\tfrac{1}{2}\) (regular SIR endpoint) or \(a_\alpha\to-\infty\) (O:saddle/C/D). Because the modal mass scales as \(m_i=2\mathcal{G}\lambda_i\) and the eigenvalues \(\lambda_i\) depend on the local magnitude of \(a_\mu\), the RG limits of \(a_\alpha\) determine the asymptotic behaviour of \(\lambda_i\), \(m_i\) and hence \(\omega_i\).

Combining \(\omega_i^2\propto a_{\rm iso}^4\) with the RG limits yields the following: at UV fixed points A and B, where \(a_{\rm iso}\to 0\), modal frequencies vanish, causing the zero-point contribution \(\tfrac12 r_\hbar\omega_i\) of an individual mode to tend to zero. The behavior of the total zero-point energy then depends on the spectral density of these low-frequency modes in the ultraviolet regime. At regular SIR endpoints (e.g. \(a_{\rm iso}\to\tfrac12\)), frequencies remain finite and \(E^{\rm UV}_0\) acquires a finite contribution proportional to \(a_{\rm iso}^2\) at that scale. On SIR branches with \(a_{\rm iso}\to\infty\), modal frequencies grow and individual zero-point contributions increase, though changes in \(\lambda_i\) and mode decoupling can modify the net vacuum energy; the final value is determined by the regulated spectral measure.

TABLE~I and the beta-flow plots (Fig.~\ref{fig3}) show transitions between discrete/static and fluctuating phases. When the flow reaches a discrete fixed point, the dual-measurement construction enforces a minimal microscopic interval (nonzero \(L^\alpha\)), truncating the high-\(k\) part in the standard quantum field theory and providing an intrinsic ultraviolet regulator that regulates the na\"{\i}ve divergence of the sum over vacuum fluctuations of all field modes.

Finally, TABLE~I highlights regimes where \(b_\alpha\) or \(a_\alpha\) diverge or vanish; such behaviour alters \(\lambda_i\), can create or remove zero modes, and may flip the sign of \(\lambda_i\). Since \(\omega_i^2\propto k_i/\lambda_i\), zeros or sign changes in \(\lambda_i\) signal instabilities (ghosts or gapless directions). The RG flow therefore selects physically admissible branches (those avoiding ghosts) and thereby determines the consistent modal set that contributes to \(E^{\rm UV}_0\). In summary, the RG dynamics govern the modal eigenvalues \(\lambda_i\), modal masses \(m_i\), frequencies \(\omega_i\), and ultimately the zero-point energy. This RG-induced phase structure supplies a natural ultraviolet regulator and a principled selection of the physical modal spectrum, linking microscopic RG behaviour to semiclassical quantities such as black-hole entropy and the cosmological constant problem. 
The zero-point energy of the scale modes, instead of being a rigid ultraviolet divergence as in conventional QFT, is dynamically tied to the geometric renormalization group flow of spacetime microstructure. Consequently, the observed small cosmological constant may be interpreted as the effect of this flow, rather than the sum over all ultraviolet modes. The contribution of high-energy modes is renormalized during the flow process.

In our scaling-based localization framework, the fundamental dynamical variables are the direction-labelled, dimensionless scale amplitudes \(a_\mu(X^\mu)\) (and their derivatives \(b_\mu\)), which encode the local stretch/compression of the underlying substrate geometry and determine all local geometric and spectral quantities. The modal decomposition of the quadratic scale-sector operator produces a discrete sum of harmonic-oscillator contributions that are constructed from these dimensionless scale factors; accordingly, the natural vacuum contribution obtained from that spectral sum is itself a dimensionless, local quantity governed by the local values of the scale amplitudes and their renormalization-group (RG) flow. Physical units are recovered only by comparing microscopic interval measurements to a fixed reference length \(d\lambda_0\), so the corresponding physical energy density is obtained by multiplying the dimensionless spectral quantity \(E_0^{\mathrm{UV}}\) by the conversion factor \(\hbar c/d\lambda_0^4\). Therefore, it is both natural and consistent within our formalism to define the modal (spectral) zero-point contribution as the dimensionless local vacuum energy \(E_0^{\mathrm{UV}}\); physical energy densities follow after restoration of units via the micro-measurement scale \(d\lambda_0\).

As a consequence, let \(E_0^{\mathrm{UV}}\) denote the dimensionless local vacuum energy density. The corresponding physical vacuum energy density is then
\begin{equation}
\rho_{\mathrm{vac}}^{(\mathrm{phys})}
= \hat{\rho}_{\mathrm{vac}}\,\frac{\hbar c}{d\lambda_0^4}
= E_0^{\mathrm{UV}}\,\frac{\hbar c}{d\lambda_0^4},
\end{equation}
where \(\hat{\rho}_{\mathrm{vac}}\) is dimensionless and equal to \(E_0^{\mathrm{UV}}\).
Using the relation between vacuum energy and the cosmological constant in general relativity gives
\begin{equation}
\Lambda \;=\; \frac{8\pi G}{c^4}\,\rho_{\mathrm{vac}}^{(\mathrm{phys})}
\;=\; E_0^{\mathrm{UV}}\,\frac{8\pi G\hbar}{c^3}\,\frac{1}{d\lambda_0^4}
\;=\; E_0^{\mathrm{UV}}\,\frac{8\pi l_p^{2}}{d\lambda_0^4},
\end{equation}
with \(l_p=\sqrt{\hbar G/c^3}\) the Planck length.
If one adopts the convenient choice \(d\lambda_0=l_p\), this expression simplifies to
\begin{equation}
\Lambda \;=\; 8\pi\,E_0^{\mathrm{UV}}\,l_p^{-2}.
\end{equation}
For comparison, the observed value of the cosmological constant is
\begin{equation}
\begin{aligned}
\Lambda_{\mathrm{obs}} &\approx 1.1\times10^{-52}\ \mathrm{m}^{-2}\\
&\approx 2.9\times10^{-122}\ l_p^{-2}\\
&\approx 8\pi\times 1.2\times10^{-123}\ l_p^{-2}.
\end{aligned}
\end{equation}
Thus, if \(E_0^{\mathrm{UV}}\) were generically \(\mathcal{O}(1)\) with \(d\lambda_0=l_p\), one would expect a Planck-scale cosmological constant; instead the observed value is suppressed by roughly \(10^{-123}\).

Within our scaling-based quantization framework, this enormous suppression can be interpreted as a geometry renormalization-group effect associated with the flow of the scale variables. In this picture \(E_0^{\mathrm{UV}}\) is not a fixed Planck-scale constant but depends on the spectral distribution of the scale modes. Near the ultraviolet fixed point, the modal frequencies scale with the isotropic scale parameter as \(\omega_i\propto a_{\rm iso}^2\), so the zero-point contribution to the vacuum energy vanishes as \(a_{\rm iso}\to 0\). By contrast, when \(a_{\rm iso}\) approaches values of order unity (for example \(a_{\rm iso}\sim 1/2\)), the same mechanism naturally produces \(E_0^{\mathrm{UV}}\sim\mathcal{O}(1)\), resulting in the enormous mismatch.

Accordingly, this framework allows for a scenario in which the extreme smallness of $\Lambda_{obs}$ emerges dynamically from the scaling flow of the spacetime microstructure, providing an alternative to ad hoc cancellations. This perspective therefore suggests a potential dynamical mechanism that could ameliorate the cosmological constant problem: instead of invoking an unnatural adjustment of vacuum energy at the Planck scale, the observed suppression may arise naturally from the RG evolution of the scale degrees of freedom and the resulting zero-point spectrum of spacetime.
We emphasize, however, that turning this qualitative RG-based suppression into a quantitative account of the \(10^{-123}\) factor requires a detailed calculation.

\section{Micro area operator and entropy of black hole}
\subsection{Micro area operator}
In the study of spacetime quantization, it is crucial to define fundamental observable operators that characterize the quantum nature of geometry itself  \cite{RovelliSmolin1995}. This section constructs a micro-area operator based on the previously established quantum theory of scaling fluctuations. The core concept is that the physical area of a spacetime surface is composed of fundamental quantum fluctuation modes, rather than being a continuous classical quantity. By defining an area element through suitable combinations of the scale operators $a_\mu$, we reveal the discrete microstructure of spacetime geometry. The introduction of this operator provides the necessary microscopic foundation for deriving thermodynamic properties such as black hole entropy from first principles, directly linking the quantum structure of spacetime to semiclassical gravitational phenomena.

Within the framework of general relativity, the classical area of a two-dimensional surface 
\(\mathcal{S}\) is determined by the induced metric,  
\begin{equation}
A_{\mathcal{S}} = \int_{\mathcal{S}} \sqrt{g}\, d^2x, 
\qquad g = \det(g_{mn}),
\label{eq:classical_area}
\end{equation}
where \( g_{mn} \) denotes the induced metric on the surface, encoding its intrinsic geometry. 
In the present scaling-based approach with scale metric $\hat g=\eta$, the determinant of the induced metric takes the form, e.g., index $\mu=2,3$,
\begin{equation}
\sqrt{g} = \frac{a_2 a_3}{ a^2}.
\label{eq:g_param}
\end{equation}
Motivated by this relation and Eq.~\eqref {eq:microeq1}, we introduce the operator associated with the microscopic area element, defined as
\begin{equation}
\hat{A}_{\min} = \frac{\hat L^2 \hat L^3 \, dx_0^2}{a^2}\, \mathcal{\hat O},
\label{eq:Amin_def}
\end{equation}
where the symmetrized product 
\begin{equation}
\mathcal{\hat O} = \tfrac{1}{2}\Big[\hat{a}_2(X^2)\hat{a}_3(X^3) 
+ \hat{a}_3(X^3)\hat{a}_2(X^2)\Big]
\label{eq:O_def}
\end{equation}
is manifestly Hermitian and can be considered as an operator for a microscopic area through appropriate regularization and spectral processing. Eqs.~\eqref{eq:Amin_def}--\eqref{eq:O_def} together provide a consistent definition of the microscopic area operator in this setting.

Below are the steps to calculate the eigenvalues of the double-mode mixed operator in the Fock state, including writing it in matrix form, performing a general Bogoliubov (BdG) transformation \cite{Bogoliubov1947,Nam2015} to diagonalize it, and finally writing the eigenvalue expression. 

Working in the spectral/modal basis \(\{s^{(i)}\}\) (indices \(i,j=0,\dots,D-1\) or an integral label for continuum modes), the operators \(\hat a_\mu\) in terms of modal canonical operators \(\hat d_i,\hat d_i^\dagger\) are 
\[
\hat a_\mu \;=\; \sum_{i=0}^{D-1} s^{(i)}_\mu\sqrt{\frac{r_\hbar}{2m_i\omega_i}}\,(\hat d_i+\hat d_i^\dagger).
\]
Therefore, the product of interest expands as
\begin{equation}\label{eq:a2a3_expand1}
\begin{aligned}
\hat a_2\hat a_3
&=\sum_{i,j}^{D-1} \mathcal{K}_{ij}\;(\hat d_i+\hat d_i^\dagger)(\hat d_j+\hat d_j^\dagger),\\
\mathcal{K}_{ij}&= s^{(i)}_2\sqrt{\frac{r_\hbar}{2m_i\omega_i}}s^{(j)}_3\sqrt{\frac{r_\hbar}{2m_j\omega_j}}.
\end{aligned}
\end{equation}
Reorder the operator product in Eq. (\ref{eq:a2a3_expand1}) into normal form (all creation operators to the left). Using \(\hat d_i\hat d_j^\dagger = \hat d_j^\dagger\hat d_i + \delta_{ij}\) we obtain
\begin{equation}\label{eq:a2a3_normal}
\hat a_2\hat a_3
=:\!(\hat a_2\hat a_3)\!:\;+\; \mathcal{C}_0,
\end{equation}
where $\mathcal{C}_0=\frac{1}{2}\sum_i\mathcal{K}_{ii}
+\mathcal{K}_{ii}^{T}$ and the normal-ordered operator equals
\[
:\!(\hat a_2\hat a_3)\!:\;=\sum_{i,j}^{D-1} \mathcal{K}_{ij}\Big(\hat d_i^\dagger\hat d_j^\dagger + \hat d_i^\dagger\hat d_j +\hat d_j^\dagger\hat d_i + \hat d_i\hat d_j\Big).
\]
thus
\begin{equation}
\begin{aligned}
\mathcal{\hat O}&= (\hat{a}_2 \hat{a}_3+\hat{a}_3 \hat{a}_2)/2\\
&=:\mathcal{\hat O}:+\mathcal{C}_0\\
&= (:\hat{a}_2 \hat{a}_3:+:\hat{a}_3 \hat{a}_2:)/2+\mathcal{C}_0.
\end{aligned}
\end{equation}
Define
\[
\mathcal{K}^{(s)} = \frac{1}{2} \left( \mathcal{K} + \mathcal{K}^T \right), \quad \mathcal{K}_{ij}^{(s)} = \frac{\mathcal{K}_{ij} + \mathcal{K}_{ji}}{2}
\]
One gets
\begin{equation}
\mathcal{\hat O}=\sum_{i,j}^{D-1} \mathcal{K}^{s}_{ij}\;(\hat d_i+\hat d_i^\dagger)(\hat d_j+\hat d_j^\dagger)+\mathcal{C}_0,
\end{equation}
Hence, the full product consists of (i) a normal-ordered quadratic operator containing number and anomalous (pairing) terms, and (ii) a \(c\)-number (contraction) \(\mathcal{C}_0\) which equals the vacuum expectation value
\(\langle0|\hat a_2\hat a_3|0\rangle\) for the modal vacuum \(\hat d_i|0\rangle=0\).
Group it according to number-conserving terms / pairing terms:
\begin{equation}
\mathcal{\hat O} = \hat{\mathbf{d}}^\dagger h \hat{\mathbf{d}} + \frac{1}{2} (\hat{\mathbf{d}}^\dagger g \hat{\mathbf{d}}^\dagger + \hat{\mathbf{d}} g^\dagger \hat{\mathbf{d}})+\mathcal{C}_0,
\end{equation}
where (unifying the subscript style for vectors and matrices) \(\hat{\mathbf{d}}\) denotes the column vector \((\hat{d}_1, \ldots, \hat{d}_n)^T\), \(\hat{\mathbf{d}}^\dagger\) the row vector, matrices
\[
h = \mathcal{K}^{s} + \mathcal{K}^{sT}=\mathcal{K} + \mathcal{K}^{T}, \quad g = \mathcal{K}^s=h/2.
\]

For compactness, define the modal coefficient
\begin{equation}\label{eq:Cdef}
\mathcal{C}^{(i)}_{\mu}(X)=\sqrt{\frac{r_\hbar}{2\,m_i\,\omega_i}}\;\Phi_{i,\,\mu}(X).
\end{equation}
Hence
\[
\hat a_\mu(X)=\sum_{i\in I_{\rm phys}}\Big[\mathcal{C}^{(i)}_{\mu}(X)\,\hat d_i
+ \mathcal{C}^{(i)*}_{\mu}(X)\,\hat d_i^\dagger\Big].
\]
Using the modal coefficients,
\begin{equation}
\begin{aligned}
\hat a_2(X)\hat a_3(Y)
&= \sum_{i,j} \Big[\mathcal{C}^{(i)}_{2}(X)\hat d_i+\mathcal{C}^{(i)*}_{2}(X)\hat d_i^\dagger\Big]\\
&\times\Big[\mathcal{C}^{(j)}_{3}(Y)\hat d_j+\mathcal{C}^{(j)*}_{3}(Y)\hat d_j^\dagger\Big]\nonumber\\[4pt]
&=\sum_{i,j}\Big\{
\mathcal{C}^{(i)}_{2}(X)\,\mathcal{C}^{(j)}_{3}(Y)\,\hat d_i\hat d_j\\
&+ \mathcal{C}^{(i)}_{2}(X)\,\mathcal{C}^{(j)*}_{3}(Y)\,\hat d_i\hat d_j^\dagger\\
&+ \mathcal{C}^{(i)*}_{2}(X)\,\mathcal{C}^{(j)}_{3}(Y)\,\hat d_i^\dagger\hat d_j\\
&+ \mathcal{C}^{(i)*}_{2}(X)\,\mathcal{C}^{(j)*}_{3}(Y)\,\hat d_i^\dagger\hat d_j^\dagger
\Big\}.\label{eq:a2a3_expand2}
\end{aligned}
\end{equation}
Collecting terms gives the decomposition
\[
\mathcal{\hat O} = : \mathcal{\hat O} : + \mathcal{C}(X, Y),
\]
where the $c$-number (contraction, vacuum expectation) is
\begin{equation}
\begin{aligned}
\mathcal{C}&=\mathcal{C}(X, Y) =\sum_i B_{ii}\\
&= \frac{1}{2} \sum_i[ \mathcal{C}_{2}^{(i)}(X) \; \mathcal{C}_{3}^{(i)*}(Y)+\mathcal{C}_{3}^{(i)}(X) \; \mathcal{C}_{2}^{(i)*}(Y)].
\end{aligned}
\label{eq:contraction}
\end{equation}
Thus 
\begin{equation}
\begin{aligned}
\mathcal{\hat O}&=:\mathcal{\hat O}:+\mathcal{C}\\
&= (:\hat{a}_2 \hat{a}_3:+:\hat{a}_3 \hat{a}_2:)/2+\mathcal{C}\\
&= \sum_{ij}(A_{ij}\, \hat d_i\hat d_j + B_{ij}\, \hat d_i\hat d^\dagger_{j} + C_{ij}\, \hat d^\dagger_{i}\hat d_j + D_{ij}\, \hat d^\dagger_{i}\hat d^\dagger_j)
\end{aligned}
\end{equation}
with the kernels
\[
A_{ij}= (\mathcal{C}^{(i)}_{2}\,\mathcal{C}^{(j)}_{3}+\mathcal{C}^{(i)}_{3}\,\mathcal{C}^{(j)}_{2})/2,
\]
\[
B_{ij}= (\mathcal{C}^{(i)}_{2}\,\mathcal{C}^{(j)*}_{3}+\mathcal{C}^{(i)}_{3}\,\mathcal{C}^{(j)*}_{2})/2,
\]
\[
C_{ij}= (\mathcal{C}^{(i)*}_{2}\,\mathcal{C}^{(j)}_{3}+\mathcal{C}^{(i)*}_{3}\,\mathcal{C}^{(j)}_{2})/2,
\]
\[
D_{ij}= (\mathcal{C}^{(i)*}_{2}\,\mathcal{C}^{(j)*}_{3}+\mathcal{C}^{(i)*}_{3}\,\mathcal{C}^{(j)*}_{2})/2.
\]
Normal order all annihilation operators to the right. Use the commutation relation
\[
\hat{d}_i \hat{d}^\dagger_j = \hat{d}^\dagger_j \hat{d}_i + \delta_{ij}.
\]
The normal-ordered operator is
\begin{equation}
\begin{aligned}
:\mathcal{\hat O}: &= \sum_{i,j} \Big\{ A_{ij} \hat{d}_i \hat{d}_j + \left(B_{ij} - \frac{1}{2} \delta_{ij} B_{ii}\right) \hat{d}^\dagger_j \hat{d}_i \\
& + C_{ij} \hat{d}^\dagger_i \hat{d}_j + D_{ij} \hat{d}^\dagger_i \hat{d}^\dagger_j \Big\},
\end{aligned}
\end{equation}
but it is clearer to reorganize into the canonical quadratic form below.
The normal-ordered operator \(\mathcal{\hat O}\) can be written as:
\begin{equation}
:\mathcal{\hat O}: = \hat{\mathbf{d}}^\dagger h \hat{\mathbf{d}} + \frac{1}{2} (\hat{\mathbf{d}}^\dagger g \hat{\mathbf{d}}^\dagger + \hat{\mathbf{d}} g^\dagger \hat{\mathbf{d}}),
\end{equation}
where 
\[h_{ij} = C_{ij} + B_{ji}
= (\mathcal{C}^{(i)*}_{2} \mathcal{C}^{(j)}_{3} + \mathcal{C}^{(i)*}_{3} \mathcal{C}^{(j)}_{2}),\]
\[g_{ij} = \frac{1}{2} (\mathcal{C}^{(i)*}_{2} \mathcal{C}^{(j)*}_{3} + \mathcal{C}^{(i)*}_{3} \mathcal{C}^{(j)*}_{2}).\]
If \(\mathcal{C}^{(2)}, \mathcal{C}^{(3)}\) are real vectors, then the above expressions simplify to \(h_{ij}=h\), and \(g_{ij} = \frac{1}{2} h\).
From the above, it is directly visible that \(h^\dagger = h\) (i.e., \(h^*_{ji} = h_{ij}\)), and \(g^T = g\) (the pairing matrix is symmetric). Therefore, \(:\mathcal{\hat O}:\) is a Hermitian operator (because the second term appears paired with its Hermitian conjugate).

Form the Nambu (doubled) vector \cite{Nambu1960} and symplectic metric \cite{Colpa1978}
\[
\Psi = \begin{pmatrix}\hat{\mathbf d}\\[4pt]\hat{\mathbf d}^\dagger\end{pmatrix},\qquad
\Sigma = \begin{pmatrix} I & 0 \\[4pt] 0 & -I \end{pmatrix}.
\]
Define the BdG block matrix
\begin{equation}\label{eq:Hbdg}
\mathcal{H}^{A} = \begin{pmatrix} h & g \\[4pt] g^\dagger & h^T \end{pmatrix},
\qquad
\mathcal{L}^{A}= \Sigma\mathcal{H}^{A} = \begin{pmatrix} h & g \\[4pt] -g^\dagger & -h^T \end{pmatrix},
\end{equation}
where the upper right index means these operators are corresponding to the micro area operator.
Up to ordering constants \(\mathcal{\hat O}=\tfrac12\,\Psi^\dagger\mathcal H^A\Psi\). The linear bosonic eigenproblem is
\begin{equation}\label{eq:bdg_ev}
\mathcal{L}^{A}\,\xi_n = \omega^A_n\,\xi_n,\qquad n=1,\dots,2D,
\end{equation}
where \(2D\) is the number of discrete modes; replace by appropriate discretization for continuum. Eigenvalues appear in \(\pm\) pairs; select the \(N\) positive-frequency eigenvalues \(\{\omega^A_n>0\}\) and their eigenvectors \(\{\xi_n\}\). Normalize them with respect to the symplectic form,
\[
\xi_n^\dagger \Sigma \xi_m = \operatorname{sign}(\omega^A_n)\,\delta_{nm}.
\]
Collect these eigenvectors into a para-unitary matrix \(T\) which satisfies
\(T\Sigma T^\dagger=\Sigma\). The canonical transformation
\[
\Psi = T\,\Phi,\qquad \Phi=(\hat{\mathbf e},\hat{\mathbf e}^\dagger)^T,
\]
defines new bosonic operators \(\hat e_n\) obeying \([\hat e_n,\hat e_m^\dagger]=\delta_{nm}\) and diagonalizes \(\mathcal{\hat O}\):
\begin{equation}\label{eq:diagO}
:\mathcal{\hat O}: \;=\; \sum_{n=0}^{D-1} \omega^A_n\,\hat e_n^\dagger \hat e_n \;+\; \text{(constant)}\; ,
\end{equation}
reveals that the spectrum of the micro-area operator is quantized, with eigenvalues determined by the occupation numbers of these fundamental ``area quanta'' excitations. This is the foundation for the state counting that follows.
Hence the full product is
\begin{equation}\label{eq:full-product}
\mathcal{\hat O} = \sum_{n=0}^{D-1}\omega^A_n\,\hat e_n^\dagger\hat e_n + \text{const} \;+\; \mathcal{C},
\end{equation}
where \(\mathcal{C}\) is the contraction Eq. \((\ref{eq:contraction})\) and ``const'' denotes the zero-point/normal-ordering shift arising from the Bogoliubov transformation.
Diagonalization yields independent bosonic normal modes \(\hat e_n\). Physically acceptable quantization requires \(\omega^A_n\in\mathbb{R}_{>0}\); complex or imaginary eigenvalues signal dynamical instability, while negative eigenvalues indicate directions where the quadratic form is unbounded below. 

The anomalous pair terms (the \(\hat d^2,\hat d^{\dagger2}\) pieces) signal that the original \(\hat d\)-basis is not the eigenbasis; they correspond to correlated pair production/annihilation processes and are removed by squeezing. Stability (real positive \(\omega^A\)) is necessary for a well-defined Fock quantization; zero or imaginary modes indicate gauge/constraint directions or dynamical instabilities that must be treated before quantization. The product \(\hat a_2\hat a_3\) decomposes into a c-number plus a quadratic normal-ordered operator. The latter can be cast as a bosonic BdG problem and diagonalized by a para-unitary (Bogoliubov) transform. 

Within the finite-mode construction presented, the operator $\hat{\mathcal{O}}$ is Hermitian by design, as a consequence of its symmetrized form and appropriate normalization. This Hermiticity is manifest in the structure of the operator: each pairing term is accompanied by its Hermitian conjugate. Upon diagonalization (provided the non-real branches of the spectrum are excluded), the resulting eigenvalues are real and non-negative. The construction relies on underlying bosonic modes satisfying the canonical commutation relations $[d_i, d_j^\dagger] = \delta_{ij}$, and the operator is built using only modes associated with the same unit, so that no non-trivial cross-commutation relations arise. Furthermore, the form of $\hat{\mathcal{O}}$ is invariant under two-dimensional diffeomorphisms on the horizon surface, as well as under local rotations in the intrinsic 2--3 plane, reflecting its geometric character. The operator depends exclusively on the intrinsic area fluctuation operators $\hat{a}_2$ and $\hat{a}_3$, with no external structural bias. In summary, $\hat{\mathcal{O}}$ constitutes a consistent and operationally meaningful candidate for a quantum micro-area operator in this framework.

\subsection{The black hole entropy}
Black hole thermodynamics provides a fundamental benchmark for microscopic quantum-gravity models: any candidate theory should reproduce the Bekenstein-Hawking relation
\begin{equation}
S_{\mathrm{BH}}=\frac{A}{4G\hbar}=\frac{A}{4l^2_p},
\end{equation}
at least in the appropriate semiclassical limit \cite{Bekenstein1973,Hawking1975}. For concreteness, we adopt the Schwarzschild geometry and count horizon microstates by partitioning the horizon into \(N\) quantum units \cite{Bombelli1986}. Each unit hosts \(D\) local modes whose excitations contribute additively to the unit area. Based on Eq. (\ref{eq:full-product}), denoting the occupation number of mode \(j\) by \(n_j\) and the per-quantum area weight by \(\omega^A_j\), we parametrize the unit area as
\begin{equation}
A_{\min}(\{n_j\})=\gamma\sum_{j=0}^{D-1} \omega^A_j\,n_j + A_0,\quad n_j\in\{0,\dots,n_{j,\max}\},
\label{eq:unit_area_param}
\end{equation}
where \(\gamma=\frac{L^2 L^3 \, dx_0^2}{ a^2}\) is the geometric prefactor and \(A_0\) collects zero-point/normal-ordering contributions. This parametrization allows modes to contribute unequally to the area, a feature important for physically faithful counting.

To count the microstates compatible with a fixed total area $A$, we employ a saddle-point evaluation in the microcanonical ensemble. This is facilitated by first considering the canonical partition function \cite{FetterWalecka2003} for a single unit 
\begin{equation}
Z_1(\beta)=\sum_{\{n_j\}} e^{-\beta A_{\min}(\{n_j\})}
= e^{-\beta A_0}\prod_{j=0}^{D-1}\Biggl(\sum_{n=0}^{n_{j,\max}} e^{-\beta\gamma\omega^A_j n}\Biggr).
\label{eq:Z1_def}
\end{equation}
The total microcanonical count for a horizon composed of \(N\) such independent units with fixed total area \(A=\sum_{i=1}^N A_i\) is given by the inverse Laplace transform
\begin{equation}
\begin{aligned}
\Omega_{\rm tot}(A)&=\frac{1}{2\pi i}\int_{c-i\infty}^{c+i\infty} d\beta\; e^{\mathcal S(\beta)},\\
\mathcal S(\beta)&= N\ln Z_1(\beta)+\beta A,
\end{aligned}
\label{eq:Omega_bromwich}
\end{equation}
which is amenable to a saddle-point evaluation for large \(N\) (and hence large \(A\)) \cite{Pathria2011}.  Denote by \(\beta_0\) the saddle satisfying \(\mathcal S'(\beta_0)=0\).  Using
\begin{equation}
\langle A_{\min}\rangle_\beta = -\partial_\beta\ln Z_1(\beta)
= A_0+\gamma\sum_{j=0}^{D-1}\omega^A_j\,\langle n_j\rangle_\beta.
\label{eq:mean_unit_area}
\end{equation}
The saddle condition is simply
\begin{equation}
A = N\langle A_{\min}\rangle_{\beta_0}.
\label{eq:saddle_condition}
\end{equation}

To Gaussian order, the saddle-point approximation yields the entropy
\begin{equation}
\begin{aligned}
S(A)&=\ln\Omega_{\rm tot}(A)\simeq S_0-\tfrac12\ln\!\bigl(2\pi N\sigma^2(\beta_0)\bigr)+\cdots,\\
S_0&= N\ln Z_1(\beta_0)+\beta_0 A,
\end{aligned}
\label{eq:entropy_saddle}
\end{equation}
where the single-unit variance
\begin{equation}
\sigma^2(\beta)=\mathrm{Var}_\beta(A_{\min})=\gamma^2\sum_{j=0}^{D-1}(\omega^A_j)^2\,\mathrm{Var}_\beta(n_j)
\label{eq:variance_def}
\end{equation}
determines the Gaussian fluctuation prefactor (the total variance is \(N\sigma^2\)).  Thus, the leading term \(S_0\) is extensive in \(A\) and reproduces an area law under the natural assumption \(N\propto A\); the subleading logarithmic correction arises from the determinant of quadratic fluctuations about the saddle.

Several aspects of this state-counting model warrant clarification. First, the statistical properties, such as $\langle n_j\rangle_\beta$ and $\mathrm{Var}_\beta(n_j)$, depend on whether the occupation numbers $n_j$ are bounded ($n_{j,\max}<\infty$) or unbounded (bosonic), which in turn influences the saddle point $\beta_0$, the leading entropy $S_0$, and the fluctuation prefactor. Second, the logarithmic correction to the entropy is sensitive to combinatorial assumptions. For example, treating the $N$ units as distinguishable typically yields a $-\tfrac12\ln A$ term (since $N \propto A$), while modeling them as indistinguishable or imposing additional projection or gauge constraints can alter this coefficient, in some cases to values such as $-1$ or those reported in certain loop quantum gravity studies \cite{Rovelli1996,Ashtekar1998,Bambi2024}. Third, the zero-point area $A_0$ generally requires regularization, e.g., via point-splitting, momentum cutoffs, or zeta-function methods, and different prescriptions affect both $\beta_0$ and the inferred microscopic scale when matching $S_0/A$ to $1/(4G\hbar)$. 

This study starts from the first principles of quantum spacetime fluctuations and derives the area law of black hole entropy through microscopic state counting. To achieve precise agreement with the Bekenstein-Hawking formula and determine the microscopic scales within the theory, a parameter fitting procedure is employed. This approach, similar to other quantum gravity approaches, represents a crucial step from microscopic quantum geometry towards fully deriving classical gravity. However, a completely parameter-free, first-principles derivation of the Planck length remains an ultimate goal for future research.
It follows that deriving the Bekenstein-Hawking area law of black hole entropy through microscopic state counting in this framework constitutes a model-dependent fitting procedure. A similar reliance on matching conditions appears in other approaches that connect entanglement or thermodynamic arguments to semiclassical gravity \cite{Jacobson2016,Bombelli1986}.

To make these statements explicit, we present two analytic examples that illustrate how the leading area scaling and the logarithmic correction arise and how they depend on modelling choices.

\vspace{4pt}\noindent\text{(i) Bosonic limit (\(n_{j,\max}\to\infty\)).}  Defining \(q_j= e^{-\beta\gamma\omega^A_j}\), the single-unit partition function becomes
\begin{equation}
Z_1(\beta)=e^{-\beta A_0}\prod_{j=0}^{D-1}(1-q_j)^{-1},
\end{equation}
and the familiar Bose expressions apply:
\(\langle n_j\rangle_\beta=(e^{\beta\gamma\omega^A_j}-1)^{-1}\) and \(\mathrm{Var}_\beta(n_j)=e^{\beta\gamma\omega^A_j}/(e^{\beta\gamma\omega^A_j}-1)^2\).  The saddle-point equation
\(\tfrac{A}{N}=A_0+\gamma\sum_j \omega^A_j (e^{\beta_0\gamma\omega^A_j}-1)^{-1}\)
determines \(\beta_0\), and the extensive contribution to the entropy is
\begin{equation}
S_0 = N\sum_{j=0}^{D-1} \Bigl[\frac{\beta_0\gamma\omega^A_j}{e^{\beta_0\gamma\omega^A_j}-1}
- \ln\bigl(1-e^{-\beta_0\gamma\omega^A_j}\bigr)\Bigr],
\end{equation}
with variance
\(\sigma^2(\beta_0)=\gamma^2\sum_j (\omega^A_j)^2\,e^{\beta_0\gamma\omega^A_j}/(e^{\beta_0\gamma\omega^A_j}-1)^2\).

\vspace{4pt}\noindent\text{(ii) Single-frequency, identical-mode model.}  If all \(M\) modes per unit share the same frequency \(\omega\), then \(q=e^{-\beta\gamma\omega}\) and
\(Z_1(\beta)=e^{-\beta A_0}(1-q)^{-D}\).  In the high-excitation (large-area) regime, \(\beta_0\gamma\omega\ll1\), one may approximate \(e^x-1\approx x\).  The saddle condition then yields
\[
\frac{A}{N}-A_0 \approx \frac{D}{\beta_0}\quad\Longrightarrow\quad
\beta_0\approx\frac{D}{A/N - A_0},
\]
and the leading entropy becomes, to leading logarithmic accuracy,
\begin{equation}
S_0 \approx N D\Bigl[1 - \ln\!\Bigl(\frac{\gamma\omega D}{A/N - A_0}\Bigr)\Bigr],
\end{equation}
so that for fixed \(D\) and \(N\propto A\) the entropy scales linearly with \(A\).  The Gaussian fluctuation term scales as \(-\tfrac12\ln(2\pi N\sigma^2)\approx -\tfrac12\ln N +\) const (with \(\sigma^2_{\rm unit}\approx\gamma^2 D/\beta_0^2\) in this regime), hence producing the familiar \(-\tfrac12\ln A\) subleading behavior in the simplest independent, distinguishable-unit model.

To compare the microscopic count with the semiclassical result one matches the 
leading information entropy \(S_0(A)\) obtained from the saddle-point evaluation to the 
Bekenstein--Hawking expression \cite{Bekenstein1973,Hawking1975}.
This matching yields the general relation 
\begin{equation}
S_0(A)=\frac{A}{4l_p^{2}},
\end{equation}
so that any explicit expression for \(S_0\) immediately gives the corresponding Planck 
length as a function of the microscopic parameters. For example, in the bosonic 
(unbounded-occupation) limit, one has 
\begin{equation}
S_0=N\sum_{j=0}^{D-1}\left[\frac{\beta_0\gamma\omega^A_j}{e^{\beta_0\gamma\omega^A_j}-1}
-\ln\!\bigl(1-e^{-\beta_0\gamma\omega^A_j}\bigr)\right],
\end{equation}
with \(\beta_0\) fixed implicitly by the saddle condition 
\begin{equation}
\frac{A}{N}=A_0+\gamma\sum_{j=0}^{D-1}
\frac{\omega^A_j}{e^{\beta_0\gamma\omega^A_j}-1}.
\end{equation}
Substitution into \(S_0=A/(4l_p^2)\) yields \(l_p\) as an implicit function of 
\(\{\gamma,\omega^A_j,A_0,D,N\}\). In the single-frequency, high-excitation approximation one sets \(\omega^A_j=\omega^A\) for all \(j\) and introduces the shorthand \(z=\beta_0\gamma\omega^A\). The saddle-point equation for the mean cell area \(\langle A_{\min}\rangle_{\beta_0}= A/N\),
\begin{equation}
\langle A_{\min}\rangle_{\beta_0} - A_0 \;=\; \gamma D\,\frac{\omega^A}{e^{z}-1},
\end{equation}
and the entropy
\begin{equation}
S_0 \;=\; N D\!\left[\frac{z}{e^{z}-1}-\ln\!\big(1-e^{-z}\big)\right],
\end{equation}
are then expanded for \(z\ll1\). Using \(e^{z}-1 = z + \tfrac{z^2}{2}+O(z^3)\) and \(\ln(1-e^{-z})=\ln z -\tfrac{z}{2}+O(z^2)\) one obtains, to leading order in \(z\),
\begin{equation}
\frac{z}{e^{z}-1}-\ln(1-e^{-z}) \approx 1-\ln z,
\end{equation}
and hence
\begin{equation}
S_0\approx N D(1-\ln z).
\end{equation}
Combining this with the Bekenstein--Hawking identification \(S_0=A/(4l_p^2)=N \langle A_{\min}\rangle_{\beta_0}/(4l_p^2)\) gives
\begin{equation}
D\big(1-\ln z\big)=\frac{\langle A_{\min}\rangle_{\beta_0}}{4l_p^2}.
\end{equation}
Using the saddle relation in the small-\(z\) limit,
\begin{equation}
z\approx\frac{\gamma\omega^A D}{\langle A_{\min}\rangle_{\beta_0}-A_0},
\end{equation}
one arrives at the compact estimates 
\begin{equation}
\gamma \;\approx\; \frac{\langle A_{\min}\rangle_{\beta_0} - A_0}{\omega^A D}\,
\exp\!\left(1 - \frac{\langle A_{\min}\rangle_{\beta_0}}{4D\,l_p^2}\right),
\end{equation}
or  
\begin{equation}
l_p^2 \;\approx\; 
\frac{\langle A_{\min}\rangle_{\beta_0}}{4D\left[1-\ln\!\left(\frac{\gamma\omega^A D}{\langle A_{\min}\rangle_{\beta_0}-A_0}\right)\right]}.
\end{equation}
These formulae show explicitly that the identification \(S_0\leftrightarrow S_{\mathrm{BH}}\) is achievable in our microscopic scaling framework. Concretely, if one fixes the effective Planck length \(l_p\) by matching it to the model's microscopic parameters (e.g. the mode-spectrum parameters \(\gamma,\omega^A\), the additive constant \(A_0\), and the effective mode count \(D\)), the Bekenstein-Hawking area law \(S_{\mathrm{BH}}=A/(4l_p^2)\) can be recovered. We emphasize that this is not a parameter-free prediction: the value of \(l_p\) inferred from the matching is sensitive to the chosen microscopic inputs, and thus obtaining explicit numerical agreement requires specifying or deriving \(\gamma,\omega^A,A_0,D\) from the underlying microphysics.

This section constructs a micro-area operator derived from scaling fluctuations and uses it to count microscopic states of spacetime geometry. By specifying the microstate spectrum parameters, the resulting state-counting accommodates the leading term of black hole entropy, thereby linking the scaling-operator framework directly to thermodynamic properties of horizons. This connection shows that quantum fluctuations of spacetime not only shape geometry but also provide a natural microscopic foundation for black hole entropy.

\section{Discussions and Comparison with previous theories}
\subsection{Summary for each key section}
In section II, the micro-measurement framework describes how microscope distances at the Planck scale cannot be treated as fixed quantities but are measured by scaling functions. Because quantum fluctuations disturb these infinitesimal intervals, measurement outcomes depend on dynamic scaling functions that capture the degree of local stretching or compression. Three regimes emerge: static measurements without fluctuations, linear fluctuations with simple proportional scaling, and nonlinear fluctuations representing complex or discrete variations. By introducing first- and second-order scale factors, the framework encodes these fluctuations directly into the geometry, turning the physical metric into a composite structure shaped by measurement-dependent deformations. A dual principle shows that the same microscopic lengths can be described either nonlinearly or by equivalent linearized variables, leading to the existence of a natural discrete structure of spacetime. In this way, the micro-measurement framework proposes a physically motivated approach to connecting fluctuations of spacetime with geometry description, suggesting a potential link among measurement, discreteness, and the microscopic structure of spacetime.

Section III develops an operator quantization scheme in which spacetime fluctuations are encoded through locally scaled position and momentum operators. Their commutators reveal a scaled, fluctuation-dependent Planck constant, leading to a generalized uncertainty principle and a discrete microstructure of spacetime. Reformulated Klein-Gordon and Dirac equations incorporate these scale factors, showing how quantum fluctuations deform field dynamics while preserving covariance, embedding microscopic spacetime fluctuations into the behavior of quantum fields. This establishes a concrete framework for calculating how quantum geometry influences matter.

Section IV establishes the covariance of the proposed scaling framework under both Lorentz transformations and general coordinate changes. It demonstrates that the first-order scale factors $a_\mu$ behave as Lorentz-invariant scalars, ensuring the physical laws formulated on the scale manifolds respect fundamental symmetries and capture intrinsic spacetime fluctuations, while the associated metric $\tilde{g}$ preserves invariance under generalized transformations. By embedding these factors into the vierbein formalism, the work provides a mathematical bridge between the anisotropic scale degrees of freedom and standard differential geometry.

Section V extends the classical concept of geodesics from free particle trajectory and geometric explanation to the quantum fluctuations themselves within the scale manifolds. It derives first-order geodesic equations governing the evolution of scaling coordinates, and second-order equations governing the evolution of the fluctuation amplitude coordinates. This provides a geometric language for describing the hierarchical and nonlinear structure of quantum spacetime fluctuations, offering a variational principle for fluctuation dynamics analogous to the least-action principle in classical physics.

Section VI presents the canonical quantization of the first-order fluctuation operators, treating them as fundamental scalar degrees of freedom rather than directly quantizing the metric tensor. The construction yields a Hamiltonian describing anisotropic "breathing modes" of spacetime, which is diagonalized via modal decomposition into a spectrum of independent bosonic excitations. This establishes a covariantly consistent framework for quantum gravity where spacetime geometry is described by more elementary quantum degrees of freedom.
The introduction of finite modal frequencies $\omega_i$, modal mass $m_i$, and a restoring kernel $K$ to form complete harmonic oscillators represents a crucial physical requirement rather than merely a mathematical convenience. This construction resolves the pathologies of a purely kinetic model by ensuring a stable vacuum state, a discrete energy spectrum, and a well-defined Fock space of particle-like excitations. Consequently, spacetime itself acquires a quantum-mechanical structure composed of quantized oscillators, providing the essential elements of discrete quantum states and vacuum fluctuations.
Spectral analysis reveals that the vacuum energy of these discrete harmonic modes is governed by scale geometry renormalization group flow between ultraviolet and infrared regimes. In ultraviolet regions, individual zero-point contributions vanish as modes become gapless, while infrared behavior exhibits discrete phases with minimal length intervals that yield finite energy. This structure provides a natural mechanism for regulating ultraviolet divergences while maintaining infrared completeness, suggesting a potential dynamical mechanism that could ameliorate the cosmological constant problem.

Section VII introduces a microscopic area operator built from scaling fluctuations to represent the quantum-geometric degrees of freedom of spacetime surfaces. By counting the operator's quantum states under the assumption of a fixed macroscopic area, the analysis demonstrates that the model can accommodate the leading Bekenstein-Hawking area law, thereby providing a consistency check that links black hole thermodynamics to the Planck-scale fluctuation structure of spacetime.

\subsection{Physical picture of the scale factors and the two-tier manifold}

The basic dynamical ingredients of the framework are the scaling families $ a_\mu$ and $ b_\mu$, which together provide a compact framework for a layered, measurement-driven description of spacetime fluctuating geometry. Each $ a_\mu$ is conceived as a direction-labelled, dimensionless scale amplitude along the coordinate direction $\mu$. A useful physical analogy is to regard spacetime at microscopic scales as an anisotropic, dynamic elastic substrate. In this picture, the factors $a_\mu$ represent local, directional stretches or compressions of this substrate, while their derivatives $b_\mu$ quantify the inhomogeneity of these deformations. In this picture $ a_\mu$ rescales an underlying substrate geometry $\hat g_{\alpha\beta}$ anisotropically and thereby modulates the physical measurement metric as $g_{\alpha\beta} = \hat g_{\alpha\beta}\,\frac{ a_\alpha  a_\beta}{ a^2}$, where the fraction is to be interpreted as a scaled, direction-dependent rescaling. Because the label $\mu$ denotes directions rather than components of a conventional Lorentz vector, $ a_\mu$ is therefore a collection of direction-labelled, scalar, dimensionless, scale factors.

The companion factors $ b_\mu$ encode the spacetime inhomogeneity of these first-order amplitudes: defined component-wise as the (negative) derivative with respect to the scale coordinates, $ b_\mu= -\partial_{X^\mu} a_\mu$, it measures how rapidly the local stretch varies across scale space.  Regions with large $ b_\mu$ correspond to strong spatial variation in the fluctuation amplitude, indicating a highly irregular microscopic structure, whereas $ b_\mu\approx0$ corresponds to nearly homogeneous fluctuations. Crucially, the inhomogeneity of the fluctuations, encoded in the derivatives of $b_{\mu}$, directly contributes to the connection and thus to the Riemann curvature $R^{\rho}_{\sigma\mu\nu}$. Within this framework, a non-trivial spacetime curvature can be described by the inhomogeneous configuration of the scale fluctuations and the underlying substrate metric $\hat{g}_{\mu\nu}$. This demonstrates that what is classically identified as curvature, at a more fundamental level, is related to the non-uniformity of the spacetime microstructure.

These elements are organized into a two-tier geometric picture. The first-order or scale manifold $\mathcal{M}^{( a)}$ carries abstract scale coordinates $X^\mu$ and the substrate metric $\hat g_{\mu\nu}$; it provides the kinematic stage on which the $ a_\mu$ fields live, with the familiar physical coordinates $x^\mu$ obtained operationally as a projection $x^\mu = L^\mu(X)dx_0^\mu + x_0^\mu$. For the second-order or amplitude manifold $\mathcal{M}^{( b)}$: its coordinates are the amplitude configurations $ a_\mu$ themselves, and its metric $\tilde g_{\alpha\beta}$ measures distances between different amplitude configurations. Geodesics on $\mathcal{M}^{( b)}$ describe the evolution of amplitude configurations, and the dynamics defined there determine the behavior of $ a_\mu$, which in turn sets the effective geometry on $\mathcal{M}^{( a)}$.

\subsection{Comparison with direct metric quantization}
As an alternative to direct metric quantization, this scaling-first approach constructs physical geometry from the scale sector; intermediate appearances of \(\hat{g}_{\alpha\beta}\) are always in combination with scale factors $ a_\mu$. The relation \(\tau^2 = g_{\alpha\beta} L^\alpha L^\beta\) defines proper length measurements in terms of the metric \(g_{\alpha\beta}\). In scale-sensitive extensions of differential geometry, it is useful to distinguish between the physical measurement metric \(g_{\alpha\beta}\), used for physical measurement, and the scaling substrate metric \(\hat{g}_{\alpha\beta}\), which encodes elastic substrate structure. These are related via a locally anisotropic scaling transformation \(g_{\alpha\beta} = \Omega^{( a)}_{\alpha\beta} \hat{g}_{\alpha\beta}\), where the scaling tensor \(\Omega^{( a)}_{\alpha\beta}\) is defined using the microscopic directional scaling \( a_\alpha\). In the isotropic limit $ a_\alpha =  a_{\rm iso}$, this relation reduces to a global conformal rescaling. More generally, variations in $ a_\alpha$ define a generalized anisotropic structure in which the scaling coordinates $X^\alpha$ and the substrate metric $\hat g_{\alpha\beta}$ provide a higher-resolution geometric description: $(\mathcal{M}^{( a)},\hat g_{\alpha\beta},X^\alpha)$ is an elastic substrate whose local rescalings reflect the operational nature of microscopic measurement.

The classical geometry and quantum fluctuations are related through a scale-sensitive extension of the manifold \((\mathcal{M}, g_{\alpha\beta}, x^\alpha)\) to \((\mathcal{M}^{( a)}, \hat{g}_{\alpha\beta}, X^\alpha)\), where the scaling coordinates \(X^\alpha\) and metric \(\hat{g}_{\alpha\beta}\) provide a refined geometric description. Thus, spacetime geometry is not a fixed classical entity but a dynamically rescaled structure tied to the operational nature of measurement at microscopic scales.

By contrast, conventional approaches to quantum gravity typically involve direct quantization of the spacetime metric \(g_{\mu\nu}\). However, this path encounters significant challenges, including the preservation of full diffeomorphism invariance, the handling of Hamiltonian and momentum constraints (as in the Wheeler-DeWitt equation), and non-renormalizability in perturbative treatments~\cite{Kiefer2007}. Moreover, the metric is not directly observable; physical measurements correspond to relational quantities such as proper distances and time intervals, which often require material reference frames for operational meaning~\cite{rovelli2004}. 

The scaling-first framework shifts attention from the direct metric degrees of freedom to the algebraic and spectral properties of the scale factors $a_\mu$ (and $b_\mu$). This shift may offer new routes to address problems encountered in metric quantization, for example, it provides alternative variables for organizing UV/IR behavior and for formulating operationally meaningful observables, but realizing these advantages requires further study.

\subsection{Comparison with previous theories}
Our approach departs from the standard perturbative quantization paradigm by developing a novel, scaling-based framework built on direction-dependent scaling variables \( a_\mu\) and \( b_\mu\), while retaining a quadratic Lagrangian structure that allows for explicit mode decomposition. In the present scaled differential geometry, the theory is formulated and quantized in terms of these fundamental scaling variables, avoiding the need for higher-derivative regulators. In this sense, the non-perturbative character of the geometry is encoded directly in the fundamental variables, in contrast to being recovered through an order-by-order perturbative expansion.

A central conceptual innovation is to reinterpret renormalization in geometric terms. Nonlinear scaling functions \(L_\alpha(X^\alpha)\) interpolate among three regimes (static, linear, and fully nonlinear) and their scale dependence is characterized by a sequence of geometric fixed points. This geometrization of renormalization provides a distinct implementation of how ultraviolet and infrared behaviors connect within a geometric framework. Crucially, instead of removing divergences with external cutoffs, the microscopic, scale-sensitive geometry itself supplies a natural suppression of short-distance modes and an intrinsic ultraviolet scale. The mechanism is therefore inherently nonperturbative, as it does not rely on a perturbative expansion, and geometric, as the renormalization group flow is embodied in the dynamics of the scale manifold itself.

Compared with other prominent approaches to quantum gravity, our construction offers a distinct technical realization while preserving core geometric structures. Unlike string theory \cite{Polchinski1998}, it operates without introducing extra dimensions or extended fundamental objects. In contrast to loop quantum gravity \cite{Rovelli1990}, it retains the differential-manifold language while encoding discreteness in the local scale measurement structure rather than in pre-geometric combinatorial structures. We promote the local, direction-dependent scale measurements and their operator representatives to the primary quantum degrees of freedom. In this picture, Planck-scale discreteness is encoded in the local measurement structure instead of being imposed as a prespecified discrete structure. The approach shares conceptual features with proposals that emphasize anisotropic scaling or minimal-length phenomenology, but it differs technically by maintaining a continuum, vierbein-compatible formulation in which the scale factors transform appropriately under coordinate changes.

To place our approach in a broader landscape of nonperturbative quantum-gravity programs, we also compare it with Causal Dynamical Triangulations (CDT) \cite{CDT2001,CDT2005,CDT2024}, which defines a path-integral over causal simplicial complexes to recover continuum geometry. CDT's computational strength is evidenced by results like dynamical dimensional reduction and emergent de Sitter-like phases. By contrast, our framework works in a continuum Lorentzian language and promotes local scale factors $a_{\mu}$ to dynamical quantum fields. These differing starting points give complementary strengths and distinct challenges: CDT's discrete, numerical setup is powerful for probing nonperturbative phase structure but requires further control of continuum limits and matter coupling, while the scale-field formulation must develop a variationally complete action, explicit propagators, and a demonstrated closure of constraint relations. CDT's numerically observed scale-dependent and potentially nonlocal phenomena may help inform possible nonlocal or scale-dependent terms in effective actions for the $a_{\mu}$.

The physical picture is that of a scale measurement manifold: scaling coordinates and rescaled metrics provide finer resolution where quantum fluctuations are important, while the effective metric for macroscopic measurements is recovered when the scale factors become trivial. This viewpoint is complementary to other strategies. For example, Ho\v{r}ava-Lifshitz models \cite{Horava2009} improve ultraviolet behavior by imposing anisotropic scaling and higher-order spatial derivatives at the cost of explicitly breaking Lorentz invariance and restricting diffeomorphisms. Our framework maintains manifest covariance while proposing a mechanism for short-distance renormalization through intrinsic, direction-dependent scale factors. Ho\v{r}ava-Lifshitz targets perturbative renormalizability via controlled symmetry breaking, whereas the present work emphasizes a covariant, scaling geometric quantization.

Our proposal also differs from asymptotic-safety and functional-RG approaches by shifting emphasis from running couplings to running measurement scales, and it contrasts with noncommutative\cite{Connes1994}, causal-set\cite{sorkin2005,Loll2019}, tensor/group-field\cite{Gurau2012}, and holographic constructions \cite{Maldacena1999,RyuTakayanagi2006} by focusing on the quantization of measurement structure rather than on alternative kinematics or dual descriptions. Technically, this means that many standard tools (e.g., spin-network bases or explicit discrete pregeometries) are replaced here by an operator algebra and representation theory for the scale variables; establishing that algebra and its representations is therefore a central technical task.

To develop this framework into a complete theory, key next steps include: constructing a variationally complete action with consistent matter couplings and boundary terms; deriving propagators and performing power-counting and renormalization-group analyses to probe ultraviolet behavior rigorously; and characterizing the operator algebra and representation theory of \(\hat a_\mu\) and \(\hat b_\mu\) to ensure a stable, positive-definite scaling geometry and to identify or exclude pathological modes. In parallel, one should develop phenomenological outputs (modified dispersion relations, signatures in cosmology and black-hole thermodynamics, and other observational discriminants) and translate them into concrete, testable predictions that connect the scale-operator paradigm with low-energy field theory and experiment.

Finally, while the framework is presented here independently, it does not preclude compatibility or integration with existing approaches to quantum gravity. Rather than positioning this construction as a rival, we view it as a potentially integrable viewpoint that can interact with and enrich other established programmes. In the subsequent research on specific topics, further detailed comparisons and analyses will be conducted.

\section{Conclusion}
We have developed a framework that promotes local, direction-dependent scale measurements \(L_\alpha(X^\alpha)\) and their associated factors \(a_\alpha, b_\alpha\) to the primary quantum degrees of freedom, serving as the fundamental object for Planck-scale spacetime. The construction employs a two-tiered hierarchy of scale manifolds, a first-order manifold \((\mathcal{M}^{( a)},\hat g)\) with coordinates \(X^\alpha\) and a second-order amplitude manifold \((\mathcal{M}^{( b)},\tilde g)\) with coordinates \( a_\alpha\). And formulates differential geometry, field equations, and canonical quantization consistently on the first-order manifolds.

This scaling-based paradigm naturally implements a geometric renormalization-group flow for scale variables and admits intrinsic spacetime discreteness while preserving general covariance and a vierbein embedding. From the formalism follow several concrete physical consequences: a generalized uncertainty relation with scale-dependent coefficients; locally scaled Klein-Gordon and Dirac equations; generalized geodesic equations governing quantum fluctuation dynamics; and first-order local operator quantization of the scaling degrees of freedom. Importantly, the microscopic area operator constructed from the fundamental scaling fluctuations accommodates the Bekenstein-Hawking area law by state counting, providing a statistical-mechanical underpinning for horizon thermodynamics.

By formulating a quadratic action for the scale sector and performing a spectral decomposition, we obtain discrete modal degrees of freedom that can be quantized as harmonic oscillators, with requiring that the kinetic matrix $M$ is positive-definite. We show that quantized spacetime scale modes produce a discrete harmonic-oscillator zero-point spectrum whose magnitude is controlled by geometric renormalization-group flows, so that UV fixed points suppress individual zero-point contributions while IR discrete fixed points furnish a natural microscopic regulator and vacuum energy, suggesting a potential dynamical mechanism that could ameliorate the cosmological constant problem. The framework yields modal masses and frequencies determined by local scale factors, and it prescribes a consistent projection onto the physical spectral sector, enabling a Fock-space construction.

As a conclusion of this work, we have developed a comprehensive scaling-based framework for spacetime quantization that provides a concrete mathematical foundation for investigating Planck-scale geometry, while acknowledging that further development is required to establish a complete theory. The paper provides a set of definitions, demonstrates internal consistency, and presents physically motivated constructions that together form a systematic starting point for further mathematical development and phenomenological exploration. Several important issues remain to be addressed in future work, including but not limited to: extension to scale-curved substrate, full operator-theoretic treatment of second-order fluctuations, coupling to matter fields, closure of constraint algebras, rigorous analysis of the classical limit, comprehensive stability studies of the quantum vacuum, and predictions for observations.

\section*{Acknowledgements}
This work is supported by the National Natural Science Foundation of China with Grants No. 12475138 and 12147101, the Strategic Priority Research Program of Chinese Academy of Sciences (No. XDB34000000), and the Science and Technology Commission of Shanghai Municipality (23590780100).

\bibliographystyle{unsrt}

\end{CJK*}
\end{document}